\documentclass[a4paper,11pt]{article}

\usepackage{./auxi/jheppub}
\usepackage[utf8]{inputenc}
\usepackage[ngerman, english]{babel}
\usepackage[T1]{fontenc}
\usepackage[titletoc]{appendix}
\usepackage{titletoc}
\usepackage{graphicx}
\usepackage{subcaption} 
\usepackage{xcolor}
\usepackage{amsmath,amssymb,amsthm}
\usepackage{bm}
\usepackage{float}
\usepackage[format=plain,labelfont={bf,it},textfont=it]{caption}
\usepackage{multirow}
\usepackage{slashed} 
\usepackage{tabularx}
\usepackage{bbold}
\usepackage[shortlabels]{enumitem}
\usepackage{tikz}
\usetikzlibrary{positioning,arrows,calc}
\usetikzlibrary{arrows.meta}
\usetikzlibrary{decorations.pathmorphing}
\usetikzlibrary{decorations.markings}
\usetikzlibrary{shapes.geometric}
\usetikzlibrary{patterns}
\tikzset{
  snake it/.style={
    decorate, 
    decoration=snake,
    segment length=3
  }
}
\usepackage{xparse}
\usepackage{pdfpages} 
\usepackage{titlesec} 
\usepackage{fancyhdr} 
\usepackage{emptypage} 
\usepackage{dsfont} 

\usepackage[table]{xcolor}
\definecolor{DarkBlueGrey}{RGB}{76,94,107}
\definecolor{MediumBlueGrey}{RGB}{110,135,153}
\definecolor{LightBlueGrey}{RGB}{134,163,184}
\definecolor{VeryLightBlueGrey}{RGB}{242,249,255}
\definecolor{WCOrange}{RGB}{242,146,29}
\definecolor{VeryLightOrange}{RGB}{255,245,233}
\definecolor{SCRed}{RGB}{179,48,48}
\definecolor{VeryLightRed}{RGB}{255,239,239}
\definecolor{VertexColor}{RGB}{242,146,29}
\definecolor{GluonColor}{RGB}{255,172,172}
\definecolor{SEColor}{RGB}{134,163,184}

\definecolor{BGBox}{RGB}{255,254,230}
\definecolor{PlaneColor}{RGB}{230,230,230}
\definecolor{BlobColor}{RGB}{190,180,230}

\newcommand{\pd}{\partial}

\newcommand{\vev}[1]{\langle\, #1 \, \rangle}

\def\veps{\varepsilon}

\newcommand{\bR}{{\beta/R}}

\def\Cm{{\mathcal{C}}}

\def\Nm{{\mathcal{N}}}
\def\Om{{\mathcal{O}}}

\usepackage{bbm}

\def\veps{\varepsilon}

\newcommand\zb{{\bar{z}}}

\newif\ifstartcompletesineup
\newif\ifendcompletesineup
\pgfkeys{
    /pgf/decoration/.cd,
    start up/.is if=startcompletesineup,
    start up=true,
    start up/.default=true,
    start down/.style={/pgf/decoration/start up=false},
    end up/.is if=endcompletesineup,
    end up=true,
    end up/.default=true,
    end down/.style={/pgf/decoration/end up=false}
}
\pgfdeclaredecoration{complete sines}{initial}
{
    \state{initial}[
        width=+0pt,
        next state=upsine,
        persistent precomputation={
            \ifstartcompletesineup
                \pgfkeys{/pgf/decoration automaton/next state=upsine}
                \ifendcompletesineup
                    \pgfmathsetmacro\matchinglength{
                        0.5*\pgfdecoratedinputsegmentlength / (ceil(0.5* \pgfdecoratedinputsegmentlength / \pgfdecorationsegmentlength) )
                    }
                \else
                    \pgfmathsetmacro\matchinglength{
                        0.5 * \pgfdecoratedinputsegmentlength / (ceil(0.5 * \pgfdecoratedinputsegmentlength / \pgfdecorationsegmentlength ) - 0.499)
                    }
                \fi
            \else
                \pgfkeys{/pgf/decoration automaton/next state=downsine}
                \ifendcompletesineup
                    \pgfmathsetmacro\matchinglength{
                        0.5* \pgfdecoratedinputsegmentlength / (ceil(0.5 * \pgfdecoratedinputsegmentlength / \pgfdecorationsegmentlength ) - 0.4999)
                    }
                \else
                    \pgfmathsetmacro\matchinglength{
                        0.5 * \pgfdecoratedinputsegmentlength / (ceil(0.5 * \pgfdecoratedinputsegmentlength / \pgfdecorationsegmentlength ) )
                    }
                \fi
            \fi
            \setlength{\pgfdecorationsegmentlength}{\matchinglength pt}
        }] {}
    \state{downsine}[width=\pgfdecorationsegmentlength,next state=upsine]{
        \pgfpathsine{\pgfpoint{0.5\pgfdecorationsegmentlength}{0.5\pgfdecorationsegmentamplitude}}
        \pgfpathcosine{\pgfpoint{0.5\pgfdecorationsegmentlength}{-0.5\pgfdecorationsegmentamplitude}}
    }
    \state{upsine}[width=\pgfdecorationsegmentlength,next state=downsine]{
        \pgfpathsine{\pgfpoint{0.5\pgfdecorationsegmentlength}{-0.5\pgfdecorationsegmentamplitude}}
        \pgfpathcosine{\pgfpoint{0.5\pgfdecorationsegmentlength}{0.5\pgfdecorationsegmentamplitude}}
}
    \state{final}{}
}

\tikzset{
corner/.style={line width=1pt,dashed,draw=black,dash pattern=on 6pt off 4pt},
scalar/.style={line width=1pt,draw=black},
gluon/.style={line width=1pt,decorate, draw=GluonColor,
    decoration={complete sines,aspect=0,amplitude=1.25mm,segment length=1.5mm,start up,end up}},
gluontwo/.style={line width=1pt,decorate, draw=GluonColor,
    decoration={complete sines,aspect=0,amplitude=.7mm,segment length=1mm,start up,end up}},
ghost/.style={line width=1pt,loosely dotted,draw=black},
wilson/.style={line width=2pt,draw=black},
 }
\NewDocumentCommand\semiloop{O{black}mmmO{}O{above}}
{%
\draw[#1] let \p1 = ($(#3)-(#2)$) in (#3) arc (#4:({#4+180}):({0.5*veclen(\x1,\y1)})node[midway, #6] {#5};)
}

\newcommand\x{.75}

\newcommand{\BlackHole}{\begin{tikzpicture}[scale = .9, baseline={([yshift=-.5ex]current bounding box.center)}]
\draw[scalar,fill=LightBlueGrey] (0,1-.5) -- (0,3+.5) -- (1,2+.5) -- (1,0-.5) -- (0,1-.5);
\draw[scalar,fill=DarkBlueGrey] (6,1) -- (6,2) -- (6.5,1.5) -- (6.5,.5) -- (6,1);
\draw[scalar] (0,1-.5) .. controls (1.5,.8) .. (6,1);
\draw[scalar] (0,3+.5) .. controls (1.5,2.5) .. (6,2);
\draw[scalar] (1,2+.5) .. controls (2.25,1.75) .. (6.5,1.5);
\draw[scalar] (1,0-.5) .. controls (2.25,.25) .. (6.5,.5);
\draw[black,fill=black] (.3,1.8) circle (.12);
\draw[black,fill=black] (.7,.8+.25) circle (.12);
\draw[scalar,very thick] (.3,1.8) .. controls (3.5,1.8) and (3.5,.8+.25) .. (.65,.8+.25);
\draw[scalar] (0,1-.5) -- (0,3+.5) -- (1,2+.5) -- (1,0-.5) -- (0,1-.5);
\node[LightBlueGrey] at (.5,-.5-.5) {CFT at finite $T$};
\node[DarkBlueGrey] at (6.5,0) {Horizon};
\end{tikzpicture}}
\newcommand{\ThermalCircle}{\begin{tikzpicture}[scale = 1.2,baseline={([yshift=-3ex]current bounding box.center)}]
\draw[scalar] (-.1,0) ellipse (.5cm and 1cm);
\draw[wilson,LightBlueGrey] (1.4,0) ellipse (.5cm and 1cm);
\draw[ghost] (3,0) ellipse (.5cm and 1cm);
\draw[scalar] (-.1,1) -- (3,1);
\draw[scalar] (-.1,-1) -- (3,-1);
\draw[scalar,-Latex] (1.5,1.2) -- (3,1.2);
\node[] at (2.25,1.5) {\footnotesize $\vec{x}$};
\draw[scalar,-Latex] (4.95-1.25,-.5) .. controls (5.05-1.25,-.16) and (5.05-1.25,.16) .. (4.95-1.25,.5);
\node[] at (5.25-1.25,0) {\footnotesize $\tau$};
\node[LightBlueGrey] at (1.4,-1.35) {\footnotesize $|\vec{x}|=0$};
\draw[fill=DarkBlueGrey,DarkBlueGrey] (1.82,.5) circle (.075);
\node[DarkBlueGrey] at (2.2,.65) {\footnotesize $\phi (\tau)$};
\draw[fill=DarkBlueGrey,DarkBlueGrey] (1.82,-.5) circle (.075);
\node[DarkBlueGrey] at (2.2,-.65) {\footnotesize $\phi (0)$};
\path[clip] (3,-1) rectangle (5,2);
\draw[scalar] (3,0) ellipse (.5cm and 1cm);
\end{tikzpicture}}

\newcommand{\BouncingSingularities}{\begin{tikzpicture}[scale = .9, baseline={([yshift=-.5ex]current bounding box.center)}]
\draw[scalar,-Latex] (0,0) -- (6.2,0);
\draw[scalar,-Latex] (3,-2) -- (3,2);
\node[] at (5.7+.2,1.7+.25) {$\tau$};
\draw[scalar] (5.4+.2,1.9+.2) -- (5.4+.2,1.45+.2);
\draw[scalar] (5.4+.2,1.45+.2) -- (6+.2,1.45+.2);
\draw[fill=SCRed,SCRed] (.5,0) circle (.075);
\draw[fill=SCRed,SCRed] (1.75,0) circle (.075);
\draw[fill=SCRed,SCRed] (3,0) circle (.075);
\draw[fill=SCRed,SCRed] (3+1.25,0) circle (.075);
\draw[fill=SCRed,SCRed] (5.5,0) circle (.075);
\draw[fill=WCOrange,WCOrange] (1.125,.625) circle (.075);
\draw[fill=WCOrange,WCOrange] (1.125+1.25,.625) circle (.075);
\draw[fill=WCOrange,WCOrange] (1.125+2.5,.625) circle (.075);
\draw[fill=WCOrange,WCOrange] (1.125+3.75,.625) circle (.075);
\draw[fill=WCOrange,WCOrange] (1.125,-.625) circle (.075);
\draw[fill=WCOrange,WCOrange] (1.125+1.25,-.625) circle (.075);
\draw[fill=WCOrange,WCOrange] (1.125+2.5,-.625) circle (.075);
\draw[fill=WCOrange,WCOrange] (1.125+3.75,-.625) circle (.075);
\node[] at (6.85,0) {$\text{Re}(\tau)$};
\node[] at (3,2.4) {$\text{Im}(\tau)$};
\end{tikzpicture}}

\newcommand{\BulkToBoundaryPropagator}{\begin{tikzpicture}[scale = 1.5, baseline={([yshift=-.5ex]current bounding box.center)}]
\draw[scalar] (0,0) circle (.5);
\draw[scalar] (.5,0) -- (0,0);
\draw[fill=white] (0,0) circle (.05);
\draw[fill=white] (.5,0) circle (.05);
\end{tikzpicture}}
\newcommand{\BulkToBulkPropagator}{\begin{tikzpicture}[scale = 1.5, baseline={([yshift=-.5ex]current bounding box.center)}]
\draw[scalar] (0,0) circle (.5);
\draw[scalar] (0,.25) -- (0,-.25);
\draw[fill=white] (0,.25) circle (.05);
\draw[fill=white] (0,-.25) circle (.05);
\end{tikzpicture}}
\newcommand{\VertexOne}{\begin{tikzpicture}[scale = 1.5, baseline={([yshift=-.5ex]current bounding box.center)}]
\draw[scalar] (0,0) circle (.5);
\draw[scalar] (0,0) -- (.2,.2);
\draw[scalar] (0,0) -- (.2,-.2);
\draw[scalar,SCRed,fill=SCRed] (0,0) circle (.035);
\node[SCRed] at (-.25,0) {\tiny $(1)$};
\end{tikzpicture}}
\newcommand{\VertexTwo}{\begin{tikzpicture}[scale = 1.5, baseline={([yshift=-.5ex]current bounding box.center)}]
\draw[scalar] (0,0) circle (.5);
\draw[scalar] (0,0) -- (.2,.2);
\draw[scalar] (0,0) -- (.2,-.2);
\draw[scalar,SCRed,fill=SCRed] (0,0) circle (.035);
\node[SCRed] at (-.25,0) {\tiny $(2)$};
\end{tikzpicture}}
\newcommand{\WittenDiagramBH}{\begin{tikzpicture}[scale = 1.5, baseline={([yshift=-.5ex]current bounding box.center)}]
\draw[scalar] (0,0) circle (.5);
\draw[scalar,fill=black] (0,0) circle (.15);
\draw[scalar] (.37,.33) .. controls (.25,.15) and (.25,-.15) .. (.37,-.33);
\draw[fill=white] (.37,.33) circle (.075);
\draw[fill=white] (.37,-.33) circle (.075);
\end{tikzpicture}}
\newcommand{\WittenDiagramOne}{\begin{tikzpicture}[scale = 1.5, baseline={([yshift=-.5ex]current bounding box.center)}]
\draw[scalar] (0,0) circle (.5);
\draw[scalar] (.37,.33) .. controls (.25,.15) and (.25,-.15) .. (.37,-.33);
\draw[fill=white] (.37,.33) circle (.075);
\draw[fill=white] (.37,-.33) circle (.075);
\end{tikzpicture}}
\newcommand{\WittenDiagramTwo}{\begin{tikzpicture}[scale = 1.5, baseline={([yshift=-.5ex]current bounding box.center)}]
\draw[scalar] (0,0) circle (.5);
\draw[scalar,SCRed,fill=SCRed] (0,0) circle (.035);
\node[SCRed] at (-.25,0) {\tiny $(1)$};
\draw[scalar] (.37,.33) -- (0,0);
\draw[scalar] (0,0) -- (.37,-.33);
\draw[fill=white] (.37,.33) circle (.075);
\draw[fill=white] (.37,-.33) circle (.075);
\end{tikzpicture}}
\newcommand{\WittenDiagramThree}{\begin{tikzpicture}[scale = 1.5, baseline={([yshift=-.5ex]current bounding box.center)}]
\draw[scalar] (0,0) circle (.5);
\draw[scalar,SCRed,fill=SCRed] (0,0) circle (.035);
\node[SCRed] at (-.25,0) {\tiny $(2)$};
\draw[scalar] (.37,.33) -- (0,0);
\draw[scalar] (0,0) -- (.37,-.33);
\draw[fill=white] (.37,.33) circle (.075);
\draw[fill=white] (.37,-.33) circle (.075);
\end{tikzpicture}}
\newcommand{\WittenDiagramFour}{\begin{tikzpicture}[scale = 1.5, baseline={([yshift=-.5ex]current bounding box.center)}]
\draw[scalar] (0,0) circle (.5);
\draw[scalar,SCRed,fill=SCRed] (0,.2) circle (.035);
\node[SCRed] at (-.225,.2) {\tiny $(1)$};
\draw[scalar,SCRed,fill=SCRed] (0,-.2) circle (.035);
\node[SCRed] at (-.225,-.2) {\tiny $(1)$};
\draw[scalar] (.37,.33) -- (0,.2);
\draw[scalar] (0,-.2) -- (0,.2);
\draw[scalar] (0,-.2) -- (.37,-.33);
\draw[fill=white] (.37,.33) circle (.075);
\draw[fill=white] (.37,-.33) circle (.075);
\end{tikzpicture}}
\newcommand{\WittenDiagramFive}{\begin{tikzpicture}[scale = 1.5, baseline={([yshift=-.5ex]current bounding box.center)}]
\draw[scalar] (0,0) circle (.5);
\draw[scalar,SCRed,fill=SCRed] (0,0) circle (.035);
\node[SCRed] at (-.25,0) {\tiny $(3)$};
\draw[scalar] (.37,.33) -- (0,0);
\draw[scalar] (0,0) -- (.37,-.33);
\draw[fill=white] (.37,.33) circle (.075);
\draw[fill=white] (.37,-.33) circle (.075);
\end{tikzpicture}}
\newcommand{\WittenDiagramSix}{\begin{tikzpicture}[scale = 1.5, baseline={([yshift=-.5ex]current bounding box.center)}]
\draw[scalar] (0,0) circle (.5);
\draw[scalar,SCRed,fill=SCRed] (0,.2) circle (.035);
\node[SCRed] at (-.225,.2) {\tiny $(1)$};
\draw[scalar,SCRed,fill=SCRed] (0,-.2) circle (.035);
\node[SCRed] at (-.225,-.2) {\tiny $(2)$};
\draw[scalar] (.37,.33) -- (0,.2);
\draw[scalar] (0,-.2) -- (0,.2);
\draw[scalar] (0,-.2) -- (.37,-.33);
\draw[fill=white] (.37,.33) circle (.075);
\draw[fill=white] (.37,-.33) circle (.075);
\end{tikzpicture}}
\newcommand{\WittenDiagramSeven}{\begin{tikzpicture}[scale = 1.5, baseline={([yshift=-.5ex]current bounding box.center)}]
\draw[scalar] (0,0) circle (.5);
\draw[scalar,SCRed,fill=SCRed] (.1,.25) circle (.035);
\node[SCRed] at (-.15,.25) {\tiny $(1)$};
\draw[scalar,SCRed,fill=SCRed] (.1,-.25) circle (.035);
\node[SCRed] at (-.15,-.25) {\tiny $(1)$};
\draw[scalar,SCRed,fill=SCRed] (0,0) circle (.035);
\node[SCRed] at (-.22,0) {\tiny $(1)$};
\draw[scalar] (.37,.33) -- (.1,.25);
\draw[scalar] (.1,-.25) -- (0,0);
\draw[scalar] (.1,.25) -- (0,0);
\draw[scalar] (.1,-.25) -- (.37,-.33);
\draw[fill=white] (.37,.33) circle (.075);
\draw[fill=white] (.37,-.33) circle (.075);
\end{tikzpicture}}

\newcommand{\AnalyticStructureTauPoles}{\begin{tikzpicture}[scale = 1, baseline={([yshift=-.4ex]current bounding box.center)}]
\draw[scalar,-Latex] (0,0) -- (6.2,0);
\draw[scalar,-Latex] (3,-2.25) -- (3,2.25);
\node[] at (5.7+.2,1.7+.25) {$\tau'$};
\draw[scalar] (5.4+.2,1.9+.2) -- (5.4+.2,1.45+.2);
\draw[scalar] (5.4+.2,1.45+.2) -- (6+.2,1.45+.2);
\draw[fill=black,black] (4+.5,1-.25) circle (.075);
\node[] at (4.7+.35,1-.25) {$\tau$};
\node[MediumBlueGrey] at (4.7-.8,1-.25) {$\Cm_0$};
\draw[scalar,MediumBlueGrey] (4+.5,1-.25) circle (.3);
\draw[scalar,MediumBlueGrey,-Latex] (3.62+.1+.5,1.00-.25-.25+.35) -- (3.62+.1+.5,0.90-.25-.25+.15);
\draw[scalar,SCRed] (1,0) circle (.3);
\draw[scalar,SCRed,-Latex] (3.62-4+1+.1,1.09-1.25+.15) -- (3.62-4+1+.1,1.08-1.25+.35);
\draw[fill=SCRed,SCRed] (1,0) circle (.075);
\draw[scalar,SCRed] (3,0) circle (.3);
\draw[scalar,SCRed,-Latex] (3.62-4+3+.1,1.09-1.25+.15) -- (3.62-4+3+.1,1.08-1.25+.35);
\draw[fill=SCRed,SCRed] (3,0) circle (.075);
\draw[scalar,SCRed] (2,0) circle (.3);
\draw[scalar,SCRed,-Latex] (3.62-4+2+.1,1.09-1.25+.15) -- (3.62-4+2+.1,1.08-1.25+.35);
\draw[fill=SCRed,SCRed] (2,0) circle (.075);
\draw[scalar,SCRed] (4,0) circle (.3);
\draw[scalar,SCRed,-Latex] (3.62-4+4+.1,1.09-1.25+.15) -- (3.62-4+4+.1,1.08-1.25+.35);
\draw[fill=SCRed,SCRed] (4,0) circle (.075);
\draw[scalar,SCRed] (5,0) circle (.3);
\draw[scalar,SCRed,-Latex] (3.62-4+5+.1,1.09-1.25+.15) -- (3.62-4+5+.1,1.08-1.25+.35);
\draw[fill=SCRed,SCRed] (5,0) circle (.075);
\end{tikzpicture}}
\newcommand{\AnalyticStructureTauPolesV}{\begin{tikzpicture}[scale = 1, baseline={([yshift=-.4ex]current bounding box.center)}]
\draw[scalar,-Latex] (0,0) -- (6.2,0);
\draw[scalar,-Latex] (3,-2.25) -- (3,2.25);
\node[] at (5.7+.2,1.7+.25) {$\tau'$};
\draw[scalar] (5.4+.2,1.9+.2) -- (5.4+.2,1.45+.2);
\draw[scalar] (5.4+.2,1.45+.2) -- (6+.2,1.45+.2);
\draw[fill=black,black] (4+.5,1-.25) circle (.075);
\node[] at (4.7+.35,1-.25) {$\tau$};
\node[MediumBlueGrey] at (4.7-.8,1-.25) {$\Cm_0$};
\draw[scalar,MediumBlueGrey] (4+.5,1-.25) circle (.3);
\draw[scalar,MediumBlueGrey,-Latex] (3.62+.1+.5,1.00-.25-.25+.35) -- (3.62+.1+.5,0.90-.25-.25+.15);
\draw[scalar,SCRed] (1,0) circle (.3);
\draw[scalar,SCRed,-Latex] (3.62-4+1+.1,1.09-1.25+.15) -- (3.62-4+1+.1,1.08-1.25+.35);
\draw[fill=SCRed,SCRed] (1,0) circle (.075);
\draw[scalar,SCRed] (3,0) circle (.3);
\draw[scalar,SCRed,-Latex] (3.62-4+3+.1,1.09-1.25+.15) -- (3.62-4+3+.1,1.08-1.25+.35);
\draw[fill=SCRed,SCRed] (3,0) circle (.075);
\draw[scalar,SCRed] (2,0) circle (.3);
\draw[scalar,SCRed,-Latex] (3.62-4+2+.1,1.09-1.25+.15) -- (3.62-4+2+.1,1.08-1.25+.35);
\draw[fill=SCRed,SCRed] (2,0) circle (.075);
\draw[scalar,SCRed] (4,0) circle (.3);
\draw[scalar,SCRed,-Latex] (3.62-4+4+.1,1.09-1.25+.15) -- (3.62-4+4+.1,1.08-1.25+.35);
\draw[fill=SCRed,SCRed] (4,0) circle (.075);
\draw[scalar,SCRed] (5,0) circle (.3);
\draw[scalar,SCRed,-Latex] (3.62-4+5+.1,1.09-1.25+.15) -- (3.62-4+5+.1,1.08-1.25+.35);
\draw[fill=SCRed,SCRed] (5,0) circle (.075);
\draw[scalar,SCRed] (1,1.5) circle (.3);
\draw[scalar,SCRed,-Latex] (3.62-4+1+.1,1.09-1.25+1.5+.15) -- (3.62-4+1+.1,1.08-1.25+1.5+.35);
\draw[fill=SCRed,SCRed] (1,1.5) circle (.075);
\draw[scalar,SCRed] (3,1.5) circle (.3);
\draw[scalar,SCRed,-Latex] (3.62-4+3+.1,1.09-1.25+1.5+.15) -- (3.62-4+3+.1,1.08-1.25+1.5+.35);
\draw[fill=SCRed,SCRed] (3,1.5) circle (.075);
\draw[scalar,SCRed] (2,1.5) circle (.3);
\draw[scalar,SCRed,-Latex] (3.62-4+2+.1,1.09-1.25+1.5+.15) -- (3.62-4+2+.1,1.08-1.25+1.5+.35);
\draw[fill=SCRed,SCRed] (2,1.5) circle (.075);
\draw[scalar,SCRed] (4,1.5) circle (.3);
\draw[scalar,SCRed,-Latex] (3.62-4+4+.1,1.09-1.25+1.5+.15) -- (3.62-4+4+.1,1.08-1.25+1.5+.35);
\draw[fill=SCRed,SCRed] (4,1.5) circle (.075);
\draw[scalar,SCRed] (5,1.5) circle (.3);
\draw[scalar,SCRed,-Latex] (3.62-4+5+.1,1.09-1.25+1.5+.15) -- (3.62-4+5+.1,1.08-1.25+1.5+.35);
\draw[fill=SCRed,SCRed] (5,1.5) circle (.075);
\draw[scalar,SCRed] (1,-1.5) circle (.3);
\draw[scalar,SCRed,-Latex] (3.62-4+1+.1,1.09-1.25-1.5+.15) -- (3.62-4+1+.1,1.08-1.25-1.5+.35);
\draw[fill=SCRed,SCRed] (1,-1.5) circle (.075);
\draw[scalar,SCRed] (3,-1.5) circle (.3);
\draw[scalar,SCRed,-Latex] (3.62-4+3+.1,1.09-1.25-1.5+.15) -- (3.62-4+3+.1,1.08-1.25-1.5+.35);
\draw[fill=SCRed,SCRed] (3,-1.5) circle (.075);
\draw[scalar,SCRed] (2,-1.5) circle (.3);
\draw[scalar,SCRed,-Latex] (3.62-4+2+.1,1.09-1.25-1.5+.15) -- (3.62-4+2+.1,1.08-1.25-1.5+.35);
\draw[fill=SCRed,SCRed] (2,-1.5) circle (.075);
\draw[scalar,SCRed] (4,-1.5) circle (.3);
\draw[scalar,SCRed,-Latex] (3.62-4+4+.1,1.09-1.25-1.5+.15) -- (3.62-4+4+.1,1.08-1.25-1.5+.35);
\draw[fill=SCRed,SCRed] (4,-1.5) circle (.075);
\draw[scalar,SCRed] (5,-1.5) circle (.3);
\draw[scalar,SCRed,-Latex] (3.62-4+5+.1,1.09-1.25-1.5+.15) -- (3.62-4+5+.1,1.08-1.25-1.5+.35);
\draw[fill=SCRed,SCRed] (5,-1.5) circle (.075);
\end{tikzpicture}}

\begin{document} 

\thispagestyle{empty}

\vspace*{-.6in}
\begin{flushright}
    DESY-25-139\\
    YITP-SB-2025-16
\end{flushright}

\vspace{1cm}
{\large
\begin{center}
    {\Large \bf Analytic thermal bootstrap meets holography
    }\\
\end{center}}

\vspace{0.5cm}

\begin{center}
    {Julien Barrat,$^{a,}\footnote{julien.barrat@desy.de}$ Deniz N. Bozkurt,$^{b,}\footnote{deniz.bozkurt@desy.de}$ Enrico Marchetto,$^{a,}\footnote{enrico.marchetto@desy.de}$ \\ Alessio Miscioscia,$^{a,c,}\footnote{alessio.miscioscia@stonybrook.edu}$ and Elli Pomoni$^{a,}\footnote{elli.pomoni@desy.de}$}\\[0.5cm] 
    { \small
    $^{a}$Deutsches Elektronen-Synchrotron DESY, Notkestr. 85, 22607 Hamburg, Germany\\
    \small $^{b}$ Institut f\"ur Theoretische Physik, Universit\"at Hamburg, Luruper Chaussee 149,
    22607 Hamburg, Germany\\
    \small $^{c}$
    C. N. Yang Institute for Theoretical Physics, Stony Brook University, Stony Brook, NY 11794, USA
    }
\vspace{1cm} 

   \bf Abstract
\end{center}

\begin{abstract}
\noindent
We compute thermal holographic correlators by combining their analytic structure with the Kubo--Martin--Schwinger (KMS) condition and multi-stress tensor OPE coefficients determined from the dual AdS description.
We focus on two-point functions of identical scalar operators with integer conformal dimensions at zero spatial separation.
In the black brane background, we show explicitly that holographic two-point functions split into three contributions: a principal one, computed exactly, plus regularized and arcs contributions, both approximated through the use of OPE coefficients asymptotics. For $\Delta_\phi=3$, we show that the principal contribution agree with good approximation with the numerical solution of the bulk wave equation.
Moreover, we demonstrate that the expansion in generalized free field correlators proposed in~\cite{Barrat:2025nvu} admits a natural interpretation in terms of Witten diagrams.
Finally, we initiate the study of thermal correlators in  the spherically symmetric black hole background, computing their principal contributions.

\end{abstract}

\newpage

\setcounter{tocdepth}{2}

\noindent

\tableofcontents

\setcounter{page}{1}

\section{Introduction}
\label{sec:Introduction}

Black holes are among the most fascinating objects in gravitational physics, providing natural laboratories for exploring the interplay between geometry, thermodynamics, and quantum theory.
Within the framework of the AdS/CFT correspondence~\cite{Maldacena:1997re}, black holes in asymptotically Anti-de Sitter (AdS) space are understood as holographic duals of thermal states in  conformal field theories (CFTs)~\cite{Witten:1998zw,Aharony:1999ti}.
In particular, planar black holes (or black branes) in AdS$_{d+1}$ spacetime correspond to CFTs at finite temperature, or equivalently CFTs on $S^1_\beta\times\mathbb{R}^{d-1}$ with inverse temperature $\beta=1/T$~\cite{Matsubara:1955ws}, while spherically symmetric black holes correspond to CFTs on $S^1_\beta\times S^{d-1}_R$, with $R$ the radius of the sphere.
A striking manifestation of this duality is the Hawking–Page transition between thermal AdS and the spherical black hole~\cite{Hawking:1982dh}, which on the CFT side corresponds to the confinement–deconfinement transition~\cite{Witten:1998zw}.\footnote{This correspondence is subtle due to the presence of complex saddles in the partition function, which can lead to \textit{delayed deconfinement}~\cite{Copetti:2020dil}.}

Since its modern revival~\cite{Rattazzi:2008pe}, the conformal bootstrap has become an indispensable tool for studying and classifying CFTs~\cite{Poland:2018epd}.
The analytic bootstrap program at zero temperature marked a major conceptual advance by revealing the underlying simplicity of CFT data in the large-spin regime~\cite{Fitzpatrick:2012yx,Komargodski:2012ek,Alday:2016njk,Caron-Huot:2017vep,Simmons-Duffin:2017nub}.
Building on these insights, recent progress has extended these techniques to settings where conformal symmetry is partially broken~\cite{Liendo:2012hy,Billo:2016cpy,Mazac:2018biw,Barrat:2022psm,Bianchi:2022ppi,Cuomo:2024vfk}.
For CFTs at finite temperature, the bootstrap strategy relies on two central principles~\cite{Iliesiu:2018fao}: (1) the operator product expansion (OPE) remains valid locally, allowing correlation functions to be expressed in terms of one-point function OPE coefficients~\cite{Katz:2014rla,Petkou:2018ynm}, (2) thermal correlators satisfy a periodicity relation formulated in the form of the Kubo–Martin–Schwinger (KMS) condition along the thermal circle~\cite{Kubo:1957mj,Martin:1959jp}, which plays the role of a crossing equation by relating the OPE data in one channel to an infinite sum of operators in the other~\cite{El-Showk:2011yvt}.
These constraints have been successfully implemented numerically, most notably in the determination of thermal one-point functions in $3d$ $\mathrm{O}(N)$ models~\cite{Iliesiu:2018zlz,Barrat:2025wbi}.
On the analytic side, a framework based on the Lorentzian inversion formula and dispersion relations has recently been developed for perturbative CFTs~\cite{Alday:2020eua,Barrat:2025nvu}.
In particular, in the zero spatial separation limit, thermal two-point functions of identical scalars can systematically be expanded in generalized free field (GFF) correlators, providing strong analytic control.

\begin{figure}[t]
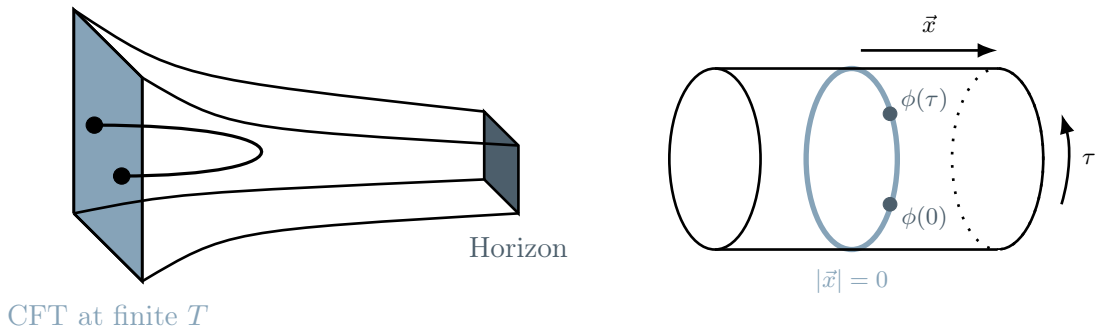

    \centering
    \begin{tabularx}{\textwidth}{lr}
        \BlackHole & \qquad \ThermalCircle
    \end{tabularx}
    \caption{Depiction of the setup considered in this paper.  
    \textbf{Left:} In the AdS/CFT framework, we study correlation functions of two scalar operators in a finite temperature CFT dual to a black hole background. The connecting line represents the propagator, which is non-trivial due to the presence of the horizon. 
    \textbf{Right:} We focus on the case where both operators lie on the same thermal circle, corresponding to zero spatial separation.}
    \label{fig:Setup}
\end{figure}

Parallel developments have taken place on the gravitational side, focusing on boundary correlators in black hole backgrounds (schematically illustrated in Figure~\ref{fig:Setup}). On the gravity side,\footnote{Weakly-coupled gravity corresponds to strong coupling on the CFT side.
In this paper we consider CFTs in the large $N$ limit with infinite 't~Hooft coupling $\lambda = g_\text{YM}^2 N$ (see Section \ref{subsec:StringyCorrectionsBB} for comments on corrections in $1\lambda$).} it has been established that for identical scalar operators that the spectrum consists of double-trace states together with an infinite tower of multi-stress tensor exchanges, corresponding to gravitons that generate the non-trivial background~\cite{Fitzpatrick:2014vua}.
In momentum space, closed-form expressions for these correlators have been obtained in terms of the $4d$ $\mathcal{N}=2$ partition functions in the Nekrasov–Shatashvili limit, allowing their analytic structure to be explored~\cite{Aminov:2020yma,Dodelson:2022yvn,Dodelson:2023vrw,Bhattacharya:2025vyi,Grozdanov:2025ulc}.
In position space, analytic progress has been made to understand the contributions of the multi-stress tensors for non-integer external dimensions $\Delta_\phi$, whereas the treatment of double-trace operators remains more delicate due to subtleties in implementing boundary conditions~\cite{Fitzpatrick:2019zqz,Rodriguez-Gomez:2021pfh,Parisini:2022wkb,Karlsson:2022osn,Parisini:2023nbd,Esper:2023jeq,Ceplak:2024bja}.

Combining the lessons from both sides of the AdS/CFT correspondence, namely by using the asymptotic form of multi-stress tensor coefficients together with the KMS condition, the double-trace sector has been approximated for non-integer $\Delta_\phi$~\cite{Buric:2025anb,Niarchos:2025cdg,Buric:2025fye}.
However, at integer values of $\Delta_\phi$ this approach breaks down because the contributions from double-trace operators and from graviton exchanges overlap in such a way that the divergent part of their coefficients must cancel precisely for consistency.
This delicate cancellation has so far obstructed a complete determination of two-point functions at integer $\Delta_\phi$.
Complementing these analytic results, numerical studies of the scalar wave equation in black hole backgrounds have been carried out both in momentum and in position space~\cite{Bonelli:2022ten,Parisini:2023nbd,Dodelson:2022yvn}.

\bigskip

In this work, we employ analytic bootstrap techniques to study two-point functions of scalar operators with integer scaling dimensions $\Delta_\phi$ in AdS black brane and spherical black hole backgrounds.
Specializing to the zero spatial separation limit depicted in Figure~\ref{fig:Setup}, we show that the correlators can be obtained by considering three contributions: (1) the KMS inversion of the low-lying multi-stress tensor operators (satisfying $\Delta<2\Delta_\phi$), (2) an infinite number of multi-stress tensor operators (satisfying $\Delta\geq\Delta_\phi$) which can be approximated using the asymptotic form of the thermal OPE coefficients, similarly to what done in ~\cite{Buric:2025anb,Buric:2025fye}, (3) and arc contributions, responsible for cancelling spurious poles created by the resummation of the infinite tower of operators in (2). These spurious poles are the bouncing singularities.
We demonstrate how all three terms can be computed (or approximated when necessary) by making use of the following expansion in terms of GFF correlators:
\small
\begin{equation}
   \langle \phi(\tau) \phi(0)\rangle_\beta
    =
    \sum_{\Delta < 2\Delta_\phi} \frac{a_\Delta}{\beta^{\Delta}} g_\text{GFF}\left(\Delta_\phi-\frac{\Delta}{2},\tau\right)
    - 2
    \sum_{\Delta \geq 2\Delta_\phi} \frac{  
    \text{Res}_{\Delta_\phi} a_\Delta}{\beta^{\Delta}} g^{(1,0)}_\text{GFF}\left(\Delta_\phi-\frac{\Delta}{2},\tau\right)
    +
    g_{\text{arcs}}(\tau) \,,
    \label{eq:IntroEq}
\end{equation}%
\normalsize
where $g_\text{GFF}^{(1,0)} (\Delta,\tau)$ is the derivative of the GFF correlator with respect to scaling dimensions while $\text{Res}_{\Delta_\phi} a_{\Delta}$ is the residue of the OPE coefficient at specific integer $\Delta_\phi$.
Importantly, the only input needed consists of multi-stress tensor OPE coefficients: a finite number for the first term and an infinite number for the second one.
The first sum encodes the information about the leading operators in the OPE, and for this reason we expect it to be dominant in real $\tau$ (imaginary time) around $\tau \sim 0$ and $\tau \sim \beta$.
We calculate this contribution exactly for the black brane and spherically symmetric black hole.
The second sum generates logarithmic terms and we treat it by approximating the multi-stress tensor OPE coefficient by its asymptotic values.
Finally, the arcs admit an elegant formulation in terms of GFF correlators with shifted $\tau$, as we show in detail in the paper.
For the black brane we compare our results to numerical solutions of the scalar wave equation and find good agreement, surprisingly revealing that the first term in~\eqref{eq:IntroEq} seems to be dominant everywhere for real $\tau$.
We also present evidence that the analytic bootstrap expansion presented above admits a natural interpretation as an expansion in graviton modes around the thermal AdS background.
Finally, we comment on how higher-curvature corrections also fit into our framework by appropriately modifying the multi-stress tensor OPE coefficients.

\bigskip

The remainder of the paper is organized as follows.  
Section~\ref{sec:BlackBraneBackground} reviews the thermal analytic bootstrap for identical scalars at zero spatial separation and applies it to black brane backgrounds.
We compare the resulting correlators with numerical holographic solutions and provide an AdS interpretation in terms of Witten diagrams, computed explicitly up to second order in graviton modes.
Additionally, we show that higher-curvature corrections can also be captured by our method.
Section~\ref{sec:BlackHoleBackground} extends the analysis to spherical black holes, deriving the first term in \eqref{eq:IntroEq} by using a new dispersion relation for CFTs on $S^1_\beta \times S^{d-1}_R$.
Finally, Section~\ref{sec:Conclusions} summarizes our results and discusses possible extensions.
A set of appendices provides technical details and supplementary material. In Appendix~\ref{app:SphericalGFF} we report the details of the derivation of the GFF correlator on the manifold $S^{1}_\beta \times S^{d-1}_R$.
In Appendix~\ref{app:ONModelAtLargeN} we provide results for the thermal correlator in the 3$d$ large $N$ vector model.
In Appendix~\ref{app:TheAsymptoticModel} we comment on the applicability of our method for the case of $\Delta_\phi$ being half-integer.
We show in particular how the results of~\cite{Buric:2025anb} for the asymptotic model can be interpreted in the analytic bootstrap formalism of~\cite{Barrat:2025nvu}.
In Appendix~\ref{app:NumericalSolutionOfTheWaveEquation} we provide details on the numerical computation of the non-perturbative thermal correlators through holography, i.e., by solving the wave equation in the AdS black brane background.
Finally, in Appendix~\ref{app:DetailsOnWittenDiagrams}, we report additional details needed to perform the Witten diagrams computation.

\bigskip

\textbf{Note added:}{ \emph{In the first version of this preprint the contributions coming from the regularization of the diverging OPE coefficients were not taken into account.
We updated this paper with a systematic way to approximate this contribution, together with arcs as a natural consequence.
We thank Simon Caron-Huot for bringing this to our attention.
}}

\section{Black brane background}
\label{sec:BlackBraneBackground}

This section is dedicated to the analysis of holographic two-point functions in the presence of a black brane background.
We start by briefly reviewing the analytic bootstrap framework for thermal correlators established in~\cite{Barrat:2025nvu} from the CFT side, as well as the specific features of the holographic setup arising on the weakly-coupled gravity side, such as the spectrum of operators and the OPE coefficients of the multi-stress tensor sector.
Specializing to the four-dimensional case, we then use these techniques to derive two-point functions of scalar operators with integer scaling dimension $\Delta_\phi$ in the zero spatial separation limit, complementing the kinematic result with dynamical input from holography.
The dominant piece in the OPE is determined exactly, while the terms arising from poles in the multi-stress tensor OPE coefficients is approximated using the asymptotic value of the latter.
We argue that the arc contributions can then be systematically determined. In particular, we show that the  framework of~\cite{Barrat:2025nvu} can be used to study the bouncing singularities of the multi-stress tensor sector, for which we find a perfect match with previous studies \cite{Fidkowski:2003nf,Ceplak:2024bja}.
Moreover, for $\Delta_\phi=3$ we compare the exactly computed first part of the thermal correlator to a numerical solution of the wave equation, following the method of~\cite{Parisini:2023nbd}, and find very good agreement.
We then provide an interpretation of our correlators in terms of Witten diagrams, performing a small temperature expansion of the bulk metric.
In this context, we find a natural splitting between perturbative and non-perturbative sectors of the correlator, with the perturbative part matching our bootstrap results for integer $\Delta_\phi$, while we associate the non-perturbative effects to the arc terms.
We conclude by showing that higher-curvature corrections can be incorporated in the bootstrap framework.

\subsection{Thermal two-point functions}
\label{subsec:ThermalTwoPointFunctions}

We study two-point functions of identical scalar operators at finite temperature.
We express the correlators in the Matsubara (or imaginary time) formalism~\cite{Matsubara:1955ws}, which amounts to considering the CFT correlator on $S^1_\beta \times \mathbb{R}^{d-1}$.
In this formalism, the signature is Euclidean and $\beta = 1/T$ is both the inverse temperature of the system and the length of the thermal circle. 
It is convenient to define 
\begin{equation}
    g(\tau,x) = \vev{\phi(0,0)\, \phi(\tau,x)}_{S^1_\beta \times \mathbb{R}^{d-1}}\,,
    \label{eq:gDefinition}
\end{equation}
with $x \equiv |\vec{x}|$.
In this work we mainly focus on the case where the two operators are inserted at zero spatial separation, i.e., $x = 0$. 
To streamline the notation we define $g(\tau) \equiv g(\tau,0)$.

Thermal two-point functions satisfy a number of structural properties, which we use as \textit{bootstrap axioms} of the problem.  We summarize them in the following.

\paragraph{OPE.} 
The OPE remains valid on the geometry $S^1_\beta \times \mathbb{R}^{d-1}$ with the limitation that it converges only within a ball of maximal radius, i.e., $\tau^2 + x^2 < \beta^2$.
Provided that these conditions are satisfied, we can expand the two-point function in thermal OPE blocks~\cite{Iliesiu:2018fao}:
\begin{equation}
    g(\tau,x)
    =
    \sum_{\mathcal O} \frac{a_{\mathcal O} }{\beta^{\Delta}} 
    C_{J}^{(\nu)}\!\left(\frac{\tau}{\sqrt{\tau^2+x^2}}\right)
    \left(\tau^2+x^2\right)^{\Delta/2-\Delta_\phi} \label{eq:thermal OPE}\,,
\end{equation}
where $\Delta$ and $J$ are respectively the scaling dimension and spin of the exchanged operators $\Om$, while $\nu=\frac{d-2}{2}$ and $a_{\mathcal O} =\frac{J!}{2^J (\nu)_J} f_{\mathcal O \phi \phi} b_{\mathcal O}$.
Here $f_{\mathcal O \phi \phi}$ is the zero temperature three-point function and $b_{\mathcal O}$ is defined as the thermal one-point function coefficient via~\cite{Iliesiu:2018fao,Marchetto:2023fcw}
\begin{equation} \label{eq: 1ptfunction}
    \langle \mathcal O^{\mu_1\ldots \mu_J}\rangle_{S^1_\beta\times \mathbb{R}^{d-1}} \equiv \langle \mathcal O^{\mu_1\ldots \mu_J}\rangle_{\beta}
    = \frac{b_{\mathcal O}}{\beta^{\Delta}} 
    \left(e^{\mu_1}\ldots e^{\mu_J}-\text{traces}\right)\,,
\end{equation}
with $e^\mu$ the unit vector whose only non-zero component is at $\mu=0$, and $C_{J}^{(\nu)}$ is a Gegenbauer polynomial.
For $x=0$, the OPE reduces to
\begin{equation}\label{eq:OPEx0}
    g(\tau) = \frac{1}{\tau^{2\Delta_\phi}} \sum_{\Delta} a_{\Delta} \left(\frac{\tau}{\beta}\right)^{\Delta}\,, 
    \qquad 
    \text{with } a_\Delta = \sum_{\mathcal O:\, \Delta_{\mathcal O} = \Delta} a_{\mathcal O} \, C_{J}^{(\nu)}(1)\,.
\end{equation}
If the coefficients $a_\Delta$ contain poles for specific $\Delta_\phi$, the series \eqref{eq:OPEx0} requires a regularization which generates logarithmic contributions.
We come back to this point in Section \ref{subsec:HolographicThermalCorrelators}.

\paragraph{KMS condition.}
The two-point function of identical scalars satisfies periodicity (or KMS condition) around the thermal circle~\cite{Kubo:1957mj,Martin:1959jp}.
Combining the KMS condition with the transformation $\tau \to -\tau$, which is always a symmetry in a CFT,\footnote{If the CFT is not parity-invariant, it may be necessary to accompany this transformation with a reflection in some spatial coordinates~\cite{Iliesiu:2018fao}.} one finds that the two-point function satisfies
\begin{equation}
    g(\tau) = g(\beta-\tau)\,.
\end{equation}

\begin{figure}[t]
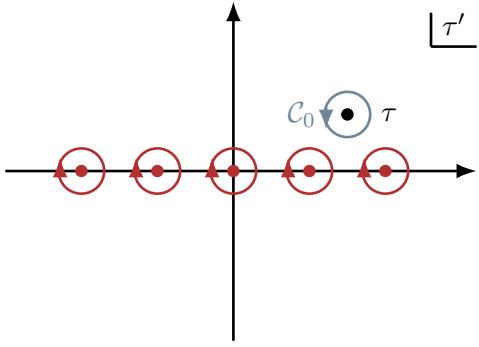

    \centering
    \AnalyticStructureTauPoles
    \caption{Analytic structure of two-point functions at finite temperature and infinite volume in the complex $\tau$-plane for integer spectrum, for the case in which the OPE coefficients do not contain poles. In this case the correlator contains poles at the locations $\beta k$, $k \in \mathbb{Z}$, along the real axis which corresponds to euclidean time.
    The corresponding correlator is then given by the sum over the residues.
    In the holographic case this corresponds to the dominant terms in the OPE.    }
    \label{fig:AnalyticStructure}
\end{figure}

\paragraph{Analytic structure and GFF expansion.}
The analytic structure of thermal correlation functions is a crucial ingredient that enabled the derivation of a thermal inversion formula~\cite{Iliesiu:2018fao} and the crossing-symmetric inversion of single operators in the OPE~\cite{Alday:2020eua,Barrat:2025nvu}. 
Here we spell out the analytic structure and its consequences in the complex $\tau$-plane for this special case.

It was shown in~\cite{Barrat:2025nvu} that thermal correlators admit the general expansion
\begin{equation}
    g(\tau) = \sum_{\Delta} \frac{a_\Delta}{\beta^{\Delta}}
    g_\text{GFF} \left(\Delta_\phi - \frac{\Delta}{2}, \tau\right)
    + g_\text{arcs}(\tau)\,,
    \label{eq:ExpansionInGFFtau}
\end{equation}
where $g_\text{GFF}$ corresponds to the GFF correlator of a scalar of conformal dimension $\Delta_\phi-\Delta/2$, defined as\footnote{We note that the convention adopted in this paper differs slightly from that of~\cite{Barrat:2025nvu}, where the GFF expansion is denoted as $g_\text{GFF}\!\left(2\Delta_\phi - \Delta, \tau\right)$ instead of $g_\text{GFF}\!\left(\Delta_\phi - \Delta/2, \tau\right)$.}
\begin{equation}
    g_{\text{GFF}}(\Delta,\tau)
    =
    \frac{1}{\beta^{2\Delta}}
    \left[
    \zeta_H\!\left(2\Delta,\frac{\tau}{\beta}\right)
    +
    \zeta_H\!\left(2\Delta,1-\frac{\tau}{\beta}\right)
    \right]\;,
    \label{eq:Gfffirstdef}
\end{equation}
where $\zeta_H$ denotes the Hurwitz $\zeta$ function.
We conclude that any two-point function at zero spatial separation can be expanded in terms of GFF correlators, up to a residual term $g_\text{arcs}(\tau)$ that we discuss later.
As already noticed in~\cite{Barrat:2025nvu}, the formula~\eqref{eq:ExpansionInGFFtau} is particularly efficient, since many operators such as the double-twist operators drop out of the expansion.

We now focus on two-point functions of identical scalars at zero spatial separation and with a spectrum consisting only of operators with integer conformal dimensions.
If all conformal dimensions are integer and the corresponding OPE coefficients are finite, then the correlator $g(\tau)$ is analytic except for the Matsubara poles, i.e., poles that are located at $\tau = \beta k$, $k \in \mathbb{Z}$, as illustrated in Figure~\ref{fig:AnalyticStructure}.
On the other hand, if the OPE coefficients have poles at specific values of $\Delta_\phi$, then~\eqref{eq:ExpansionInGFFtau} generates derivatives of Hurwitz $\zeta$ functions with respect to the first argument.
In this case the analytic structure consists of the branch cuts already discussed in~\cite{Barrat:2025nvu}.
In Section \ref{subsec:HolographicThermalCorrelators} we show that both cases are relevant and can be treated separately in the holographic case.

\paragraph{Boundedness conditions.}
The correlator is bounded in the complex $\tau$-plane by its value on the real axis~\cite{Fidkowski:2003nf},
\begin{equation}\label{eq:boundess}
   |g(\tau)| \;\geq\; |g(\tau + i \eta)|\,, 
   \qquad \tau, \eta \in \mathbb{R}\,,
\end{equation}
and also by its value on the imaginary axis~\cite{Barrat:2025nvu},
\begin{equation}\label{eq:boundess2}
   |g(i \eta)| \;\geq\; |g(\tau+i \eta)|\,, 
   \qquad \tau, \eta \in \mathbb{R}\,.
\end{equation}

\paragraph{Clustering condition.}
At infinite volume, the two-point function is expected to satisfy the clustering conditions
\begin{equation}
    g(\tau,x) \;\xrightarrow[x \to \infty]{}\; \langle \phi \rangle_{\beta}^2 \,, 
    \qquad 
    g(t = i\tau,x) \;\xrightarrow[t \to \infty]{}\; \langle \phi \rangle_{\beta}^2 \,. \label{eq:clustering}
\end{equation}
In~\cite{Barrat:2025nvu} it was shown that, once a candidate thermal two-point function exhibits the correct analytic structure and satisfies the conditions given above, it is defined up to a real constant which can be interpreted as related to the squared one-point function $\vev{\phi}_\beta^2$.

\subsection{Holographic thermal correlators}
\label{subsec:HolographicThermalCorrelators}

We specialize our study to the case of holographic two-point functions in the presence of a black brane background in asymptotic AdS$_{d+1}$ spacetime. We will consider a simplified problem where the gravity dual is given by a free massive scalar field $\Phi$ in the bulk, assuming that other matter fields decouple and $\Phi$ does not self-interact. This amounts to studying the (Euclidean) action
\begin{equation}\label{eq:gravityaction}
    S = \int \text{d}^{d+1}y \,\sqrt{g}\, \left(\mathcal{L}_{\Phi} + \mathcal{L}_\text{grav}\right) + S_\text{bdry}\,,
\end{equation}
with
\begin{equation}
    \mathcal{L}_\Phi = \frac{1}{2} g^{MN} \partial_{M} \Phi \partial_{N}\Phi + \frac{1}{2} m^2 \Phi^2\,,
    \qquad 
    \mathcal{L}_\text{grav} = R + \Lambda + \ldots \,,
\end{equation}
where the ellipsis denotes possible higher-derivative curvature corrections and $S_\text{bdry}$ denotes the Gibbons--Hawking boundary term.
Here $y^{M=0,1,\ldots,d} = \lbrace z, \tau, \vec{x} \rbrace$.
The bulk field $\phi$ is dual to the scalar operator in the boundary theory.
The holographic dictionary relates the mass of the scalar field in AdS to the conformal dimension of the scalar field on the boundary $\phi(\tau,\vec{x})$ through
\begin{equation}
    m^2 = \Delta_\phi (\Delta_\phi - d)\,.
\end{equation}
Here we set the AdS radius $\ell$ to unity.
The rotationally invariant and stationary metric in Euclidean signature corresponding to a black brane in AdS can be written as
\begin{equation}
    \text{d}s^2 =
    \frac{1}{z^2}
    \left(
    \frac{\text{d}z^2}{h(z)}
    + f(z)\, \text{d}\tau^2
    + \text{d}x^2
    \right)\,.
    \label{eq:metric}
\end{equation}
The asymptotic AdS boundary conditions further constrain the form of the functions $f(z)$ and $h(z)$, in particular
\begin{equation}
    f(z) = 1 - f_0 z^d +\ldots  \ , \hspace{1 cm}h(z) = 1- h_0 z^d +\ldots  \ .\label{eq:fandh}
\end{equation}
The ellipsis are related to higher-derivative curvature corrections and are discussed in Section~\ref{subsec:StringyCorrectionsBB}.
In~\cite{Fitzpatrick:2019zqz} it was shown that conformal invariance on the boundary necessarily imposes $f_0 = h_0$.
Restricting~\eqref{eq:fandh} to $f_0$ corresponds to the case of the AdS-Schwarzschild black brane, for which the horizon $z_\text{h}$ is related to $f_0$ and to the inverse temperature of the black brane via
\begin{equation}
    f_0 = \frac{1}{z_\text{h}^d} = \left(\frac{4 \pi}{d \beta} \right)^d\,.\label{eq:f0}
\end{equation}
The two-point functions in the boundary theory can be computed by solving the equations of motion for the scalar propagator $\Phi(z,\tau,x)$ in the AdS black brane background, namely,
\begin{equation}\label{eq:EOM}
    \left(\Box - m^2\right) \Phi(z,\tau,x) = 0 \,,
\end{equation}
and taking the limit
\begin{equation}
    g(\tau,x) = \lim_{z \to 0} \frac{\Phi(z,\tau,x)}{z^{\Delta_\phi}}\,.
\end{equation}
Solving this equation analytically in the presence of a black hole background is challenging; We refer to~\cite{Aminov:2020yma,Dodelson:2022yvn,Dodelson:2023vrw} and references therein for further discussions and results. 
In the following we discuss the spectrum of the OPE $\phi \times \phi$ in the black brane background, as it plays an essential role in our bootstrap method.

\paragraph{Black brane spectrum from gravity.}
At leading order in $1/C_T$ (with $C_T$ the central charge of the CFT), the gravity action~\eqref{eq:gravityaction} is effectively classical. This means that quantum gravity effects are absent and the only fields appearing in the bulk are bound states of two scalar fields and multi-particle states of gravitons.
On the dual CFT side, the former correspond to the sector of operators appearing in a (generalized) free scalar OPE -- namely operators with conformal dimensions $\Delta = 2\Delta_\phi+2n+J$, corresponding to double-twist (or \textit{double-trace}) operators; the latter to multi-stress tensor operators in the CFT at finite temperature.
The expected spectrum appearing in the OPE can therefore be summarized to:
\begin{align}
    &\mathds{1}: && \Delta=0,\ J=0\,, \notag \\
    &[\phi \phi]_{n,J}: && \Delta_{n,J} = 2\Delta_\phi + 2n + J\,, \label{eq:BlackBraneSpectrum} \\
    &[T^n]: && \Delta_n = dn,\quad J_n = 0,2,\ldots, dn/2 \qquad (n \geq 1)\,. \notag
\end{align}
It was pointed out in~\cite{Fitzpatrick:2019zqz} that, from the bulk point of view, the equation of motion \eqref{eq:EOM} can be solved at small $z, \tau, x$ by inputting the form of the OPE \eqref{eq:thermal OPE} and in particular the spectrum \eqref{eq:BlackBraneSpectrum}.
This is, however, only possible in the case where double-trace and multi-stress tensor operators do not mix, i.e., when $\Delta_\phi$ is not an integer.
This method was applied to obtain the following multi-stress tensor OPE coefficients in four dimensions, assuming no higher-order curvature corrections, i.e., $f_{k} = h_{k} = 0$ except $k=0$: 
\begin{align}
    a_{4,2} &= \frac{\tilde f_0 \Delta_\phi}{120}\,, \label{eq:a42} \\
    a_{8,4} &= \frac{\tilde f_0^2 \Delta_\phi  \left(7 \Delta_\phi ^2+6 \Delta_\phi+4\right)}{201600 (\Delta_\phi -2)}\,, \label{eq:a84} \\
    a_{8,2} &= \frac{\tilde f_0^2 \Delta_\phi  \left(7 \Delta_\phi ^3-23 \Delta_\phi ^2+22 \Delta_\phi +12\right)}{201600 (\Delta_\phi -3) (\Delta_\phi -2)}\,, \label{eq:a82} \\
    a_{8,0} &= \frac{\tilde f_0^2 \Delta_\phi  \left(7 \Delta_\phi ^4-45 \Delta_\phi ^3+100 \Delta_\phi ^2-80 \Delta_\phi +48\right)}{201600 (\Delta_\phi -4) (\Delta_\phi -3) (\Delta_\phi -2)}\,, \label{eq:a80} \\
    & \hspace{-0.75em}\vdots \notag
\end{align}
where the coefficients are labeled by the conformal dimension and the spin of the operator contributing in the OPE as $a_{\Delta, J}$.
We have also defined $\tilde f_{0} = \beta^{d} f_0$.
These coefficients do not admit a known closed form in terms of the scaling dimensions and spins of the multi-stress tensor operators.
The OPE coefficients associated with double-trace operators remain elusive, as one needs to impose additional boundary conditions on the bulk propagator $\Phi(z,\tau,x)$.
These results are consistent with the eigenstate thermalization hypothesis (ETH), which states that a thermal two-point function can be computed as a zero-temperature heavy-light-light-heavy correlator \cite{Fitzpatrick:2015zha}.
See for instance~\cite{Li:2019zba} for bootstrap results on such correlation functions.
In this setup, higher-curvature corrections correspond to $1/\lambda$ corrections.
They alter the form of the OPE coefficients but do not induce anomalous dimensions for any operators of the spectrum \eqref{eq:BlackBraneSpectrum}.
Quantum corrections come into play at the order $1/N^2$, with $C_T \sim N^2$ \cite{Alday:2016htq,Alday:2017xua}.

Note that the OPE coefficients in~\eqref{eq:a42}-\eqref{eq:a80} have poles at specific integer values of $\Delta_\phi$.
As anticipated in Section \ref{subsec:ThermalTwoPointFunctions}, this crucial point requires us to read the OPE and~\eqref{eq:ExpansionInGFFtau} carefully in order to derive the associated correlators.
In the following we show how to reconstruct the holographic correlators and how to handle these poles.

\subsection{Two-point functions for integer $\Delta_\phi$ in $4d$}
\label{subsec:ExactTwoPointFunctionsForIntegerDeltaBB}

We now study correlators in the black brane background for the case $d=4$.
According to the AdS/CFT dictionary, the parameters of the $4d$ CFT (the coupling constant $\lambda=g_\text{YM}^2 N$ and the rank of the gauge group $N$) are mapped to the parameters of the $5d$ gravity dual via
\begin{align}
    \lambda = \frac{\ell^4}{2\ell_s^4}\,,
    \qquad\qquad
    N = \frac{\ell^4}{4 \pi g_s \ell_s^4}\,,
\end{align}
such that $\ell_s$ is the string length, $\ell$ the radius of curvature, and $g_s$ the dimensionless string coupling.
Additionally, Newton's constant behaves as $G_N \sim 1/N^2$.
Keeping this map and the OPE data derived from the gravity side in mind, we now turn to the finite temperature CFT perspective.
Our approach consists in applying the formula in \eqref{eq:ExpansionInGFFtau}, using the spectrum specified in \eqref{eq:BlackBraneSpectrum}.
Two distinct cases arise: 
\begin{itemize}
    \item[$\star$] \textbf{Non-integer external dimension $\Delta_\phi$:}  
    In this case the expansion in~\eqref{eq:ExpansionInGFFtau} contains an infinite tower of contributions, corresponding to all multi-stress tensor operators.
    Since we do not have analytic control over all the (thermal) OPE coefficients of these operators, exact results are not available.
    Nevertheless, one can approximate their contributions by using the asymptotic values at large dimensions, as first shown in~\cite{Buric:2025anb,Buric:2025fye}.
    In Appendix~\ref{app:TheAsymptoticModel} we show how to reproduce these results with the methods developed in this paper;
    \item[$\star$] \textbf{Integer external dimension $\Delta_\phi$:}  
    {
    In this case the expansion in~\eqref{eq:ExpansionInGFFtau} still contains an infinite number of contributions, but they can be organized into two sets: one finite sum containing the lowest-lying multi-stress tensor operators that satisfy $\Delta < 2\Delta_\phi$ and do not contain poles, and one infinite sum coming from the regularization of the poles.
    We show in the following that the first term can be evaluated exactly, while the second one can be approximated using the OPE coefficients of the asymptotic model.
    Doing so introduces spurious poles associated to the bouncing singularities that must be removed by adding arc contributions.}
\end{itemize}
In what follows we focus exclusively on the second case and show how to obtain the correlators.
When $\Delta_\phi$ is an integer, solving the equation of motion in \eqref{eq:EOM} analytically becomes a harder task due to the mixing between double-trace and multi-stress tensor operators.
Moreover, as mentioned above, some of the multi-stress tensor OPE coefficients develop poles at integer values of $\Delta_\phi$.
For instance, all the operators of dimension $\Delta = 8$ given in~\eqref{eq:a84}--\eqref{eq:a80} diverge at $\Delta_\phi = 2$.
Similarly, $a_{8,2}$ and $a_{8,0}$ diverge for $\Delta_\phi = 3$, while $a_{8,0}$ diverges for $\Delta_\phi = 4$.
A similar pattern is observed for higher operators.
{
We identify two possibilities:
\begin{itemize}
    \item[$\star$] The coefficient $a_\Delta$ in~\eqref{eq:ExpansionInGFFtau} is not divergent for the chosen $\Delta_\phi$.
    This is the case for a finite number of multi-stress operators satisfying $\Delta < 2\Delta_\phi$.
    In the OPE, these operators are not degenerate in $\Delta$ with any double-trace operator, and the OPE contribution reads
    \begin{equation}
        a_{\Delta} \frac{\tau^{\Delta-2 \Delta_\phi}}{\beta^{\Delta}} \,.
    \end{equation}
    The dispersion relation converts it into the following contribution to the correlator:
    \begin{equation}
        \sum_{\Delta < \Delta_\phi} a_{\Delta} \, g_\text{GFF}\left(\Delta_\phi-\frac{\Delta}{2}, \tau\right) \,,
    \end{equation}
    as discussed in Section~\ref{subsec:ThermalTwoPointFunctions}.
    \item[$\star$] The coefficient $a_\Delta$ \emph{diverges}\footnote{It is worth noting that a similar situation can arise in extremal three-point functions at zero temperature~\cite{Castro:2024cmf}.
    In this case the logarithms can be reinterpreted as arising from mixing between single- and higher-trace operators~\cite{Bobev:2025gzu}.
    It would be interesting to understand if a similar mechanism exists at finite temperature.
    We thank Nikolay Bobev for bringing this to our attention.} for the chosen $\Delta_\phi$, with a pole structure
    \begin{equation}
        a_{\Delta} \sim \frac{1}{\Delta_\phi-n} \ , \quad n \in \mathbb{N} \ . 
    \end{equation}
    This is the case for an \emph{infinite} number of multi-stress operators of dimension $\Delta \geq 2 \Delta_\phi$.
    In the OPE, these operators are degenerate in $\Delta$ with an amount of double-trace operators, and the OPE contribution needs to be regularized.
    It can be shown that this reads:\footnote{We thank Simon Caron-Huot for pointing this to us.}
    \begin{equation}
        -2  \frac{\text{Res}_{\Delta_\phi} \, a_{\Delta}}{\beta^\Delta} \tau^{\Delta-2 \Delta_\phi} \log \left(\frac{\tau}{\beta}\right)\,, 
    \end{equation}
    where the residue of $a_\Delta$ is taken for the external dimension $\Delta_\phi \to n$.
    Due to the logarithm, these OPE contributions produce a cut in the complex $\tau$-plane, and consequently a finite contribution to the expansion in GFF correlators.
    This can be easily determined by employing the following regularization of the generic GFF block:
    \begin{equation}
        \frac{1}{\Delta_\phi-n}  \, g_\text{GFF}\left(\Delta_\phi-\frac{\Delta}{2},\tau\right) \to \frac{1}{\Delta_\phi+\varepsilon-n} g_\text{GFF}\left(\Delta_\phi+\varepsilon-\frac{\Delta}{2}, \tau\right) \ .
    \end{equation}
    If we define 
    \begin{equation}
        g^{(1,0)}_\text{GFF}\left(\Delta,\tau\right)=\frac{1}{\beta^{2 \Delta}} \frac{\partial}{\partial s} \left[ \zeta_{H}\left(s,\frac{\tau}{\beta}\right)+ \zeta_{H}\left(s,1-\frac{\tau}{\beta}\right) \right]_{s=2\Delta} \,,
    \end{equation}
    then the finite contribution is given by 
    \begin{equation}
       -2 \frac{\text{Res}_{\Delta_\phi} \, a_{\Delta}}{\beta^{\Delta}} \, g^{(1,0)}_{\text{GFF}}\left(\Delta_\phi-\frac{\Delta}{2} , \tau\right) \,.
    \end{equation}
    Note that even if this scenario escapes the analytic structure of Figure~\ref{fig:AnalyticStructure}, its contribution consists nonetheless in (derivatives of) GFF blocks.
    Such expressions have already been observed in perturbative setups, see for instance~\cite{Barrat:2025nvu} for the case of the $\mathrm O(N)$ model in the $\varepsilon$-expansion.
\end{itemize}
In summary, the thermal two-point function of two scalars can be expressed as
\begin{equation}
    g(\tau)=\sum_{\Delta < 2 \Delta_\phi} \frac{a_\Delta}{\beta^{\Delta}}  g_\text{GFF}\left(\Delta_\phi-\frac{\Delta}{2}, \tau \right)- 2\sum_{\Delta \geq 2 \Delta_\phi} \frac{\text{Res}_{\Delta_\phi} \, a_{\Delta}}{\beta^{\Delta}}  g^{(1,0)}_{\text{GFF}}\left(\Delta_\phi-\frac{\Delta}{2}, \tau \right)+g_{\text{arcs}}(\tau) \,.
    \label{eq:masterform}
\end{equation}
In the following, we split the three contributions into a \emph{principal} part, a \emph{regularized} part, and an \emph{arc} part:
\begin{equation}
     g(\tau)= g_{\text{pr}}(\tau)+g_{\text{reg}}(\tau)+g_{\text{arcs}}(\tau) \,.
\end{equation}
We now focus on applying the dynamical data extracted from the holographic computation to the kinematic result~\eqref{eq:masterform}, extracting the principal, the regularized, and the arcs contributions.

{\paragraph{Principal contributions.}
As pointed out above, the principal contribution coming from the non-singular $a_{[T^n]}$ contributions does not require any regularization.
It can be computed by considering 
\begin{equation}
    g_{\text{pr}}(\tau)
    =
    g_\text{GFF}(\Delta_\phi,\tau)
    +
    \sum_{n=1}^{\lfloor (\Delta_\phi-1)/2 \rfloor} \frac{a_{[T^n]}}{\beta^{4n}} g_\text{GFF}(\Delta_\phi-2 n,\tau) 
    \,,
    \label{eq:ExpansionGFF2}
\end{equation}
where the GFF correlator is defined as \eqref{eq:Gfffirstdef} and with
\begin{equation}
    a_{[T^n]} = \sum_{J = 0}^{2n}C_{J}^{(1)}(1)\, a_{4 n,J} =\sum_{J = 0}^{2n} (J+1)\, a_{4 n,J}\,,
    \label{eq:MultiStressTensorCoefficients}
\end{equation}
with the coefficients $a_{4n,J}$ given in~\eqref{eq:a42}--\eqref{eq:a80} and more generally in~\cite{Fitzpatrick:2019zqz}.

For completeness we write down the principal contributions to the correlators corresponding to $\Delta_\phi = 1,2,\ldots, 5$ as a function of $\textcolor{orange}{a_{T}}$ and $\textcolor{purple}{a_{[T^2]}}$:
{\allowdisplaybreaks
\begin{align}
    &\Delta_\phi = 1: && g_{\text{pr}}(\tau) = \frac{\pi^2}{\beta^2} \csc^2\left(\frac{\pi \tau}{\beta}\right) \,,
    \label{eq:ExpansionGFFDelta1ff} \\[10pt] 
    &\Delta_\phi = 2: && g_{\text{pr}}(\tau) = \frac{\pi^4}{3\, \beta^4} \csc^4\left(\frac{\pi \tau}{\beta}\right) \left[ \cos\left( \frac{2 \pi \tau}{\beta} \right) + 2 \right]\,,
    \label{eq:ExpansionGFFDelta2ff} \\[10pt] 
    &\Delta_\phi = 3: && g_{\text{pr}}(\tau) = \frac{\pi^6}{60\, \beta^6} \csc^6\left(\frac{\pi \tau}{\beta}\right) \left[ \cos\left( \frac{4 \pi \tau}{\beta} \right) + 26 \cos\left( \frac{2 \pi \tau}{\beta} \right) + 33 \right] \notag \\
   & &&\phantom{g(\tau) =\ } + {\textcolor{orange}{a_T}} \frac{\pi ^2}{\beta^6} \csc ^2\left(\frac{\pi  \tau }{\beta }\right)\,,
    \label{eq:ExpansionGFFDelta3ff} \\[10pt] 
    &\Delta_\phi = 4: && g_{\text{pr}}(\tau) = \frac{\pi^8}{2520\, \beta^8} \csc^8\left(\frac{\pi \tau}{\beta}\right) \left[ \cos\left( \frac{6 \pi \tau}{\beta} \right) + 120 \cos\left( \frac{4 \pi \tau}{\beta} \right) \right. \notag \\
    & &&\phantom{g(\tau) =\ } \left. + 1191 \cos \left( \frac{2 \pi \tau}{\beta} \right) + 1208 \right] \notag \\
    & &&\phantom{g(\tau) =\ } + {\textcolor{orange}{a_T}} \frac{\pi^4}{3\, \beta^8} \csc^4\left(\frac{\pi \tau}{\beta}\right) \left[ \cos\left( \frac{2 \pi \tau}{\beta} \right) + 2 \right]\,,
    \label{eq:ExpansionGFFDelta4ff} \\[10pt] 
    &\Delta_\phi = 5: && g_{\text{pr}}(\tau) = \frac{\pi^{10}}{181440\, \beta^{10}} \csc^{10}\left(\frac{\pi \tau}{\beta}\right) \left[ \cos\left( \frac{8 \pi \tau}{\beta} \right) + 502 \cos\left( \frac{6 \pi \tau}{\beta} \right)\right. \notag \\
    & &&\phantom{g(\tau) =\ } \left. + 14608 \cos\left( \frac{4 \pi \tau}{\beta} \right) + 88234 \cos\left( \frac{2 \pi \tau}{\beta} \right) + 78095 \right] \notag \\
    & &&\phantom{g(\tau) =\ } + {\textcolor{orange}{a_T}} \frac{\pi^6}{60\, \beta^{10}} \csc^6\left(\frac{\pi \tau}{\beta}\right) \left[ \cos\left( \frac{4 \pi \tau}{\beta} \right) + 26 \cos\left( \frac{2 \pi \tau}{\beta} \right) + 33 \right] \notag \\
    & &&\phantom{g(\tau) =\ } + {\textcolor{purple}{a_{[T^2]}}} \frac{\pi^2}{\beta^{10}} \csc^2\left(\frac{\pi \tau}{\beta}\right)\,.
    \label{eq:ExpansionGFFDelta5ff}
\end{align}
}%
We explicitly see the implications of \eqref{eq:ExpansionGFF2}: From $\Delta_\phi\geq3$ the multi-stress tensor operators start contributing with additional GFF blocks.
The values of $\textcolor{orange}{a_{T}}$ and $\textcolor{purple}{a_{[T^2]}}$ can be read from~\eqref{eq:a42}--\eqref{eq:a80} with~\eqref{eq:MultiStressTensorCoefficients}, yielding
\begin{equation}
    {\textcolor{orange}{a_T}} = \frac{\pi^4 \Delta_\phi}{40}\,, \quad 
    {\textcolor{purple}{a_{[T^2]}}}
    =
    {\textcolor{orange}{a_T}}^2
    \frac{63 \Delta_\phi ^4-413 \Delta_\phi ^3+672 \Delta_\phi ^2-88 \Delta_\phi +144}{126 \Delta_\phi(\Delta_\phi -4) (\Delta_\phi -3) (\Delta_\phi -2) }\,,
\end{equation}
from which we obtain explicit results in~\eqref{eq:ExpansionGFFDelta1ff}--\eqref{eq:ExpansionGFFDelta5ff}.
In principle one can produce the principal part of the correlators for arbitrarily high $\Delta_\phi$, to the condition that sufficiently many multi-stress tensor OPE coefficients are known.
Note that the limit $\Delta_\phi \to \infty$ in~\eqref{eq:ExpansionGFF2} trivially reproduces the geodesic approximation~\cite{Rodriguez-Gomez:2021pfh} by excluding all the regularized contributions.\footnote{It is easy to see that the other contributions in~\eqref{eq:masterform} drop in this limit.}

It can be argued from the OPE point of view that the principal contributions, being associated with the lightest operators in the OPE, constitute the \emph{dominant} component of the holographic correlator around $\tau \sim 0$ and $\tau \sim \beta$.
This can be tested numerically, as shown in Section \ref{subsec:NumericalChecksBB}.
Hence, the principal contribution can be considered a good approximation of the full holographic two-point function.
Note however that this term is clearly not complete as it does not capture the log contributions in the OPE nor the correct asymptotic OPE expansion at large frequency.

{\paragraph{Regularized contributions.}
We now come to the analysis of the regularized contributions, encoding the dynamical information of an infinite number of multi-stress tensors:
\begin{equation} \label{eq: greg}
    g_{\text{reg}}(\tau)=-2\sum_{\Delta \geq 2 \Delta_\phi} \frac{\text{Res}_{\Delta_\phi} \, a_{\Delta}}{\beta^{\Delta}}  g^{(1,0)}_{\text{GFF}} \left(\Delta_\phi-\frac{\Delta}{2}, \tau \right) \,.
\end{equation}
An exact computation can only be performed with the knowledge of an infinite number of coefficients $\text{Res}_{\Delta_\phi} \, a_{\Delta}$. In principle, we can compute as many OPE coefficients as we desire from the wave equation as in~\cite{Fitzpatrick:2019zqz}. However, the analytic form (or the closed form) of the multi-stress OPE coefficients for any scaling dimension is currently unknown, to the best of our knowledge.

\begin{figure}[t]
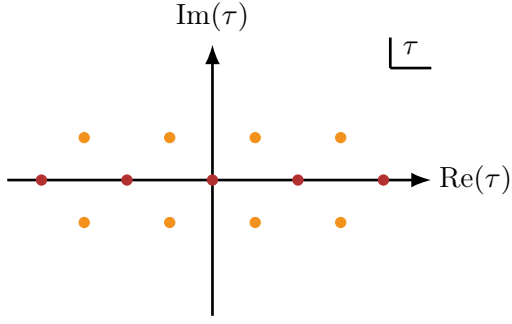

    \centering
    \BouncingSingularities
    \caption{The regularized part of the correlator, $g_{\text{reg}}(\tau)$, has additional poles in the complex $\tau$-plane called bouncing singularities.
    The analytic structure of the regularized part therefore consists of KMS poles (colored red) and bouncing singularities (colored orange).
    The latter are removed from the full correlator by adding the arc contributions.}
    \label{fig:AnalyticStructureBounce}
\end{figure}

{It is nevertheless possible to produce an \emph{approximated} resummation\footnote{To refine this approximation, one may replace the OPE coefficients of operators with low scaling dimensions by their exact values as computed in~\cite{Fitzpatrick:2019zqz}, rather than using their large–scaling-dimension approximations. A systematic analysis of this improvement is left for future work.}, which reveals the presence of new poles in the strip $0<\tau<\beta$ in the complex $\tau$-plane and they are depicted in Figure~\ref{fig:AnalyticStructureBounce}.
These poles correspond precisely to the so-called \textit{bouncing singularities} in AdS~\cite{Fidkowski:2003nf,Festuccia:2005pi}, which were first identified through the geodesic approximation.
Such singularities are unphysical for the Euclidean correlator, in the sense that a generic two-point function should only exhibit the Matsubara poles, as discussed in Section~\ref{subsec:ThermalTwoPointFunctions}.
The presence of the bouncing singularities implies that the contribution from the arcs is no longer constant:\footnote{Note that the appearance of this pole violates the analytic structure of the correlator and thus escapes the theorem presented in~\cite{Barrat:2025nvu} to show that the arc terms are constant.} Instead, they must cancel the poles, as explicitly shown for the asymptotic model in~\cite{Ceplak:2024bja,Buric:2025anb} and reviewed in Appendix~\ref{app:TheAsymptoticModel}.

Since the bouncing singularities are produced by the resummation of an infinite number of multi-stress tensor contributions, we can substitute the coefficients with their asymptotic behaviour in $\Delta=4n$, as derived in \cite{Ceplak:2024bja}: }}
\begin{equation}\label{eq:asymptoticformula}
    a_{4n} = \frac{\pi}{20}(-4)^n n^{2 \Delta_\phi -3} \left(\frac{4^{2\Delta_\phi}\Delta_\phi^2
 (\Delta_\phi -1)}
    {\Gamma \left(2 \Delta_\phi +\tfrac{3}{2}\right)}\right)\csc (\pi  \Delta_\phi) \,,
\end{equation}
{leading to the residue}
\begin{equation}
    - 2 \text{Res}_{\Delta_\phi}\, a_{4n} =  \frac{1}{40} (-4)^{\Delta_\phi+n+1} n^{2 \Delta_\phi -3} \left(\frac{4^{\Delta_\phi}\Delta_\phi^2
 (\Delta_\phi -1)}
    {\Gamma \left(2 \Delta_\phi +\tfrac{3}{2}\right)}\right) \,.
\end{equation}
{We write explicitly the regularized contributions for the correlators with $\Delta_\phi<4$.
The expressions for $\Delta_\phi\geq 4$ are easily obtainable, but we omit them for the sake of clarity:
{\allowdisplaybreaks
\begin{align}
    &\Delta_\phi = 1: && g_{\text{reg}}(\tau) = 0 \,,
    \label{eq:ExpansionGFFDelta1ffcorr} \\[10pt] 
    &\Delta_\phi = 2: && g_{\text{reg}}(\tau) = \sum_{m=-\infty}^\infty k_2 \log |m \beta+\tau |
   \frac{ (m \beta+\tau )^4 \left(2 (m \beta+\tau )^4+\beta^4\right) }{  \beta^4 \left(4 (m \beta+\tau )^4+\beta^4\right)^2}
   \,,
    \label{eq:ExpansionGFFDelta2ffcorr} \\[10pt] 
    &\Delta_\phi = 3: && g_{\text{reg}}(\tau) = \sum_{m = -\infty}^\infty k_3   \log |\beta  m+\tau |
     \bigg[\frac{ 27 (\beta  m+\tau )^6 }{\beta^3 
    \left(4 (\beta  m+\tau )^4+\beta^4\right)^4} \notag \\
    & &&\phantom{g_{\text{reg}}(\tau) =\, } +\frac{ 16 \left(32 (\beta  m+\tau )^8+31 (\beta  m+\tau
    )^4 \beta^4+11 \beta^8\right) (\beta  m+\tau )^{10} }{\beta^3 
    \left(4 (\beta  m+\tau )^4+\beta^4\right)^4}\bigg]
   \,,
    \label{eq:ExpansionGFFDelta3ffcorr}
\end{align}}
{where 
\begin{equation}
    k_2=-\frac{524288}{4725 \sqrt{\pi}} \ , \quad k_3=-\frac{33554432}{75075 \sqrt{\pi}} \ .
    \label{eq:alphas}
\end{equation}%
Note that the correlator corresponding to $\Delta_\phi = 1$ reproduces the free scalar two-point function as expected.
In this framework, this can be seen as a consequence of the absence of poles at $\Delta_\phi=1$ in the OPE coefficients.\footnote{More generally one can show that any correlator with $\Delta_\phi = 1$ in four spacetime dimensions must satisfy the free equation of motion.
Further details are provided in Appendix~\ref{app:NumericalSolutionOfTheWaveEquation}, where this case is used as a benchmark to test the numerical solution of the wave equation in AdS.}
The expressions for $\Delta_\phi=2,3$ exhibit instead new poles $\tau_c = \pm\frac{\beta}{2}\pm i \frac{\beta}{2}$, previously identified in the literature as encoding the bouncing singularities~\cite{Ceplak:2024bja}.
Their position depends only on the number of spacetime dimensions and not on the conformal dimension of the external operators.
Around $\tau_c$ we have:
\begin{equation}
     g_{\text{reg}}(\tau) \sim \frac{1}{(\tau-\tau_c)^{2 \Delta_\phi-2}} + \text{subleading poles} \,,
\end{equation}
The degree of divergence agrees with the prediction of~\cite{Ceplak:2024bja}, while the number of subleading poles depends explicitly on the value of $\Delta_\phi$. Moreover, we observe that the (regularized) sum for $\Delta_\phi = 2$, displayed in~\eqref{eq:ExpansionGFFDelta2ffcorr}, exhibits a linear growth along the real-time direction (i.e., for $t = i\tau \to \infty$). This behaviour is inconsistent with the clustering decomposition of the correlator, as expressed in~\eqref{eq:clustering}. The discrepancy can be traced back to the approximation used for the OPE coefficients in~\eqref{eq:asymptoticformula}.

One might attempt to restore the correct large-$t$ behaviour by introducing an additional arc contribution. However, such a term would, by construction, have vanishing discontinuity in the complex plane (except at $|\tau| \to \infty$) and must also satisfy KMS invariance. It is straightforward to verify that no function with these properties exists. Consequently, the only viable resolution is to modify the discontinuity of the two-point function itself. Equivalently, this requires correcting the asymptotic coefficients appearing in~\eqref{eq:asymptoticformula} so as to recover the correct late-time behavior.\footnote{For example, including a subleading $1/n$ correction in~\eqref{eq:asymptoticformula} appears to be sufficient to restore the correct large-$t$ asymptotics.}

A detailed analysis of these corrections lies beyond the scope of the present work. Nevertheless, recent progress in this direction has been made in~\cite{Buric:2025uqt,Afkhami-Jeddi:2025wra}, suggesting that a systematic computation of such corrections is a promising avenue for future investigation.

{\paragraph{Arc contributions.}
Arcs are not captured by the dispersion relation and thus by the expansion in GFF correlators.
Their contribution can be fixed by imposing the full correlator to satisfy the analytic structure and then invoking the uniqueness of the correlator, as explained in Section~\ref{subsec:ThermalTwoPointFunctions}.
Indeed, it was proven in Appendix~A of~\cite{Barrat:2025nvu} that if the two-point function satisfies all the analytic structure expected by the thermal correlator, then the correlator is fixed up to a real constant; the clustering property then fixes this constant.
The problem of computing the arcs is therefore well-posed.

In the analysis of the regularized contributions, we highlighted how the infinite tower of multi-stress tensor operators generates an extra set of poles in the complex $\tau$-plane.
As shown in~\cite{Buric:2025fye} and reviewed in Appendix~\ref{app:TheAsymptoticModel} for the case of half-integer $\Delta_\phi$, it is natural to take as a candidate for the arcs the sum over images of the bouncing singularity pole.
This appears to be the only possibility to restore the correct analytic properties of the two-point function.

The regularized contribution behaves as follows in the neighbourhood of the bouncing singularity pole $\tau_c=\frac{\beta}{2}( 1 + i)$:
\begin{equation}
    g_{\text{reg}}(\tau) \sim -\sum_{\ell = 1}^{2\Delta_\phi-2}\frac{\alpha_{\Delta_\phi}^{(\ell)}}{\beta^{2\Delta_\phi-\ell}}\frac{1}{(\tau-\tau_c)^{\ell}}+\text{regular} \,,
\end{equation}
while similar poles also appear in $\overline \tau_c$ with complex conjugate coefficients, $\overline \alpha_{\Delta_\phi}^{(\ell)}$, as well as in all the images of the strip $0<\tau<\beta$.
Here the $\alpha_{\Delta_\phi}^{(\ell)}$ are constants determined by the regularized contribution discussed before.
Therefore, in order to cancel the poles we need to consider\footnote{It is worth highlighting that the absolute value holds on the real axis only, while the outcome of the resummation is analytically continued to the whole complex $\tau$-plane.}
\begin{equation}
     h(\tau) \sim \sum_{\ell = 1}^{2\Delta_\phi-2}\sum_{m = -\infty}^\infty \frac{\alpha_{\Delta_\phi}^{(\ell)}}{\beta^{2\Delta_\phi-\ell}} \frac{1}{|\tau+m-\tau_c|^{\ell}}+
     (\tau_c\to \overline \tau_c \ , \ \alpha_{\Delta_\phi}^{(\ell)}\to \overline \alpha_{\Delta_\phi}^{(\ell)})
     \,.
\end{equation}
We recognize in this expression the definition of the GFF correlators, however shifted by $\tau_c$.
We thus have
\begin{equation}
   h(\tau) = \frac{1}{\beta^{2\Delta_\phi}} \sum_{\ell = 1}^{2\Delta_\phi-2}  \alpha_{\Delta_\phi}^{(\ell)} \, g_\text{GFF} \left( \frac{\ell}{2},\tau-\tau_c \right)
    + (\tau_c\to \overline \tau_c \ , \ \alpha_{\Delta_\phi}^{(\ell)}\to \overline \alpha_{\Delta_\phi}^{(\ell)})
    \,. \label{eq: haux}
\end{equation}
Note that, because of the shift $\tau \to \tau_c$, we need to express these GFF correlators in terms of Hurwitz $\zeta$ functions, valid in the strip $\operatorname{Re}(\tau) \in (0,\beta)$, with more attention.
In particular we consider the arc contributions to be given by the manifestly parity-invariant combination of the expression~\eqref{eq: haux}
\begin{equation}
    g_{\text{arcs}}(\tau)=h(\tau)+h(-\tau) 
    \,,
\end{equation}
where the purpose of the second term is to take into account all the poles in $\tau \in (-\beta,\beta)$ and images thereof.

This method provides closed forms for the arc contributions for arbitrary $\Delta_\phi$.
In particular, the only relevant quantities to determine the arcs are the coefficients $\alpha_{\Delta_\phi}^{(\ell)}$.
We give here the coefficients relevant for $\Delta_\phi < 4$: 
\begin{align}
    \alpha_2^{(1)} &= - \frac{2048 \left(1+i\right)}{4725 \sqrt{\pi }} \left(-1+3 \log
   \tau_c \right) \ , & 
   \alpha_2^{(2)} &=\frac{2048 \, i}{4725 \sqrt{\pi }}\log \tau_c \,,  \\[10pt]
    \alpha_3^{(1)} &= \frac{8192 \left(1-i\right)}{3003\sqrt{\pi }} \left(-3+5 \log
   \tau_c \right) \ , & \alpha_3^{(2)} &=\frac{8192}{75075\sqrt{\pi }} \left(-27+61 \log \tau_c \right) \,, \notag \\[4pt]
    \alpha_3^{(3)} &=\frac{8192 \left(1+i\right)}{25025\sqrt{\pi }}  \left(-1+4 \log
   \tau_c \right) \ , & \alpha_3^{(4)} &=-\frac{8192 \, i}{25025 \sqrt{\pi }}  \log \tau_c \ .
\end{align}
To conclude this discussion, we observe that the leading pole depends only on the asymptotic value of the OPE coefficients $a_{[T^n]}$.
In other words, we expect this contribution in the arcs to remain the same even if the regularized contributions are modified to account for the exact coefficients.
We reserve to future work a systematic analysis of how corrections of the OPE coefficients affect the subleading poles.

\begin{figure}[t]
    \centering
    \begin{minipage}{.50\textwidth}
        \includegraphics[scale=.20]{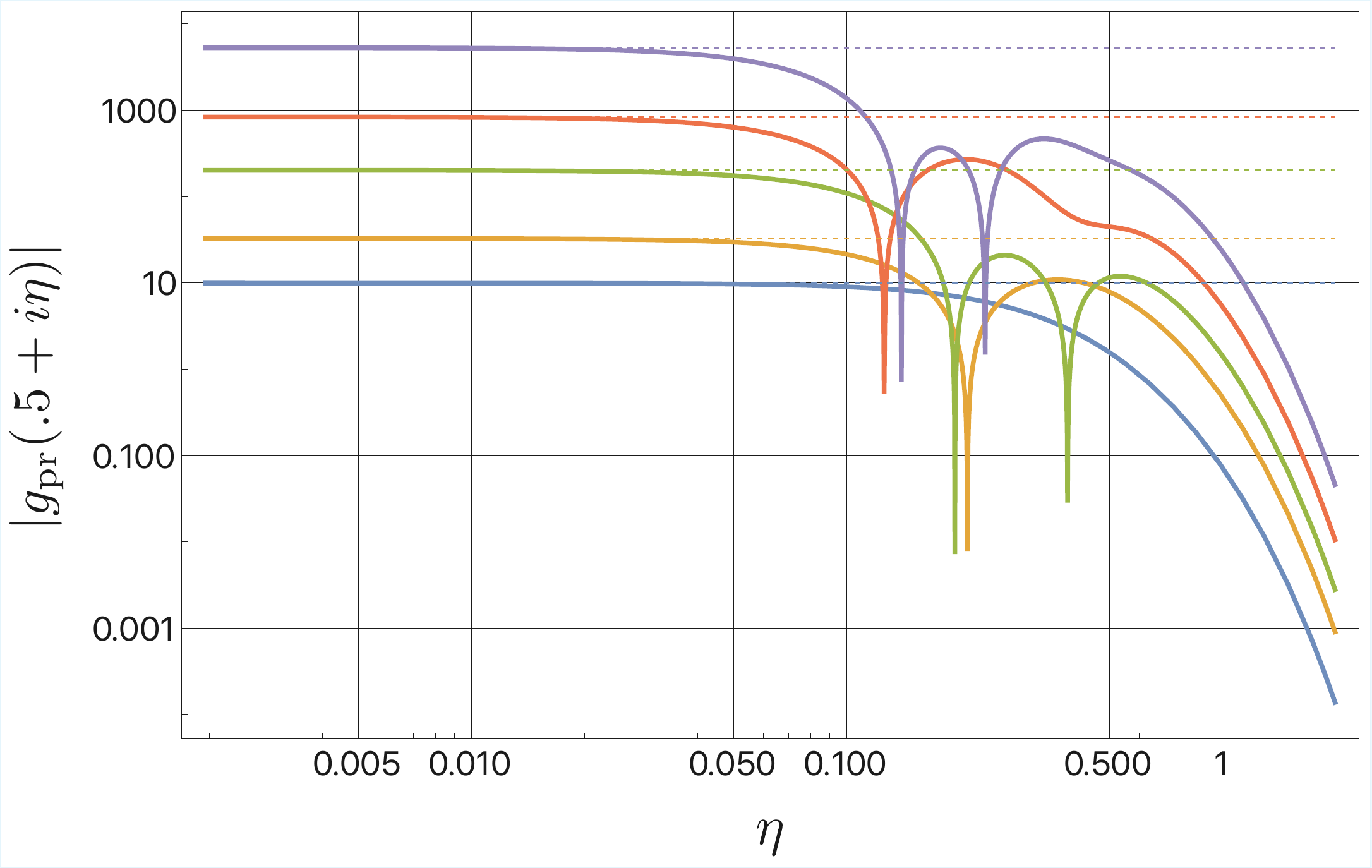}
    \end{minipage}%
    \begin{minipage}{.50\textwidth}
        \includegraphics[scale=.197]{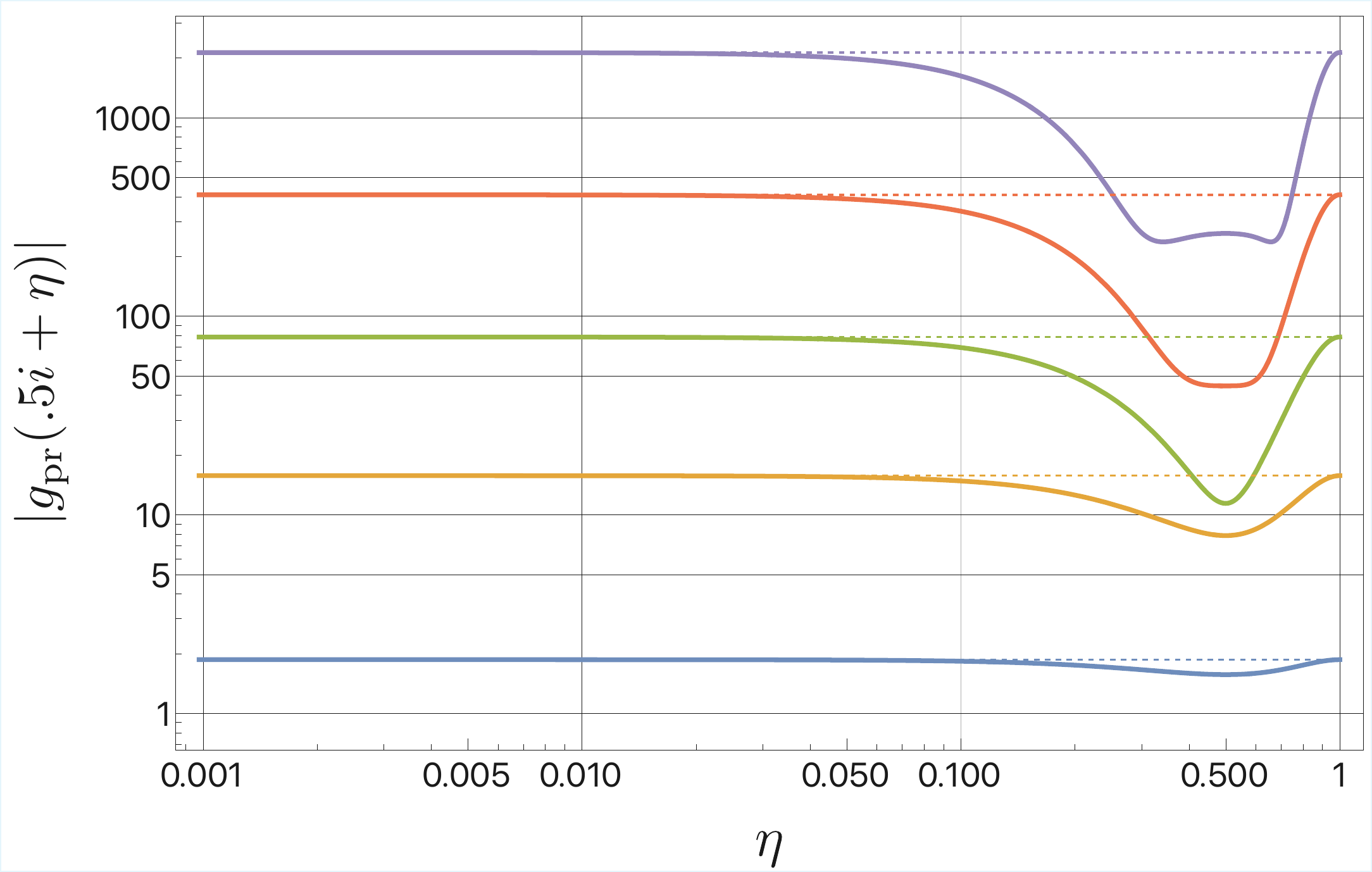}
    \end{minipage}
    \caption{
    Analysis of the boundedness conditions~\eqref{eq:boundess}--\eqref{eq:boundess2} for the principal contributions and $1 \leq \Delta_\phi \leq 5$.
    \textbf{Left:} Absolute value of $g_\text{pr}(\tau)$ evaluated along the line $\tau = 1/2 + i \eta$.
    They are consistently bounded by their values on the real axis. 
    \textbf{Right:} Absolute value of $g_\text{pr}(\tau)$ for the same range of $\Delta_\phi$, evaluated along the line $\tau = \eta +i/2$. 
    They are consistently bounded by their values on the imaginary axis. 
    Both panels use log--log scaling with $\beta=1$ without loss of generality.
    Note that these terms obey the KMS condition by construction as we vary the real part of $\tau$ and this generates bumps in the log-log scale.}
    \label{fig:Boundness}
\end{figure}

The results above are expected to represent an approximation for the holographic correlators at finite temperature.  
Since the OPE consistency, the KMS condition, and the required analytic properties are built-in by construction, the only non-trivial bootstrap check is the boundedness condition in~\eqref{eq:boundess} and \eqref{eq:boundess2}. Figure~\ref{fig:Boundness} shows the behavior of the principal contribution for different $\Delta_\phi$ along the imaginary axis for $\text{Re}(\tau) = 1/2$ and along the real axis for $\text{Im}(\tau) = 1/2$.
In both cases, the boundedness is surprisingly satisfied for all values of $\text{Re}(\tau)$ and for $\text{Im}(\tau)$ respectively, in spite of ignoring the regularized and arc contributions.
Figure~\ref{fig:Boundness2} shows the behavior of the full correlator for $\Delta_\phi=3$.
It reveals that the approximated result does not satisfy the boundedness condition~\eqref{eq:boundess2}, which can again be interpreted as an effect of the asymptotic approximation of the OPE coefficients.\footnote{This violation was already observed in~\cite{Buric:2025anb} for non-integer $\Delta_\phi$.}

Furthermore, we argue that the sum rules derived in~\cite{Marchetto:2023xap} and successfully applied in~\cite{Barrat:2025wbi,Buric:2025fye,Buric:2025anb} are automatically satisfied by the holographic correlators studied in this section.
As a result, the OPE coefficients of the multi-stress tensor operators remain input data of the problem.
The argument is the following. Consider, as an example in which the sum rules are predictive, a strongly-coupled theory such as the $3d$ Ising model. 
In this case, one is allowed to expand the correlator in terms of GFF blocks, as we do in this paper; however, they would  immediately encounter spurious contributions associated with the classical, i.e., without anomalous dimensions, double-twist operators. These operators are excluded from the OPE spectrum, which is an input of the problem.
Eliminating these unphysical terms requires tuning the thermal OPE coefficients to specific values -- this is an alternative way of explaining the logic underlying~\cite{Barrat:2025wbi}: in fact, these specific values constitute the solutions of the sum rules for the given spectrum.
An even clearer illustration arises in the $\mathrm O(N)$ model at large~$N$: the GFF expansion naturally generates a contribution from the operator $\phi^2$ with dimension $\Delta_{\phi^2} = 1$, which is absent from the correct spectrum and must instead be replaced by the Hubbard–Stratonovich field $\sigma$ with $\Delta_\sigma = 2$. 
Only a precise combination of thermal OPE coefficients -- or, equivalently, a specific value of the thermal mass $m_\text{th}$ -- cancels the $\phi^2$ contribution, enabling one to solve for the correct thermal OPE data (either analytically or numerically).

\begin{figure}[t]
    \centering
    \begin{minipage}{.50\textwidth}
        \includegraphics[scale=.14]{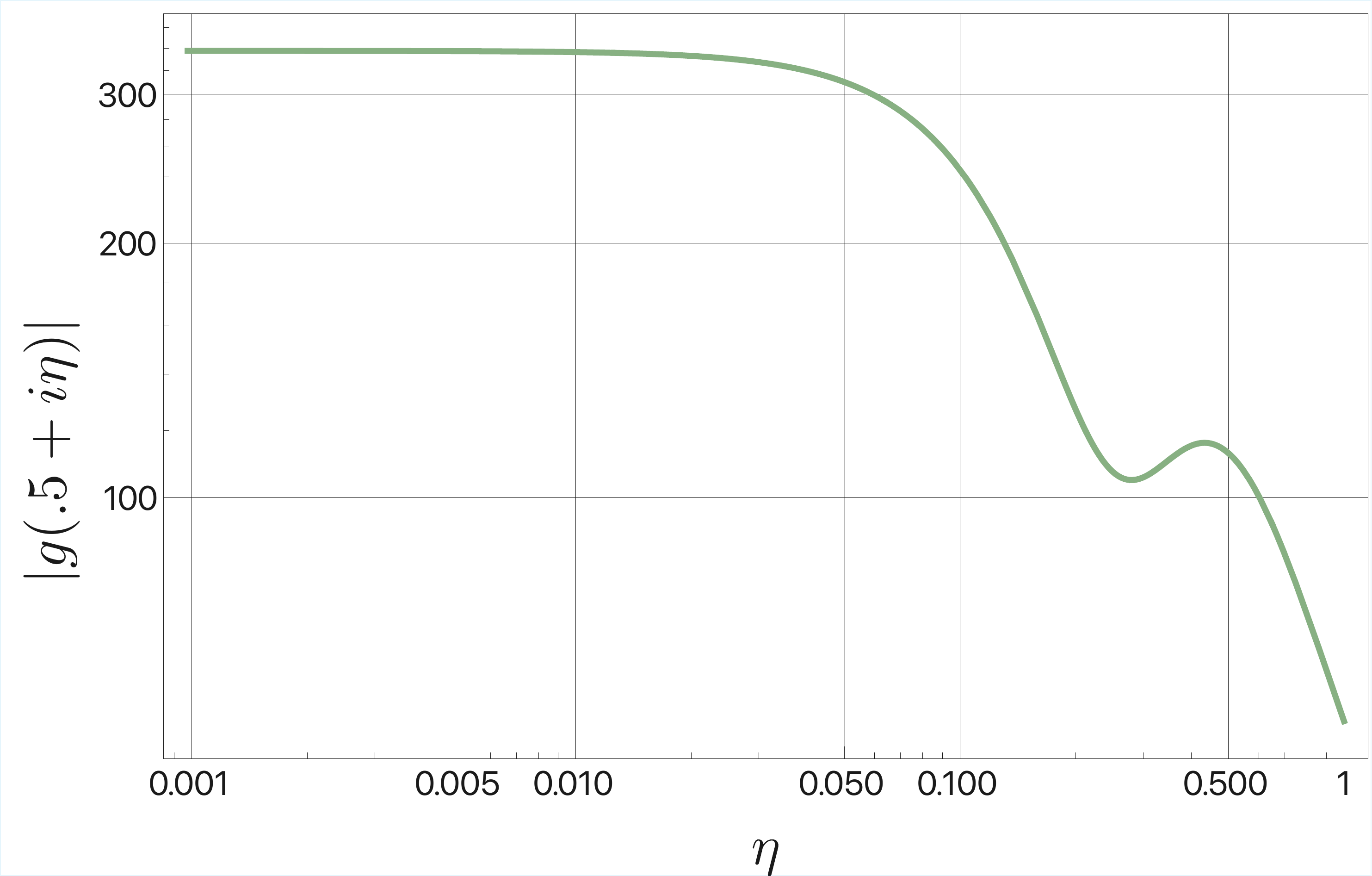}
    \end{minipage}%
    \begin{minipage}{.50\textwidth}
        \includegraphics[scale=.14]{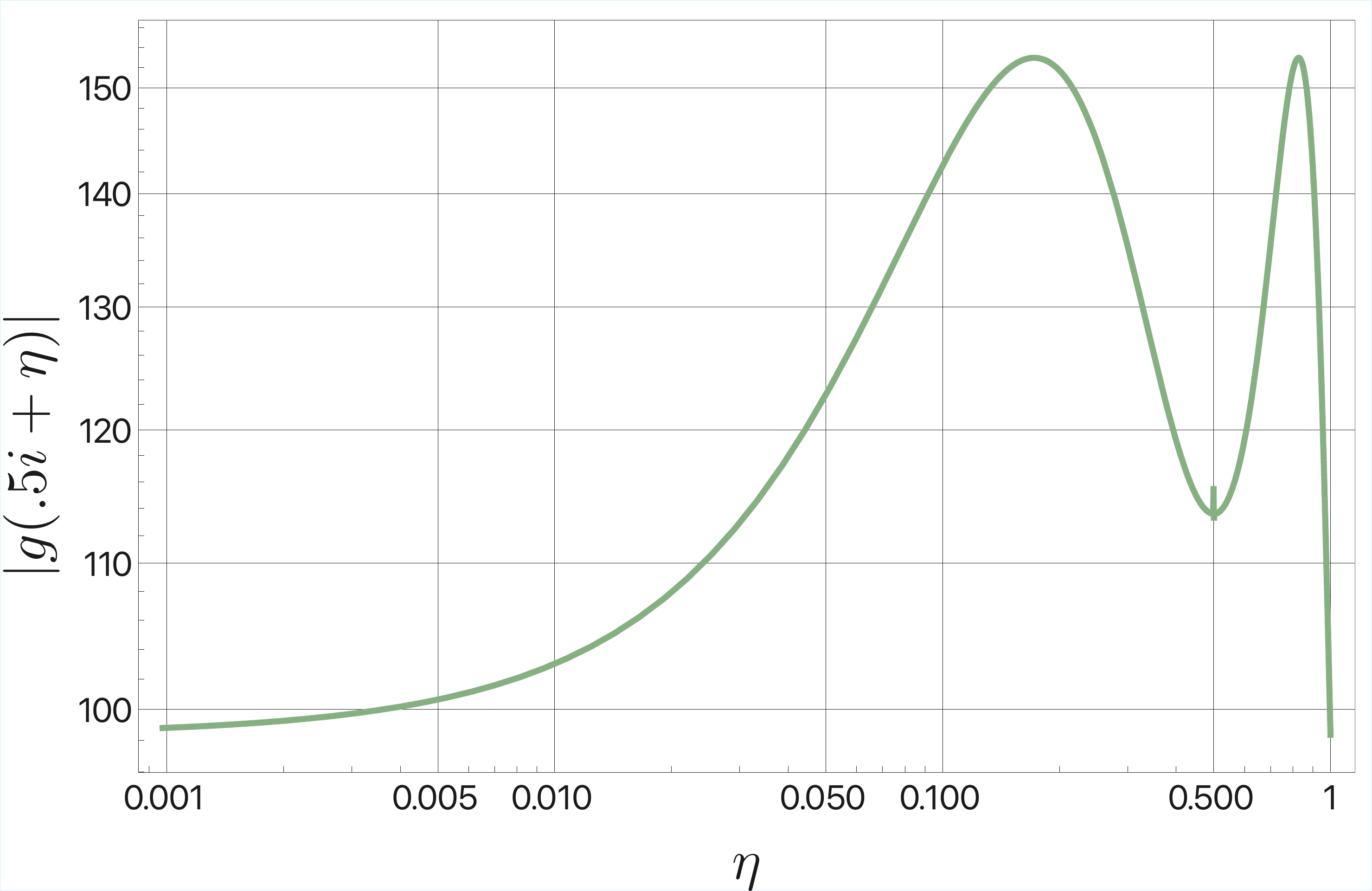}
    \end{minipage}
    \caption{
    Analysis of the boundedness conditions~\eqref{eq:boundess}--\eqref{eq:boundess2} for the approximated correlator with $\Delta_\phi=3$.
    \textbf{Left:} Absolute value of $g(\tau)$ evaluated along the line $\tau = 1/2 + i \eta$.
    The correlator is consistently bounded by its values on the real axis. 
    \textbf{Right:} Absolute value of $g(\tau)$ evaluated along the line $\tau = \eta +i/2$.
    The correlator is not bounded by its values on the imaginary axis.
    We associate this inconsistency with the fact that we use the asymptotic value of the OPE coefficients instead of their exact one.
    Both panels use log--log scaling with $\beta=1$ without loss of generality.}
    \label{fig:Boundness2}
\end{figure}

In contrast, for the holographic correlators studied in this work, the situation is different. The GFF expansion, even when modified to cure the poles in the OPE coefficients of the multi-stress tensor operators, already reproduces all the physical operators in the theory.
There is therefore no spurious term to cancel and no fine tuning occurs. This leads to the conclusion that in this case the sum rules are an indeterminate problem.
In fact, it is easy to see that there is not an unique solution: While there exist specific values of the coefficients $a_{[T^n]}$ corresponding to the holographic setup, setting $a_{[T^n]} = 0$ still yields a consistent (pure GFF) solution, for example.  
Hence it is a difficult task to further constrain the thermal OPE coefficients within the present framework.
However we noted that the asymptotic expression for $a_{[T^n]}$ does not produce a consistent correlator for $\Delta_\phi = 2,3$, as it is either not bounded for large $t = i \tau$ or by its value on the imaginary axis. 
Additional constraints might emerge by considering correlators at non-zero spatial separation or by including $1/N^2$ corrections beyond the classical large $N$ limit.

\subsection{Numerical checks for $\Delta_\phi=3$}
\label{subsec:NumericalChecksBB}

\begin{figure}[t]
    \centering
        \includegraphics[scale=.25]{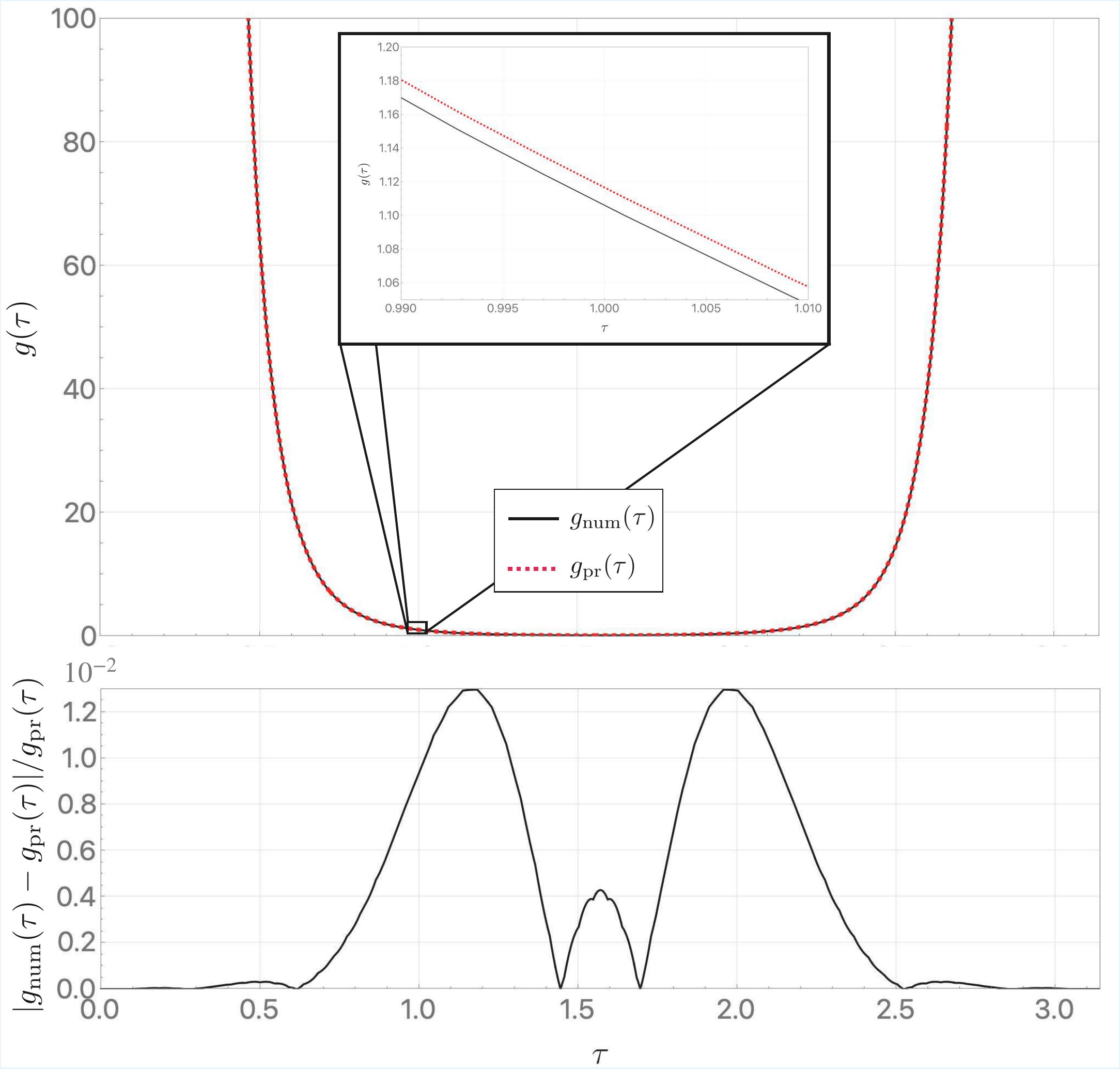}
    \caption{Comparison between the bootstrapped principal part of the two-point function for $\Delta_\phi = 3$, $g_{\text{pr}}(\tau)$, and the boundary limit of the numerical solution of the AdS$_5$ wave equation, $g_{\text{num}}(\tau)$.
    The plot below displays the relative discrepancy between $g_{\text{pr}}(\tau)$ and $g_{\text{num}}(\tau)$ for $\beta=\pi$. 
    The discrepancy is always below $1.3\%$ for $\tau \in (0,\beta)$. }
    \label{fig:comparisontp}
\end{figure}

Since all bootstrap conditions are satisfied and the solution is expected to be unique~\cite{Barrat:2025nvu}, given the input, the correlators bootstrapped in~\eqref{eq:ExpansionGFFDelta1ff}--\eqref{eq:ExpansionGFFDelta5ff} should approximately coincide with the solution of the bulk wave equation~\eqref{eq:EOM} in AdS in the appropriate limit.\footnote{We gratefully acknowledge Ilija Burić for sharing material that proved to be useful for the results of this section.}
While the latter is difficult to solve analytically in position space (see, e.g.,~\cite{Fitzpatrick:2019zqz,Parisini:2022wkb,Parisini:2023nbd} for some analytical results), it can be addressed numerically.
In this section, we solve the scalar wave equation in AdS for a bulk field dual to a boundary operator of dimension $\Delta_\phi = 3$.
Since the first correlator to involve the stress tensor coefficient in the principal contribution corresponds to $\Delta_\phi = 3$, we focus on this specific example. 
Further details on this computation are provided in Appendix~\ref{app:NumericalSolutionOfTheWaveEquation}.

Figure~\ref{fig:comparisontp} shows that the principal contribution to the two-point function from our bootstrap construction and the numerical bulk solution differ by at most $\sim 1.3\%$.
As argued in Appendix~\ref{app:NumericalSolutionOfTheWaveEquation}, this small discrepancy can be attributed to numerical accuracy.
We expect that a better numerical estimation of the two-point function could in principle also show the (subleading) regularized and arc contributions presented in Section~\ref{subsec:ExactTwoPointFunctionsForIntegerDeltaBB}.
For $\Delta_\phi = 3$, the principal contribution is composed of only two blocks: the identity and the stress tensor blocks, as can be seen from~\eqref{eq:ExpansionGFFDelta3ff}. 
In Figure~\ref{fig:blocks} we compare the numerical two-point function with the identity block (left panel), and with the stress tensor block after subtracting the identity contribution (right panel).
The apparent agreement supports the fact that the principal contributions \eqref{eq:ExpansionGFFDelta1ff}--\eqref{eq:ExpansionGFFDelta5ff} represent the dominating component of the holographic correlators.

In Appendix~\ref{app:NumericalSolutionOfTheWaveEquation} we also show the case of $\Delta_\phi = 1$ as a testing playground, since the prediction for the associated correlator is exact.

\begin{figure}[t]
    \centering
    \begin{minipage}{.48\textwidth}
        \includegraphics[scale=.20]{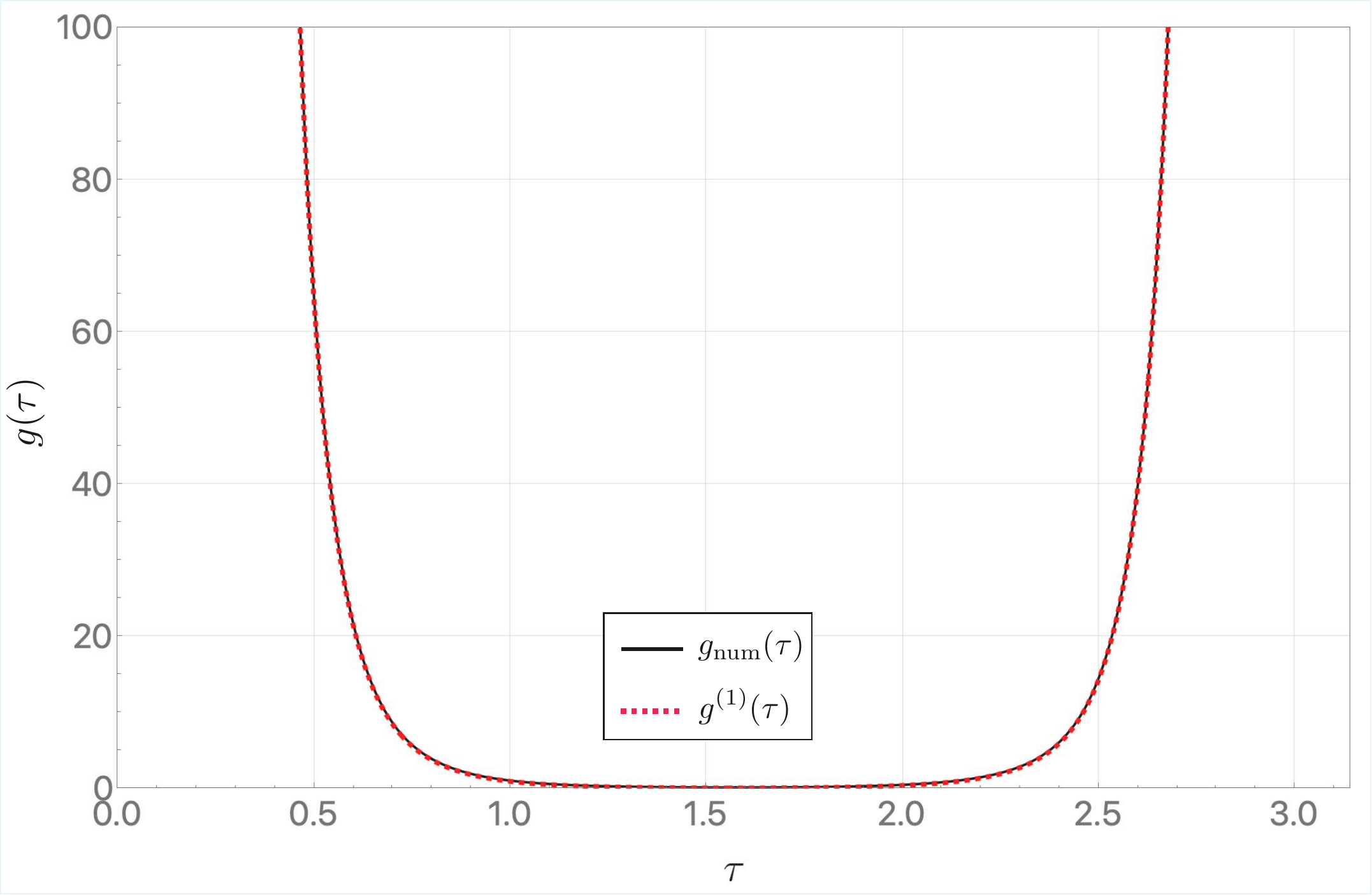}
    \end{minipage}%
    \begin{minipage}{.48\textwidth}
        \includegraphics[scale=.20]{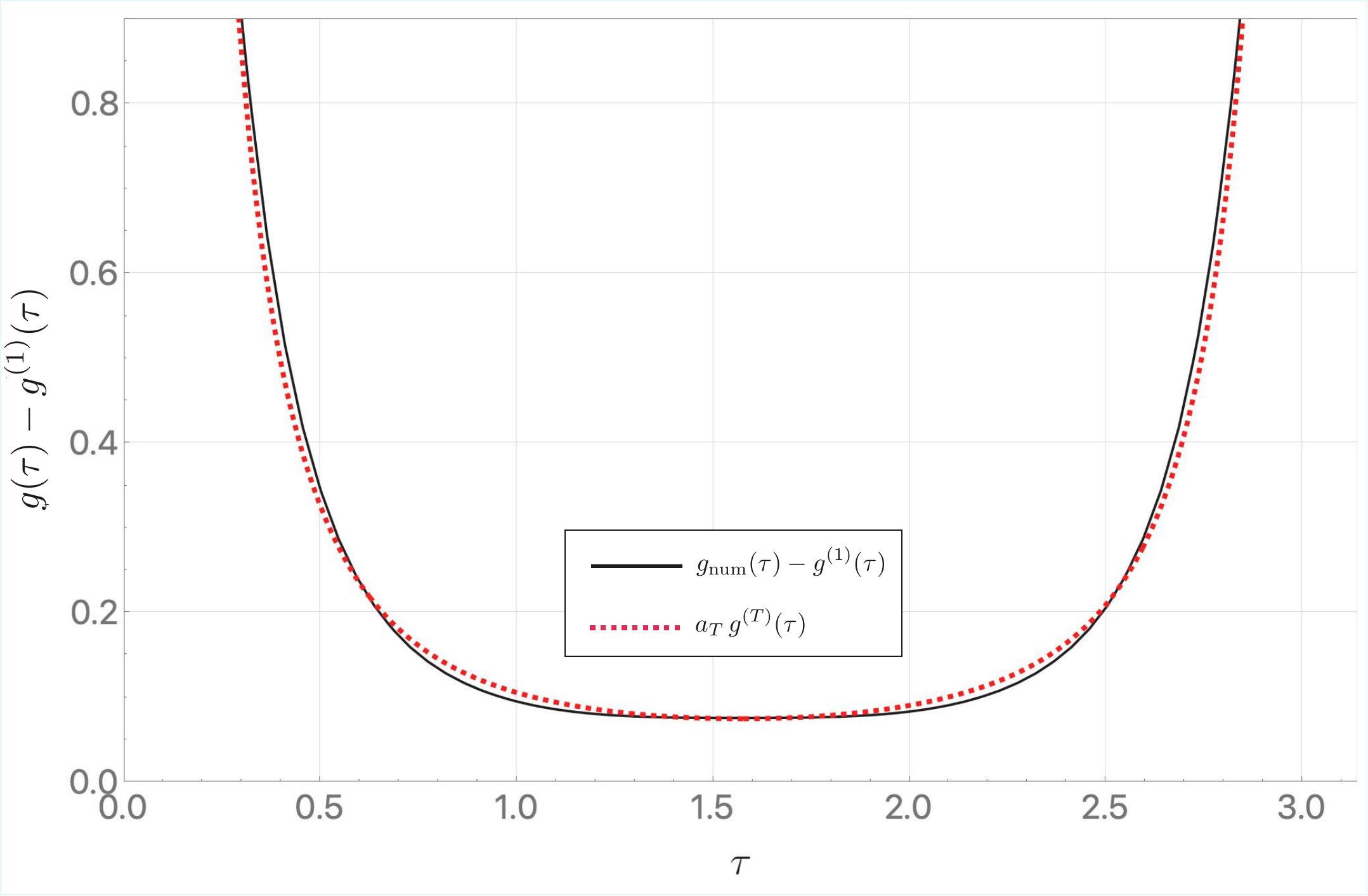}
    \end{minipage}
    \caption{\textbf{Left:} Comparison between the numerical two-point function $g_{\text{num}}(\tau)$ for $\Delta_\phi = 3$ and the identity block $g^{(\mathds{1})}(\tau)$. 
    \textbf{Right:} Comparison between $g_{\text{num}}(\tau)$ after subtraction of the identity block and the stress tensor block $a_T g^{(T)}(\tau)$. }
    \label{fig:blocks}
\end{figure}

\subsection{Bulk interpretation through Witten diagrams}
\label{subsec:BulkInterpretationInWittenDiagramsBB}

We now discuss an interpretation of our bootstrap results in terms of Witten diagrams.
In particular, we show a one-to-one correspondence between our expansion in GFF correlators (regularized to their derivatives, in the case of higher multi-stress tensors contributions) and an expansion of the action around the thermal AdS background for the case of the AdS-Schwarzschild black brane.

\paragraph{Perturbative expansion around thermal AdS.}
Inspired by the form of our correlators, we perform an expansion of the action around thermal AdS.
We define the expansion parameter
\begin{equation}
    \veps 
    =
    \frac{\ell}{\beta}\,,
    \label{eq:ExpansionParameter}
\end{equation}
with $\ell$ the AdS radius, which we keep explicit to allow dimensional analysis.
For small $\veps$, the black brane metric can be expanded in powers of $\veps^d$ around the thermal AdS metric as follows:\footnote{Note that $h^{(d)}$ is traceless.}
\begin{align}
    g_{MN}
    &=
    g_{\text{AdS}\, MN}
    +
    \veps^d
    h_{MN}^{(d)}
    +
    \veps^{2d}
    h_{MN}^{(2d)}
    + \ldots\,, \label{eq:ExpandedMetric}\\
    g^{MN}
    &=
    g_{\text{AdS}}^{MN}
    +
    \veps^d
    h^{(d)\, MN}
    +
    \veps^{2d}
    \left(
    h^{(d)\, M}_{\phantom{(d)\, M}P} h^{(d)\, PN}
    -
    h^{(2d)\, MN}
    \right)
    + \ldots\,.
    \label{eq:ExpandedInverseMetric}
\end{align}
Under this expansion the action \eqref{eq:gravityaction} takes the form
\begin{equation}
    S
    =
    S_\text{thAdS}
    +
    \veps^{d} \delta S^{(d)}
    +
    \veps^{2d} \delta S^{(2d)}
    + \ldots\,,
    \label{eq:ExpandedAction}
\end{equation}
with the thermal AdS action given by
\begin{equation}
    S_\text{thAdS}
    =
    \frac{1}{2} \int_\beta \text{d}^{d+1}y\,
    \sqrt{g_\text{AdS}} \left( g_\text{AdS}^{MN} \pd_M \phi \pd_N \phi + m^2 \phi^2 \right)\,.
    \label{eq:ActionThermalAdS}
\end{equation}
Here we defined the shorthand notation
\begin{equation}
    \int_\beta \text{d}^{d+1}y\,
    =
    \int_0^\infty \text{d}z \int_0^\beta \text{d}\tau \int \text{d}^{d-1}x\,.
    \label{eq:IntShorthand}
\end{equation}
Notice that in order to use the thermal AdS propagators we have changed the integration boundaries of $z$ to $[0, \infty)$ instead of $[0, z_\text{h}]$.
This manipulation leaves a residual \textit{non-perturbative} term which we consider later.
From the action \eqref{eq:ActionThermalAdS} we can determine the scalar propagators in the thermal background:
\begin{align}
    \text{Bulk-to-boundary:} \qquad
    & K_\Delta (z_1, \tau_{12}, x_{12})
    = 
    \sum_{m=-\infty}^\infty K^{(\text{AdS})}_\Delta (z_1, \tau_{12} + m\beta, x_{12})\,, \label{eq:ThermalBulkToBoundaryPropagator} \\
    \text{Bulk-to-bulk:} \qquad
    & G_\Delta (z_{12}, \tau_{12}, x_{12})
    =
    \sum_{m=-\infty}^\infty G^{(\text{AdS})}_\Delta (z_{12}, \tau_{12} +m\beta, x_{12})\,,
    \label{eq:ThermalBulkToBulkPropagator}
\end{align}
which are given in terms of zero-temperature propagators given in~\eqref{eq:AdSBulkToBoundaryPropagator} and~\eqref{eq:AdSBulkToBulkPropagator}.
Further details about the expansion of the metric at small $\veps$ are provided in Appendix~\ref{app:DetailsOnWittenDiagrams}.
The corrections to the action then take the general form
\begin{align}
    \delta S^{(d)}
    &=
    - \frac{1}{2} \int_\beta \text{d}^{d+1}y\,
    \sqrt{g_\text{AdS}}\, h^{(d)\, MN} \pd_M \phi \pd_N \phi\,,
    \label{eq:ExpandedAction_FirstOrder} \\
    \delta S^{(2d)} &=
    \frac{1}{2} \int_\beta \text{d}^{d+1}y\,
    \sqrt{g_\text{AdS}}
    \left[
    \left( \frac{1}{2} h^{(2d)} - \frac{1}{4} h^{(d)}_{MN} h^{(d)\, MN} \right)
    \left( g_\text{AdS}^{MN} \pd_M \phi \pd_N \phi + m^2 \phi^2 \right) \right. \notag \\
    &\phantom{=\ } \left.
    +
    \left( h^{(d)\, M}_{\phantom{(d)\, M}P} h^{(d)\, PN} - h^{(2d)\, MN} \right) \pd_M \phi \pd_N \phi
    \right]\,,
    \label{eq:ExpandedAction_SecondOrder} \\
    & \hspace{-0.75em} \vdots \notag
\end{align}
We can therefore express the boundary thermal correlator perturbatively in $\veps$ as
\begin{equation}
    g(\tau,x)
    =
    g_\text{thAdS} (\tau,x)
    +
    \veps^d g_T (\tau,x)
    +
    \veps^{2d} g_{[T^2]} (\tau,x)
    + \ldots\,,
    \label{eq:ExpandedCorrelator}
\end{equation}
where the choice of notation $g_{[T^n]}$ will soon be justified.
Under this expansion the boundary correlator can be represented conveniently using Witten diagrams.
The insertion rules associated with this action can be gathered as follows:
{\allowdisplaybreaks
\begin{align}
    &\BulkToBoundaryPropagator =\ K_\Delta (z_1, \tau_{12}, x_{12})\,, \\[.4em]
    &\BulkToBulkPropagator\ =\ G_\Delta (z_{12}, \tau_{12}, x_{12})\,, \\[.4em]
    &\VertexOne\ =\ - \frac{1}{2} \int_\beta \text{d}^{d+1}y\,
    \sqrt{g_\text{AdS}}\, h^{(d)\, MN}\, \pd_{1 M} \pd_{2 N}\,, \\[.4em]
    &\VertexTwo\ =\
    \frac{1}{2} \int_\beta \text{d}^{d+1}y\,
    \sqrt{g_\text{AdS}}
    \left[
    \left( \frac{1}{2} h^{(2d)} - \frac{1}{4} h^{(d)}_{MN} h^{(d)\, MN} \right)
    \left( g_\text{AdS}^{MN} \pd_{1 M} \pd_{2 N} + m^2 \right) \right. \notag \\[-.65em]
    & \phantom{\VertexTwo\ =\ } \left. +
    \left( h^{(d)\, M}_{\phantom{(d)\, M}P} h^{(d)\, PN} - h^{(2d)\, MN} \right) \pd_{1 M} \pd_{2 N}
    \right]\,, \label{eq:VertexTwo} \\
    & \qquad \vdots \notag
\end{align}
}%
The explicit expressions for $h^{(d)}$ and $h^{(2d)}$ are given in Appendix~\ref{app:DetailsOnWittenDiagrams}.
Since the action consists only of $\phi^2$ vertices, the correlator takes the form
\begin{equation}
    \begin{split}
    \WittenDiagramBH\ 
    &=\
    \WittenDiagramOne \\
    &\phantom{=\ }
    + \veps^d\ \WittenDiagramTwo \\
    &\phantom{=\ }
    + \veps^{2d}
    \Biggl(\
    \WittenDiagramThree\ 
    +\, \WittenDiagramFour\
    \Biggr)\\
    &\phantom{=\ }
    + \veps^{3d}
    \Biggl(\
    \WittenDiagramFive\ 
    +\, \WittenDiagramSix\
    +\, \WittenDiagramSeven\
    \Biggr)\\
    &\phantom{=\ }
    + \ldots\,.
    \end{split}
    \label{eq:WittenDiagrams}
\end{equation}
The first diagram of each line is a contact diagram, for which the graviton mode is indicated by the number between parenthesis.
The other (exchange) diagrams involve bulk-to-bulk propagators and follow from multiplicative terms in the expansion of $e^{-S}$.
Note that no loops are allowed in this expansion.

In Appendix~\ref{app:DetailsOnWittenDiagrams} we perform an explicit computation of the diagrams up to order $\veps^{2d}$.
We observe that the resulting expression matches with the bootstrap result for the principal and regularized terms.
We expect that this connection holds for any order in $\veps^d$ as a consequence of the following observations: 
\begin{enumerate}
    \item Each diagram can be reduced to a \textit{single} sum of images.
    This can be shown by using the standard trick of shifting the $\tau$-integrals and the sums.
    Generally, any Witten diagram has a number of sums of images given by\footnote{Incidentally, this is also true for Feynman diagrams.}
    \begin{equation}
        \text{number of sums}
        =
        \text{number of propagators} - \text{number of vertices}\,,
        \label{eq:NumberOfSums}
    \end{equation}
    which is always $1$ in this case, as follows from the fact that only $\phi^2$ vertices exist;
    \item For each order, one can analyze what are the operators that contribute to the zero mode by reading the power of $\beta$ and assuming that the correlator follows the principles of CFT at finite temperature.
    It is straightforward to see that, here, the zero mode is given exactly by the multi-stress-tensor contributions, since they are the only universal operators to have dimension $\Delta = n d$ independently of $\Delta_\phi$.
    This means in turn that the zero mode of each order $n$ is given by a sum of the thermal blocks associated to the operators $[T^n]$.
\end{enumerate}
Therefore we conclude that, in this expansion, the $n$-th order has to be given exactly by
\begin{equation}
    g_{[T^n]} (\tau,x)
    =
    \sum_{J=0}^{n d/2} a^{(T)}_{n d,J} \sum_{m=-\infty}^\infty f_{n d, J} (\tau + m\beta, x)\,,
    \label{eq:GMIFromWittenDiagrams}
\end{equation}
where $f_{n d, J} (\tau + m\beta, x)$ are the thermal blocks defined through \eqref{eq:thermal OPE} and $a^{(T)}_{n d,J}$ are multi-stress tensor OPE coefficients.
This expression matches the generalized sum of images discussed in~\cite{Barrat:2025nvu}.
If we set the spatial separation to zero, we recover
\begin{equation}
    g_{[T^n]} (\tau)
    =
    \frac{a_{[T^n]}}{\beta^{2\Delta_\phi}} \left[ \zeta_H \left( 2\Delta_\phi-n d, \frac{\tau}{\beta} \right) + \zeta_H \left( 2\Delta_\phi-n d, 1-\frac{\tau}{\beta} \right) \right]\,,
    \label{eq:HZFromWittenDiagrams}
\end{equation}
in perfect agreement with our bootstrap framework up to the arc contributions.\footnote{The regularized part is also included in this expression after suitably expanding the divergent coefficient multiplying a vanishing GFF block.}
Hints at this structure were already -- directly or indirectly -- observed in \cite{Fitzpatrick:2019zqz,Parisini:2022wkb,Buric:2025anb}. It is important to note that this expansion is not \textit{per se} an expansion of the correlator at small temperature; instead it corresponds to an expansion in the graviton field.
In fact, each term in \eqref{eq:ExpandedCorrelator} admits its own expansion in powers of $\beta$ due to the sum of images.

To better understand this statement, let us view the computation from another perspective, namely as an expansion around thermal AdS interpreted as a high-frequency expansion. 
The effective parameter in the correlators is in fact $\omega z_\text{h}$, which is here taken to be large, similarly to Section 5.4 of \cite{Parisini:2023nbd}. 
Perturbatively at large frequencies, one therefore expects to recover the OPE in momentum space~\cite{Caron-Huot:2009ypo,Manenti:2019wxs}. 
Double-trace operators do not appear in this OPE, in agreement with \eqref{eq:ExpansionInGFFtau}, and this matches the fact that multi-stress tensor operators are zero modes in the expansion around thermal AdS.
This can also be seen by considering the frequency space form of \eqref{eq:ExpansionInGFFtau}, as discussed in~\cite{Barrat:2025nvu}. 
While the framework of this section provides a holographic interpretation for the expansion in GFF correlators~\eqref{eq:ExpansionInGFFtau}, note that the arc contributions are not captured by this high-frequency expansion.
From the connection between the momentum space OPE and the GFF expansion of the correlators, these contributions can be understood as the piece of the correlator that governs the late-time dynamics \cite{Policastro:2002se,Amado:2008ji}.\footnote{We thank Tom Hartman for useful discussions on this point.} 

\paragraph{Non-perturbative effects and arcs.}
It is well known that the OPE in momentum space needs to be corrected with non-perturbative contributions, which become relevant at low frequencies while being exponentially suppressed at high frequencies~\cite{Caron-Huot:2009ypo,Manenti:2019wxs}.
Since the connection between the momentum space OPE and the dispersion relation is pointed out in \cite{Barrat:2025nvu}, here we focus on the relation between the arc contributions and the non-perturbative corrections at small frequencies.

Since the expansion in thermal AdS discussed above admits the interpretation of a high-frequency expansion, we expect these contributions to correspond to non-perturbative effects in that setup. 
Indeed during the calculation we made the crucial approximation of extending the integration region from $z_\text{h}$ to $\infty$ in the Witten diagrams. 
If $z_\text{h}$ is large but finite, this approximation is justified because the corrections are exponentially suppressed.
More precisely, consider the full action
\begin{align}
        S
        &= \int_0^{z_\text{h}} \! \mathrm{d}z \int_0^\beta \! \mathrm{d}\tau \int \! \mathrm{d}^{d-1} x\, \mathcal{L} \notag \\
        &= \int_0^\infty \! \mathrm{d}z \int_0^\beta \! \mathrm{d}\tau \int \! \mathrm{d}^{d-1} x\, \left(\mathcal{L}_\text{thAdS} + \ldots\right)
        - \int_{z_\text{h}}^\infty \! \mathrm{d}z \int_0^\beta \! \mathrm{d}\tau \int \! \mathrm{d}^{d-1} x\, \mathcal{L}\,.
\end{align}

The first term corresponds to our calculation, while the second term is difficult to evaluate exactly and encodes the non-perturbative effects at high frequencies. 
It is possible however to give an estimation of its contribution.
The simplest way is to observe that
\begin{equation}
      \int \mathcal D[\phi]\exp\left(-\int_{z_\text{h}}^\infty \! \mathrm{d}z \int_0^\beta \! \mathrm{d}\tau \int \! \mathrm{d}^{d-1} x\, \mathcal{L}\right)
      \;\sim\; e^{-\left(F_\text{thAdS}-F\right)} 
      \;\sim\; e^{-\mathcal{S}} 
      \;\sim\; \mathcal{O} \bigl(e^{-\veps^{d-1}}\bigr) \, ,
\end{equation}
where here $\mathcal{S}$ is the entropy of the black brane.
This estimation highlights the fact that this contribution is non-perturbative in $\veps$.
In different words, the expansion around thermal AdS that reproduces the GFF expansion~\eqref{eq:ExpansionInGFFtau} does not (perturbatively) take into account horizon effects.

A yet more accurate estimation of the non-perturbative contributions can be obtained by considering
\begin{equation}
      \int_{z_\text{h}}^\infty \! \mathrm{d}z \int_0^\beta \! \mathrm{d}\tau \int \! \mathrm{d}^{d-1} x\, \mathcal{L}\,,
\end{equation}
expanded near the horizon, i.e., at $z \sim z_\text{h}$. 
It is straightforward to estimate the leading order contribution.
At this order, the equation of motion reduces to
\begin{equation}\label{eomWKB}
    \psi_{\omega,k} '(z)+ \frac{\ell^2}{4 z_\text{h}}\left(\frac{m^2 \ell^2}{ z_\text{h}^2}+k^2+\omega^2\right)\psi_{\omega,k}(z) = 0 \,,
\end{equation}
where we have defined
\begin{equation}
    \Phi(z,\tau,x) = T\sum_{n} \int \frac{\text{d}^{d-1}k}{(2\pi)^{d-1}} e^{ \omega_n \tau + i k x} \psi_{\omega,k} (z)\, , \quad \omega_n=\frac{2 \pi n}{\beta} \, .
\end{equation}
The solution of~\eqref{eomWKB} is
\begin{equation}
   \psi_{\omega,k}(z) =\psi_c \exp \!\left[-\frac{z \ell^2}{4 z_\text{h}}\left(\frac{m^2 \ell^2}{ z_\text{h}^2}+k^2+\omega^2\right)\right] \,,
\end{equation}
with $\psi_c$ a constant fixed by boundary conditions.  
The contribution of the bulk-to-boundary propagator given above to the thermal two-point function of the boundary CFT is therefore 
\begin{equation}
    g_\text{n.p.}(k,\omega) \sim e^{-(k^2+\omega^2)} \,.
\end{equation}
As a consequence, $g_\text{n.p.}$ precisely captures the non-perturbative terms in momentum space as expected from the arguments above. 
Since the perturbative expansion in $\veps$ matches the GFF expansion in~\eqref{eq:ExpansionInGFFtau}, we identify
\begin{equation}
    g_\text{n.p.}(\tau,x) \;=\; g_\text{arcs}(\tau,x)\,.
\end{equation}
This gives a novel interpretation of $g_\text{arcs}(\tau,x)$ as encoding horizon effects, which are not captured by the GFF expansion.  

Since in this paper we have been interested in the zero spatial separation limit, an interesting direction, currently under investigation, is the extension of the present analysis to non-zero spatial separation, $x \neq 0 $, as proposed in~\cite{Barrat:2025nvu}.
In this case, we expect $g_{\text{arcs}}(x,\tau)$ to encode extra dynamical information, like the thermal mass of the theory for instance.
This expectation can in principle be tested by employing exact momentum-space formulae~\cite{Aminov:2020yma,Dodelson:2022yvn,Dodelson:2023vrw} and/or by developing a more precise version of the near-horizon expansion, which we have only sketched here.
We hope to report our results in future work.

\subsection{Higher-curvature corrections}
\label{subsec:StringyCorrectionsBB}

We now conclude our study of the black brane background by considering higher-order curvature corrections.
Such extensions may correspond to different types of gravitational theories~\cite{Lovelock:1971yv,Myers:2009ij,Camanho:2009vw,deBoer:2009gx} as well as stringy corrections with $\alpha' \sim 1/\sqrt{\lambda}$ as a parameter.
In order to preserve conformal symmetry, they can only modify the functions $f(z)$ and $h(z)$ presented in~\eqref{eq:fandh}.
We parameterize them as an expansion around the black brane metric,
\begin{align}
    f(z) &= 1 - f_0 z^4 - f_4 z^8 - f_8 z^{12} - \ldots\,, \label{eq:frFullBG} \\
    h(z) &= 1 - h_0 z^4 - h_4 z^8 - h_8 z^{12} - \ldots\,, \label{eq:hrFullBG}
\end{align}
with $f_0 = h_0$ needed as before for consistency with conformal symmetry.
There is no such constraint, however, on $f_4$, $h_4$, and higher-order terms. 
It is also important to note that, from the perspective of the thermal CFT, those corrections do not induce anomalous dimensions. 
Thus only the thermal OPE coefficients are expected to be modified.

In order to obtain the principal contributions to the correlators with modified thermal OPE coefficients, it is sufficient to consider~\eqref{eq:ExpansionGFFDelta1ff}--\eqref{eq:ExpansionGFFDelta5ff} and insert the corrected multi-stress tensor OPE coefficients~\cite{Fitzpatrick:2019zqz}, namely,
\begin{align}
    {\textcolor{orange}{a_T}} &= \frac{\pi^4 \Delta_\phi}{40}\,, \\
    {\textcolor{purple}{a_{[T^2]}}} &= {\textcolor{orange}{a_T}}^2
    \frac{63 \Delta_\phi ^4-413 \Delta_\phi ^3+672 \Delta_\phi ^2-88 \Delta_\phi +144}{126 \Delta_\phi(\Delta_\phi -4) (\Delta_\phi -3) (\Delta_\phi -2) } \notag \\
    &\phantom{=\ } - \frac{\Delta_\phi  (\Delta_\phi +1) (4 \Delta_\phi (5-2 \Delta_\phi )  \tilde{f}_4 + (\Delta_\phi (\Delta_\phi -4) + 24) \tilde{h}_4)}{5040 (\Delta_\phi
   -4) (\Delta_\phi -3) (\Delta_\phi -2)} \,,
\end{align}
where we set $f_0 = \pi^4/\beta^4$ and defined the dimensionless quantities $\tilde f_4 = f_4 \beta^8$ and $\tilde h_4 = h_4 \beta^8$.
Note that the first affected coefficients are those of the $[T^2]$ sector, specifically $a_{8,2}$ and $a_{8,0}$, while $a_{8,4}$ belongs to the universal lowest-twist family that depends only on $\tilde{f}_0$.
Since $a_{T}$ depends solely on $\tilde{f}_0$, the correlators for $\Delta_\phi \leq 4$ are identical across all backgrounds.
The correlator with $\Delta_\phi = 5$ is the first to contain higher terms ($\tilde{f}_4$ and $\tilde{h}_4$).
Correlators with higher $\Delta_\phi$ involve higher and higher corrections $\tilde{f}_k$ and $\tilde{h}_k$; they can be derived systematically following the same procedure.
Corrections to the regularized contributions (and consequently, to the arcs contributions) can be derived following a similar strategy, even though (as already highlighted before) a full prediction requires the knowledge of an infinite number of OPE coefficients.

\section{Black hole background}
\label{sec:BlackHoleBackground}

We now apply the analytic bootstrap framework to holographic thermal correlators in spherically symmetric black hole backgrounds.
We first extend the analysis of the analytic properties of thermal two-point functions to the case of correlators at finite temperature and finite volume, namely on the geometry $S_\beta^1 \times S_R^{d-1}$, which has seen interesting developments in recent years from a finite temperature point of view \cite{Shaghoulian:2016gol,Gobeil:2018fzy,Alkalaev:2024jxh,Allameh:2024qqp,Buric:2024kxo,David:2024pir,Belin:2025nqd,Buric:2025uqt}. 
We show that the two-point function in the zero spatial separation limit admits an expansion in finite volume GFF correlators, which is the analogue of~\eqref{eq:ExpansionInGFFtau}. 
This decomposition is valid for theories whose spectrum contains only operators of integer conformal dimension.
We specialize this formula to the study of holographic correlators and obtain analytic results for the principal contribution in the case $d=4$.

\subsection{Thermal two-point functions}
\label{subsec:ThermalTwoPointFunctions2}

We study two-point functions of identical scalar operators at finite temperature and finite volume.
More precisely, we consider correlators on the geometry $S^1_\beta \times S^{d-1}_R$.
We denote by $R$ the radius of the $(d-1)$-dimensional sphere. 
As in the infinite volume case it is convenient to define 
\begin{equation}
    g_{\bR}(\tau,x)=\vev{\phi(0,0)\, \phi(\tau,x)}_{\beta/R} \equiv \vev{\phi(0,0)\, \phi(\tau,x)}_{S_\beta^1 \times S_R^{d-1}}\,.
    \label{eq:gDefinitionR}
\end{equation}
We focus on the limit in which the two operators have zero spatial separation and denote the correlator by $g_{\bR}(\tau) \equiv g_{\bR}(\tau,0)$. 

\bigskip

In this section we discuss the adaptation of the bootstrap axioms of Section~\ref{subsec:ThermalTwoPointFunctions} to the finite volume case. 

\paragraph{OPE.} 
The OPE remains valid on the geometry $S^1_\beta \times S^{d-1}_R$ but converges only when $\tau^2 + x^2 < \min\!\left(\beta^2,\, 4 \pi^2 R^2\right)$.
In the limit $x=0$, the two-point function admits the following expansion in thermal blocks:
\begin{equation}
    g_{\bR}  (\tau)
    =
    \frac{1}{\beta^{2\Delta_\phi}}
    \sum_{\Delta} a_\Delta (\beta/R)\left[\frac{1}{2}\left( \frac{\beta}{R} \right)\text{csch}\left(\frac{\tau}{2 R}\right)\right]^{2 \Delta_\phi-\Delta}\,,
    \label{eq:OPEFiniteV}
\end{equation}
with
\begin{equation}
    a_\Delta (\beta/R) = \sum_{\Om:\, \Delta_\Om = \Delta}  f_{\phi\phi \Om} b_\Om (\beta/R) \left(\frac{J!}{2^J (\nu)_J}\right) C^{(\nu)}_{J}(1) \,.
\end{equation}
The decomposition is justified by the fact that~\eqref{eq: 1ptfunction} generalizes as follows 
\cite{Iliesiu:2018fao}:
\begin{equation}
    \vev{\mathcal{O}^{\mu_1\ldots \mu_J}}_{\beta/R}=\frac{b(\beta/R)}{\beta^{\Delta}}\left(e^{\mu_1}\ldots e^{\mu_J}-\text{traces}\right)\,.
\end{equation}
Starting from this expression, it is possible to express the OPE similarly to the infinite volume case. Then, using finite volume variables leads to the expansion~\eqref{eq:OPEFiniteV}.

\paragraph{KMS condition.}
The two-point function of identical scalars satisfies the KMS condition which, associated with time reversal symmetry, reads~\cite{Kubo:1957mj,Martin:1959jp}
\begin{equation}
    g_\bR(\tau) = g_\bR(\beta-\tau)\,.
\end{equation}
At finite volume there is an additional periodicity coming from the sphere geometry $S_R^{d-1}$.
In this work we focus on correlators at zero-spatial coordinates and thus leave the implication of this condition for future studies.

\begin{figure}[t]
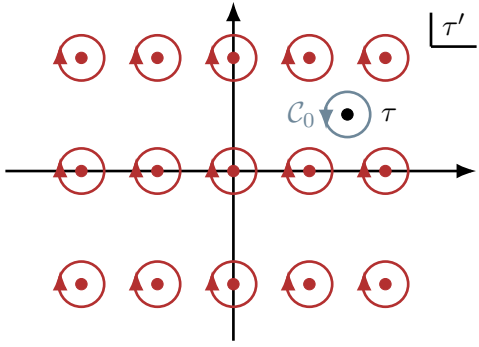

    \centering
    \AnalyticStructureTauPolesV
    \caption{
    Analytic structure at finite temperature and finite volume for integer spectrum, when the OPE coefficients do not contain poles.
    Due to the periodicity, poles are now located at every $m \beta$, $m \in \mathbb{Z}$, in the real direction and every $2\pi n R$, $n \in \mathbb{Z}$, in the imaginary direction.
    The principal part of the correlator is given by the sum of residues of these poles.
    }
    \label{fig:AnalyticStructureV}
\end{figure}

\paragraph{Analytic structure and GFF expansion.}
In~\cite{Barrat:2025nvu} the authors discuss the analytic structure of finite temperature correlators at infinite volume.
Here we extend the conclusions to include the case in which the geometry is $S_\beta^1 \times S_R^{d-1}$.
We focus on the case in which the spectrum consists only of operators of integer dimensions, i.e., we restrict here our analysis to the principal contribution of the correlators, in the sense defined in Section~\ref{subsec:ExactTwoPointFunctionsForIntegerDeltaBB}.
In the following we describe the associated analytic structure in the complex $\tau$-plane and its consequences.

The restriction to integer conformal dimensions (and no pole in the OPE coefficients) implies that only poles are allowed in the $\tau$-plane.
Because of the compact space, the poles are replicated not only along the real $\tau$-axis but also along the imaginary $\tau$-axis, together with their Matsubara images.
This can be seen by noticing that the zero-temperature finite volume two-point function is given by 
\begin{equation}
    \langle \phi(\tau)\phi(0) \rangle_{\mathbb{R} \times S_R^{d-1}} 
    = \left[\frac{1}{2R}\text{csch}\left( \frac{\tau}{2 R}\right)\right]^{2\Delta_\phi}\,,
\end{equation}
which has poles on the imaginary $\tau$-axis corresponding to UV images located at $i \tau = 2 \pi R k$, $k \in \mathbb{Z}$. 
Equivalently, this follows from the observation that
\begin{equation}
    \text{d}s^2_{\mathbb{R}^{d}} = \text{d}r^2+ r^2\, \text{d}s^2_{S_R^{d-1}} 
    = \frac{e^{2\tau/R}}{R^2} \left(\text{d}\tau^2+ R^2 \text{d}s^2_{S_R^{d-1}}\right) 
    = \frac{e^{2\tau/R}}{R^2}\, \text{d}s^2_{\mathbb{R}\times S_{R}^{d-1}}\,.
\end{equation}
We conclude that the analytic structure of the principal part of thermal correlators at finite volume for integer $\Delta_\phi$ is the one presented in Figure~\ref{fig:AnalyticStructureV}.

It is now straightforward to apply Cauchy's integral formula to rewrite the correlator as
\begin{equation}
    g_{\bR}(\tau) = \oint_{\mathcal{C}_0} \frac{\text{d}\tau'}{2\pi i}\, \frac{g_{\bR}(\tau')}{\tau'-\tau}\,,
\end{equation}
which can then be expressed as a sum over residues:
\begin{equation}
    g_{\bR}(\tau)
    =
    - \sum_{m=-\infty}^\infty \sum_{n=-\infty}^\infty 
    \underset{\tau'=0}{\text{Res}} \left[ \frac{g_{\bR}(\tau')}{\tau'-\tau+m\beta+2\pi i nR} \right] 
    + g_\text{arcs}(\tau)\,.
\end{equation}
The next step is to substitute the OPE into this expression, which yields
\begin{align}
    g_{\bR}(\tau)
    &=
    -\frac{1}{\beta^{2\Delta_\phi}}\sum_{\Delta} a_\Delta (\beta/R)\left( \frac{\beta}{2R} \right)^{2 \Delta_\phi-\Delta} \sum_{m=-\infty}^\infty \sum_{n=-\infty}^\infty 
    \underset{\tau'=0}{\text{Res}} \left[ \frac{\sinh^{2 \Delta_\phi-\Delta}\left(\frac{\tau'}{2 R}\right)}{\tau'-\tau+m\beta+2\pi i nR} 
    \right] \notag \\
    & \phantom{=\ } + g_\text{arcs}(\tau)\,.
\end{align}
The residues reduce to propagators in the complex $\tau$-plane, shifted along both the real and imaginary directions.
Moreover, they give a non-zero contribution only for $\Delta \leq 2\Delta_\phi-1$. 
After performing the sum over images, we identify each block accompanying an OPE coefficient with a GFF correlator, producing the decomposition
\begin{equation}
    g_\text{pr} (\tau)
    =
    \sum_{\Delta} \frac{a_\Delta}{\beta^{\Delta}} g_{\text{GFF}}\left(\Delta_\phi - \frac{\Delta}{2}, \tau\right) 
    +
    g_\text{arcs}(\tau)\,,
    \label{eq:ExpansionInGFFtauV}
\end{equation}
where the GFF blocks are defined as (see Equations~\eqref{eq: GFF result} and~\eqref{eq: coeff c})
\begin{equation} \label{eq:Gfffinitevol}
    g_{\text{GFF},\bR} (\Delta,\tau)
    =
     \frac{1}{\beta^{2\Delta}\Gamma(2\Delta)} \sum_{\ell=1}^{\Delta} \left(\frac{\beta^2}{R^2} \right)^{\Delta-\ell}c(\Delta, \ell) \left[ \psi_{e^{- \beta/R}}^{(2 \ell-1)}\left(\frac{\tau}{\beta}\right)+\psi_{e^{- \beta/R}}^{(2 \ell-1)}\left(1-\frac{\tau}{\beta}\right) \right]  \ ,
\end{equation}
where $\psi^{(n)}_q (z)$ is the $n$-th derivative of the $q$-Polygamma function defined as
\begin{equation}
    \psi^{(n)}_q (z)
    =
    \frac{\pd^n \psi_q (z)}{\pd z^n}\,, \qquad
    \psi_q (z)
    =
     \frac{\pd }{\pd z}\log \Gamma_q (z)\,,
\end{equation}
with $\Gamma_q (z)$ the $q$-Gamma function.\footnote{The $q$-Gamma function can be defined as $\Gamma_q(x) = (1-q)^{1-x}\prod_{i}\left(1-q^{i+1}\right)/(1-q^{i+x})$.} The derivation of the thermal finite volume GFF correlators and their basic properties can be found in Appendix~\ref{app:SphericalGFF}.
As a consistency check, in Appendix~\ref{app:ONModelAtLargeN}, we show that the two-point function of lightest scalars in the large $N$ limit of the $3d$ $\mathrm O(N)$ model satisfies the expansion in~\eqref{eq:ExpansionInGFFtauV}.

The black hole correlators are expected to additionally consist of regularized and arc contributions, as discussed in Section~\ref{subsec:ExactTwoPointFunctionsForIntegerDeltaBB} for the black brane.
We reserve the study of these contributions to future work, however we note that there is no apparent obstacle in developing a strategy similar to the one used for the black brane background.

\paragraph{Boundedness conditions.}
As in the infinite volume case, the correlator is bounded in the complex $\tau$-plane by its value on the real axis,
\begin{equation}\label{eq:boundessBH}
   |g(\tau)| \;\geq\; |g(\tau + i \eta)|\,, 
   \qquad \tau, \eta \in \mathbb{R}\,,
\end{equation}
and also by its value on the imaginary axis,
\begin{equation}\label{eq:boundessBH2}
   |g(i \eta)| \;\geq\; |g(\tau + i \eta)|\,, 
   \qquad \tau, \eta \in \mathbb{R}\,.
\end{equation}
The proofs are available in the literature for the infinite volume case~\cite{Fidkowski:2003nf,Barrat:2025nvu} and can be straightforwardly adapted to the finite volume case.

\paragraph{Clustering condition.}
At finite volume the clustering condition no longer holds.
Additionally the real-time correlators (such as the Wightman correlator) is periodic in $t=i\tau$ and thus the limit  $t\to\infty$ is not sufficient to fix the arc contributions.
Fixing the arcs is a very interesting problem on its own and we leave this to future work.\footnote{For the sake of clarity, note that the arcs are unique up to a constant with respect to the position of the operator.
This constant is however a function of the dimensionless radius $\beta/R$.}

\subsection{Holographic thermal correlators}
\label{subsec:HolographicThermalCorrelators2}

We consider, as before, a free massive scalar field in the bulk, described by the action \eqref{eq:gravityaction}, with the difference that the spatial integral is now over the spatial sphere.
The black hole background corresponding to a spherical horizon can be written as
\begin{equation}
    \text{d}s^2 =
    \left(1+\frac{f(z)}{z^2}\right) \text{d}\tau^2
    + \frac{\text{d}z^2}{z^4 \bigl(1+h(z)/z^2\bigr)}
    + \frac{\text{d}\Omega_{d-1}^2}{z^2}\,,
    \label{eq:metricBH}
\end{equation}
where $\text{d}\Omega_{d-1}^2$ is the metric on the $(d-1)$-dimensional sphere.  
The asymptotic AdS boundary conditions again constrain the functions $f(z)$ and $h(z)$ as
\begin{equation}
    f(z) = 1 - f_0 z^d + \ldots, 
    \qquad 
    h(z) = 1 - h_0 z^d + \ldots\,.
    \label{eq:fandhBH}
\end{equation}
In~\cite{Fitzpatrick:2019zqz} it was shown that conformal invariance on the boundary requires $f_0 = h_0$ as for the black brane.

\paragraph{Black hole spectrum.}
The operator spectrum in the black hole background differs slightly from the black brane case presented in~\eqref{eq:BlackBraneSpectrum}. 
In addition to the multi-stress tensor operators $[T^n]$, operators of the schematic form $\partial^\ell T^{n}$ must also be included, where the derivatives are arranged such that they produce conformal primaries. 
A list of the lightest such operators can be found in Table~6 of~\cite{Fitzpatrick:2019zqz}. 
Starting at $\Delta = 10$, additional operators of this type appear in the spectrum of the black hole background with respect to the black brane.

For the low-lying spectrum, i.e., $\Delta \leq 8$, the OPE coefficients coincide with the black brane ones given in~\eqref{eq:a42}--\eqref{eq:a80} when expressed in terms of $f_0$.
However, to make contact with the finite temperature and volume CFT, in the black hole background the dimensionless quantity $\tilde{f}_0$ should be replaced by
\begin{align} 
    \tilde{f}_0 = \frac{1}{4} \left(-\frac{\beta^4}{R^4} 
    - 2\pi^2 \frac{\beta^2}{R^2}
    \pm 2\pi^3 \sqrt{\pi^2 - 2\frac{\beta^2}{R^2}} 
    + 2\pi^4\right)\,,
    \label{eq:newtildef}
\end{align}
which we obtain by resolving the black hole singularity, i.e.,
\begin{align}
    \frac{R}{\beta}
    =
    \left.
    \frac{1}{4\pi}
    \frac{\mathrm{d}(f(z)/z^2)}{\mathrm{d}z}
    \right|_{z=z_\text{H}}\;,
\end{align}
and ensuring that $\tilde f_0$ reduces to the black brane expression given in \eqref{eq:f0} as we take $R\to\infty$. In Equation \eqref{eq:newtildef} the two signs correspond to the two possible solutions, associated to a \textit{small} black hole ($-$) and a \textit{large} black hole ($+$).
Although in the following we are agnostic about the choice of solution, we mention that the small black hole is known to be unstable thermodynamically while the large one is stable~\cite{Hawking:1982dh,Witten:1998zw,Jokela:2015sza}.

It should be noticed that, while in the black brane case $\tilde{f}_0$ (and thus the one-point functions of multi-stress tensor operators) is a constant, in the black hole case $\tilde{f}_0$ becomes a \textit{function} of the dimensionless ratio $\beta/R$. 
Consequently, the thermal one-point functions of the multi-stress tensor operators also depend non-trivially on $\beta/R$.

In the following we apply the analytic bootstrap framework (of which Equation~\eqref{eq:ExpansionInGFFtauV} is the keystone) to holographic setups in the strong-coupling regime in order to derive the principal contribution to two-point functions of scalar operators.
As already mentioned, additional operators of the schematic form $\partial^\ell T^{n}$ with conformal dimensions $\Delta = dn + \ell$ also appear. 
The method of~\cite{Fitzpatrick:2019zqz} can be extended to compute the associated OPE coefficients.\footnote{See in particular Appendix~A therein, in which exact expressions were obtained for the operators up to dimension $\Delta = 10$ (see Equations~(5.25)--(5.30)), which consist of two stress tensors and two derivatives.}

\subsection{Two-point functions for integer $\Delta_\phi$ in $4d$}
\label{subsec:ExactTwoPointFunctionsForIntegerDeltaBH}

As noted in~\cite{Fitzpatrick:2019zqz} and reviewed above, the OPE coefficients for the multi-stress tensor sector for the black brane and spherical black hole can be computed directly from the equations of motion in AdS.
We can therefore immediately expand in GFF blocks, following the same logic as in Section~\ref{sec:BlackBraneBackground}, in order to derive the principal contributions to the associated correlators.
We focus again on the integer $\Delta_\phi$ case.\footnote{For the non-integer case it would be interesting to develop a finite-volume approximation similar to the asymptotic model of~\cite{Buric:2025fye}.}

Using the finite-volume GFF correlator given in Appendix~\ref{app:SphericalGFF} we define to streamline the notation the object
\begin{equation}
    \mathcal{B}_{\beta/R}^{(\ell)}(\tau)=\psi _{e^{-\beta/R}}^{(\ell)}\left(\frac{\tau}{\beta} \right)+\psi _{e^{-\beta/R}}^{(\ell)}\left(1-\frac{\tau}{\beta} \right) \,.
\end{equation}
We readily obtain the following expressions: 
{\allowdisplaybreaks
\begin{align}
    &\Delta_\phi = 1: \quad && g_{\text{pr},\bR} (\tau) = \frac{1}{\beta^2} \mathcal{B}_{\beta/R}^{(1)}(\tau) 
    \,, \label{eq:CorrelatorFiniteV1} \\[6pt]
    &\Delta_\phi = 2: && g_{\text{pr},\bR} (\tau) =  \frac{1}{\beta^4} \left[\frac{1}{6} \mathcal{B}_{\beta/R}^{(3)}(\tau) -\frac{1}{6} \left(\frac{\beta}{R}\right)^2 \mathcal{B}_{\beta/R}^{(1)}(\tau)\right] 
    \,, \\[6pt]
    &\Delta_\phi = 3: && g_{\text{pr},\bR} (\tau) = \frac{1}{\beta^6} \left[\frac{1}{120} \mathcal{B}_{\beta/R}^{(5)}(\tau)-\frac{1}{24} \left(\frac{\beta}{R}\right)^2 \mathcal{B}_{\beta/R}^{(3)}(\tau)+ \frac{1}{30} \left(\frac{\beta}{R}\right)^4 \mathcal{B}_{\beta/R}^{(1)}(\tau)\right] \notag \\
    & && \phantom{g_{\bR} (\tau) =\ } + {\textcolor{orange}{a_T\left(\beta/R\right)}} \frac{1}{\beta^6}  \mathcal{B}_{\beta/R}^{(1)}(\tau)  
    \,, \\[6pt]
    &\Delta_\phi = 4: && g_{\text{pr},\bR} (\tau) = \frac{1}{\beta^8}\left[\frac{1}{5040}\mathcal{B}_{\beta/R}^{(7)}(\tau)-\frac{1}{360} \left(\frac{\beta}{R}\right)^2 \mathcal{B}_{\beta/R}^{(5)}(\tau)+\frac{7}{720} \left(\frac{\beta}{R}\right)^4 \mathcal{B}_{\beta/R}^{(3)}(\tau)\right. \notag \\
    &  && \phantom{g_{\bR} (\tau) =\ } \left. -\frac{1}{140} \left(\frac{\beta}{R}\right)^6 \mathcal{B}_{\beta/R}^{(1)}(\tau)\right] + {\textcolor{orange}{a_T\left(\beta/R\right)}} \frac{1}{\beta^8}  \left[\frac{1}{6} \mathcal{B}_{\beta/R}^{(3)}(\tau)\right. \notag\\
    & && \phantom{g_{\bR} (\tau) =\ } \left. -\frac{1}{6} \left(\frac{\beta}{R}\right)^2 \mathcal{B}_{\beta/R}^{(1)}(\tau)\right]
    \,, \\
    &\Delta_\phi = 5: && g_{\text{pr},\bR} (\tau) = \frac{1}{\beta^{10}}
    \left[
    \frac{1}{362880}\mathcal{B}_{\beta/R}^{(9)}(\tau)
    - \frac{1}{12096} \left(\frac{\beta}{R}\right)^2 \mathcal{B}_{\beta/R}^{(7)}(\tau) \right. \notag \\
    &  && \phantom{g_{\bR} (\tau) =\ } \left.
    +\frac{13}{17280} \left(\frac{\beta}{R}\right)^4 \mathcal{B}_{\beta/R}^{(5)}(\tau)
    -\frac{41}{18144} \left(\frac{\beta}{R}\right)^6 \mathcal{B}_{\beta/R}^{(3)}(\tau) \right. \notag \\
    &  && \phantom{g_{\bR} (\tau) =\ } \left.
    +\frac{1}{630} \left(\frac{\beta}{R}\right)^8 \mathcal{B}_{\beta/R}^{(1)}(\tau)
    \right]
    + {\textcolor{orange}{a_T\left(\beta/R\right)}} \frac{1}{\beta^{10}} \left[\frac{1}{120} \mathcal{B}_{\beta/R}^{(5)}(\tau)-\frac{1}{24} \left(\frac{\beta}{R}\right)^2 \right. \notag \\
    &  && \phantom{g_{\bR} (\tau) =\ } \left. \mathcal{B}_{\beta/R}^{(3)}(\tau)+ \frac{1}{30} \left(\frac{\beta}{R}\right)^4 \mathcal{B}_{\beta/R}^{(1)}(\tau)\right]
    + {\textcolor{purple}{a_{[T^2]}\left(\beta/R\right)}} \frac{1}{\beta^{10}} \mathcal{B}_{\beta/R}^{(1)}(\tau) \notag \\
    &  && \phantom{g_{\bR} (\tau) =\ }.
    \label{eq:CorrelatorFiniteV5}
\end{align}
}%
This reproduces the pattern observed for the black brane background in~\eqref{eq:ExpansionGFFDelta1ff}--\eqref{eq:ExpansionGFFDelta5ff}. 
The final closed form of these principal contributions can be found by using~\eqref{eq:newtildef} to obtain
\begin{align}
    {\textcolor{orange}{a_T\left(\beta/R\right)}}
    &=
    \frac{\Delta_\phi }{160} \left(-\frac{\beta^4}{R^4} 
    - 2\pi^2 \frac{\beta^2}{R^2}
    \pm 2\pi^3 \sqrt{\pi^2 - 2\frac{\beta^2}{R^2}} 
    + 2\pi^4\right)\,, \\
    {\textcolor{purple}{a_{[T^2]}\left(\beta/R\right)}}
    &=
    {\textcolor{orange}{a_T\left(\beta/R\right)}}^2
    \frac{63 \Delta_\phi ^4-413 \Delta_\phi ^3+672 \Delta_\phi ^2-88 \Delta_\phi +144}{126 \Delta_\phi(\Delta_\phi -4) (\Delta_\phi -3) (\Delta_\phi -2) }\,.
\end{align}
The principal contributions of the correlators can be obtained by inserting these values in~\eqref{eq:CorrelatorFiniteV1}--\eqref{eq:CorrelatorFiniteV5}.
Results for higher $\Delta_\phi$ can be derived in the same way, provided sufficient knowledge of multi-stress tensor OPE coefficients.

It would be of fundamental importance to compute the regularized and arc contributions for the black hole case as well, similarly to the black brane case in Section~\ref{subsec:ExactTwoPointFunctionsForIntegerDeltaBB}.
However, the lack of an asymptotic form for the OPE coefficients prevents us from studying the arc contributions, whose starting point is the location of the bouncing singularity in the complex $\tau$-plane.
Secondly, although we expect the higher multi-stress tensor contributions to be regularized as in Equation~\eqref{eq: greg}, the formula~\eqref{eq:Gfffinitevol} prevents an exact computation of the derivative of the GFF block for the time being.
Finally, finite volume correlators do not satisfy the clustering property~\eqref{eq:clustering} and, as a result, the arc contributions cannot be completely fixed.
Future work will be dedicated to a deeper understanding of correlators in the black hole background.

A non-trivial check is the fact that the two-point function at finite volume should reproduce the one at infinite volume, i.e., the black brane, in the infinite volume limit.
Since our thermal two-point functions are made by a finite sum over GFF correlators it is enough to show that this is the case for the GFF correlator.
This is discussed in~Appendix \ref{app:SphericalGFF}, and indeed the correlators in this section reduce to the ones listed in \eqref{eq:CorrelatorFiniteV1}-- \eqref{eq:CorrelatorFiniteV5}.
This limit additionally provides a constraint or a check on the expected regularized and arc contributions.

In the case of the black brane, we successfully compared our approximate analytic solution with the numerical solution of the wave equation. 
However, for the spherical black hole case, there are additional limitations which we discuss now. 
First, obtaining an accurate numerical solution is technically more challenging due to the increased complexity of the wave equation.
Second, the analytic solutions for the black hole background contains an ambiguity encoded in an arbitrary function $\kappa_{\Delta_\phi}(\beta/R)$. 
By fixing $\beta$ and $R$, one could in principle tune this function to match the numerical result; however, such procedure reduces the discrepancy in an (almost) arbitrary way. 
For this reason, we do not consider the numerical solution of the wave equation as a meaningful consistency check in this case.\footnote{It would be interesting to develop a more stable and precise way to solve the wave equation numerically.} Note that the fact that the solution is unique up to a constant in the kinematics is already a check of our correlators.

The Witten diagram interpretation discussed in Section~\ref{subsec:BulkInterpretationInWittenDiagramsBB} also differs in the black hole case. 
The expansion proposed in Section~\ref{subsec:BulkInterpretationInWittenDiagramsBB} effectively corresponds to a large $\beta$ expansion, which is less meaningful in the black hole background since solutions exist only for $\beta^2 < \pi^2 R^2 / 2$.\footnote{This can be seen from the fact that the square root in Equation~\eqref{eq:newtildef} is real only for these values of $\beta/R$.}
Whether an alternative bulk interpretation of these blocks exists remains an open question, which we plan to revisit in future work.

Let us also comment that, according to the ETH, the thermal two-point function is related to a zero-temperature four-point function of the form
\begin{equation}
    g_{\bR}(\tau,x) \sim \vev{\mathcal{O}_H (\infty) \phi(z,\zb) \phi(1) \mathcal{O}_H(0)}\,,
\end{equation}
where the dimension of the heavy operator $\Delta_H$ is taken to be large and $z, \zb$ are flat four-point cross-ratios. 
The holographic realization of this correspondence has been extensively studied in~\cite{Kulaxizi:2018dxo,Karlsson:2019dbd,Li:2019zba,Karlsson:2019qfi,Parnachev:2020fna,Li:2020dqm,Karlsson:2020ghx,Karlsson:2021duj,Dodelson:2022yvn}.
However, the case in which the dimension of the light scalar operator $\Delta_\phi$ is an integer turns out to be more subtle and technically challenging, and only limited data are currently available~\cite{Li:2019zba,Li:2020dqm}.
Having analytic expressions for thermal two-point functions may thus prove useful, as they can be directly compared with heavy–heavy–light–light correlator data. 
This provides an additional motivation to further investigate the thermal two-point function at non-zero spatial separation, where the degeneracy in spin is lifted.
In this regime, one could also identify which contributions are perturbative and which are non-perturbative in the large-spin expansion, potentially allowing new predictions for these correlation functions.

Finally, as discussed in Section~\ref{subsec:StringyCorrectionsBB}, it is also possible to include higher-curvature corrections in the computation of the correlation functions. 
As in the black brane case, the operator spectrum remains unchanged; the only difference is that the OPE coefficients of the multi-stress tensor operators now depend on the Wilson coefficients multiplying the higher-curvature terms, or equivalently on the parameters $f_4, h_4, \ldots$. 
Since the procedure to compute these corrections is identical to that described in Section~\ref{subsec:StringyCorrectionsBB}, and the resulting expressions are not particularly illuminating, we do not report them here.

\section{Conclusions}
\label{sec:Conclusions}

In this work we analyzed thermal two-point functions of scalar operators with integer scaling dimension $\Delta_\phi$, both at infinite volume and at finite volume on a sphere.
These two settings are holographically dual to AdS black brane and spherical black hole backgrounds, respectively.
Our approach is based on expanding thermal correlators at zero spatial separation in terms of GFF correlators,
\begin{equation}
    g(\tau)
    =
    \sum_{\Delta} \frac{a_\Delta}{\beta^{\Delta}} g_{\text{GFF}} \left(\Delta_\phi-\frac{\Delta}{2}, \tau\right) + g_{\text{arcs}}(\tau) \,,
    \label{eq:gff}
\end{equation}
a representation that was rigorously established for finite temperature CFTs at infinite volume in~\cite{Barrat:2025nvu}.
In the present work we extended its validity to finite volume systems, under the condition that both external and exchanged operator dimensions are integer and focusing on the terms that do not involve poles in the OPE coefficients.

For holographic correlators, the first term in the expansion~\eqref{eq:gff} splits into two contributions: a \textit{principal contribution}, to which only a finite set of multi-stress tensor operators (associated with $\Delta < 2\Delta_\phi$) contributes directly, and a \textit{regularized part}, encoding an infinite number of multi-stress tensor operators.
The regularized term require to appropriately employ derivatives of the GFF blocks.
Similarly to the non-integer $\Delta_\phi$ case, the arcs are non-trivial functions but their contributions can be computed and they also admit an expansion in terms of GFF correlators.
These tools allow the derivation of analytical results for any integer value of $\Delta_\phi$, whose explicit evaluation requires using the asymptotic form of the multi-stress tensor OPE coefficients.
To the best of our knowledge, this provides the first approximate expressions for thermal holographic correlators in $d>2$ as a function of the imaginary time in the case of integer $\Delta_\phi$.
We corroborate these results by solving the scalar wave equation numerically and find convincing agreement for the black brane case, which additionally surprisingly suggests that the principal contributions dominate everywhere.
We also provide an AdS interpretation of \eqref{eq:gff} as an expansion of the metric in graviton modes around thermal AdS, and we verify this picture through explicit Witten diagram computations up to second order in the expansion parameter.
Finally, curvature corrections to the thermal correlators are also included in \eqref{eq:gff} as anomalous dimensions are suppressed with respect to $N$, i.e., only the OPE coefficients $a_\Delta$ need to be corrected.

\bigskip

This work opens several directions for future research, which we list here:
\begin{itemize}
    \item[$\star$] The results obtained in this work admit several natural extensions.
    First, it would be interesting to understand how to better approximate the regularized contribution.
    This boils down to improving the approximation of the asymptotic OPE coefficients, for instance by including $1/n^{\#}$ corrections as in~\cite{Afkhami-Jeddi:2025wra} for instance.
    Second, beyond the kinematic regime $|\vec{x}|=0$ the OPE contributions become disentangled in spin.
    Moreover, as noted in \cite{Ceplak:2024bja}, the large $\Delta$ behavior of the multi-stress tensor sector involves alternating signs that spoil analyticity in $\Delta$.
    It would be interesting to explore the analyticity in spin $J$~\cite{Ceplak:2024bja,Fitzpatrick:2019zqz}, which would imply to study the correlator at non-vanishing spatial separation.
    Additionally, it might be possible to determine an approximate solution of the wave equation, in analogy with the procedure outlined in this paper;
    \item[$\star$] An important check for the non-zero spatial separation correlator would be to compare it with the exact momentum-space correlators computed in~\cite{Dodelson:2022yvn}. 
    The $\tau$-plane correlators studied in this paper correspond to integrating over all momentum modes, which is analytically challenging. 
    Nevertheless, it is worth noting that integrating out the momentum modes reduces the correlator of~\cite{Dodelson:2022yvn} to the Fourier transform of GFF blocks, as discussed in~\cite{Barrat:2025nvu}.
    The relation between these two exact results should be valuable to understand the IR behaviour of the correlator (such as $\omega\to\infty$) and deserves to be explored more deeply;
    \item[$\star$] As mentioned in Section \ref{subsec:ExactTwoPointFunctionsForIntegerDeltaBH}, thermal correlation functions can be viewed as zero-temperature four-point functions where two of the operators are taken to be heavy.
    For holographic theories, this correspondence was partially verified perturbatively (and at large spin) in~\cite{Dodelson:2022yvn}. 
    It would be interesting to further investigate this connection in light of the results presented in this paper.
    The case in which $\Delta_\phi$ is an integer is intricate from the perspective of four-point functions~\cite{Li:2019zba,Li:2020dqm} and the thermal bootstrap may provide useful insights into the structure of these correlators;
    \item[$\star$] A deeper understanding of the geometry $S^1_\beta \times S^{d-1}_R$ would be valuable, particularly regarding how modifications of the analytic structure might affect Equation~\eqref{eq:gff}.
    Establishing this would enable the use of the expansion beyond holographic theories, for instance in models such as the $3d$ Ising CFT, in close analogy with the infinite volume analysis of the thermal $3d$ Ising model performed in~\cite{Barrat:2025nvu}.
    \item[$\star$] The results presented in this paper correspond to the leading term in the $1/N$ expansion. Including quantum effects requires going to next-to-leading order in the same expansion. 
    The main difficulty in this context is that, once CFT operators acquire anomalous dimensions, additional operators -- potentially an infinite number -- contribute non-trivially to the dispersion relation. Nevertheless, it was shown in~\cite{Barrat:2025wbi} that the expansion in~\eqref{eq:gff} is extremely effective in perturbative setups (in that work, the case of the $\mathrm O(N)$ model in the $\varepsilon$-expansion).
    Furthermore, anomalous dimensions happen to make numerical methods more predictive: in this context we expect to being able to compute unknown OPE coefficients by making use of both analytical and numerical methods in tandem.
    Incorporating $1/N$ corrections therefore represents one of the key future directions for the thermal analytic and numerical bootstrap;
    \item[$\star$] In recent years, the thermal bootstrap has mainly focused on scalar two-point functions~\cite{Iliesiu:2018zlz,Barrat:2025wbi,Buric:2025anb,Buric:2025fye,Barrat:2025nvu}.
    Meanwhile, correlators of currents and of the stress tensor have played a central role in the study of holographic correlators~\cite{Balasubramanian:1999re,Skenderis:2002wp,Bhattacharya:2025vyi} as well as in the zero-temperature bootstrap~\cite{Chang:2024whx}.
    Furthermore, zero temperature data, such as the conformal anomalies, enter the strtwo-point function: for this reason, we expect more constraints on finite temperature data (e.g. the specific condition $a = c$ for $\mathcal N = 4$ SYM).
    It would be natural to extend the results of this paper to these correlators as well.
    Furthermore, it would be desirable to formulate a bootstrap problem for fermionic two-point functions~\cite{David:2023uya} ;
    \item[$\star$] Many of the features of the thermal correlators studied here persist -- and in some respects even simplify -- when considering correlators of defect operators inserted on a Polyakov loop, i.e., a line defect wrapping the thermal circle.
    A preliminary investigation of the corresponding thermal defect CFT was carried out in \cite{Barrat:2024aoa}, where it was shown that the relevant data reduce to one-point functions of defect operators.
    A natural question is to identify the spectrum of exchanged operators in the holographic setting.
    The analytic bootstrap can then be applied in close analogy with the present work to determine defect correlators~\cite{future2}.
    Moreover, setups such as the supersymmetric Wilson line have attracted considerable recent attention \cite{Giombi:2017cqn,Liendo:2018ukf,Cavaglia:2021bnz,Ferrero:2021bsb,Barrat:2021tpn,Artico:2024wnt,Artico:2024wut,Bonomi:2024lky,Cavaglia:2024dkk}, and the wealth of available results can be directly used as input for both analytical and numerical bootstrap analyses;
    \item[$\star$] Holographic CFTs at finite temperature and volume, such as $\Nm=4$ SYM, undergo a confinement–deconfinement transition that is holographically realized as the Hawking–Page transition between thermal AdS and the spherical black hole phase.
    From the bootstrap perspective, the two phases differ in their spectra: in the thermal AdS phase scalar two-point functions do not involve multi-stress tensor exchanges.
    Although the framework developed here does not directly probe the operator spectrum (and instead uses it as an input), it does reveal distinct correlator behavior above and below the Hawking–Page temperature.
    An interesting direction would be to devise bootstrap approaches that are sensitive to the phase transition itself.
\end{itemize}

\acknowledgments

We are particularly grateful to Nikolay Bobev, Ilija Burić, Simon Caron-Huot, Matthew Dodelson, Ivan Gusev, Tom Hartman, Manuela Kulaxizi, Andrei Parnachev, Miguel Paulos, Fedor Popov, Leonardo Rastelli, David Simmons-Duffin, Kostas Skenderis, Balt van Rees, Benjamin Withers for valuable discussions about this work.
JB and EP are supported by ERC-2021-CoG - BrokenSymmetries 101044226. AM, DB, and EP have benefited from the German Research Foundation DFG under Germany’s Excellence Strategy – EXC 2121 Quantum Universe – 390833306. EM and EP’s work is funded by the German Research Foundation DFG – SFB 1624 – “Higher structures, moduli spaces and integrability” – 506632645.

\appendix

\section{GFF thermal correlators at finite volume}
\label{app:SphericalGFF}

We provide here the derivation of the two-point function of a generalized free field of integer dimension $\Delta_\phi$ as given in Equation~\eqref{eq:Gfffinitevol}.
It can be obtained via the method of images from the correlator on the geometry $\mathbb{R}\times S^{d-1}_{R}$:
\begin{equation}
    g_{\text{GFF},\bR} (\Delta_\phi, \tau,\vec{\theta})
    =
    \frac{1}{2^{\Delta_\phi} R^{2\Delta_\phi}} \sum_{m=-\infty}^\infty \left[\frac{1}{\cosh((\tau+m \beta)/R) - \cos |\vec{\theta}|}\right]^{\Delta_\phi}\,.
    \label{eq:FreeTheoryFiniteVolume_SumOverImages}
\end{equation}
We will set zero spatial separation $\vec{\theta}=0$ to reduce to the problem treated in this work. 
 
In what follows it will be convenient to introduce the ratio $\xi=\frac{\beta}{2 \pi R}$: after setting $\beta=1$, we can rewrite the correlator as
\begin{equation} \label{eq: finit vol csch}
    g_{\text{GFF},\xi} (\Delta_\phi,\tau)
    =
    \left(\pi \xi \right)^{2\Delta_\phi} \sum_{m=0}^\infty \text{csch}\left[\pi \xi (\tau+m) \right]^{2\Delta_\phi}+ (\tau \to 1-\tau) \ .
\end{equation}
At this stage, we can make progress by choosing an integer external dimension $\Delta_\phi\in \mathbb{N}$. In terms of the parameter $q=e^{-2 \pi \xi}$, standard computations allow to reshape the series \eqref{eq: finit vol csch} as
\begin{equation} \label{eq: reshaped}
    \sum_{m=0}^\infty \text{csch}\left[\pi \xi (\tau+m) \right]^{2 \Delta_\phi}=
    \frac{4^{ \Delta_\phi}}{\Gamma(2\Delta_\phi)} \sum_{s=0}^{2 \Delta_\phi-1} \frac{c( \Delta_\phi, s)}{\log(q)^s}\frac{\text{d}^s}{\text{d}\tau^s}\sum_{\ell=1}^{\infty}  \frac{q^{\ell \tau}}{1-q^{\ell}} \ ,
\end{equation}
where we defined the $c(\Delta_\phi, s)$ coefficients, involving the Stirling numbers of the first kind $S_{n}^{(m)}$,
\begin{equation}
    c(\Delta_\phi, s)=\sum_{r=0}^{2\Delta_\phi-1} \binom{r}{s} S_{2\Delta_\phi-1}^{(r)} (\Delta_\phi-1)^{r-s} \ .
\end{equation}
The only thing left to do is to evaluate the generalized Lambert series appearing in the right hand side of Equation \eqref{eq: reshaped}. It can easily be reshaped as follows, where $q$-Polygamma function $\psi_{q}$ is employed:
\begin{equation} \label{eq: operative}
    \sum_{\ell=1}^{\infty}  \frac{q^{\ell \tau}}{1-q^{\ell}}=\sum_{\ell=1}^{\infty} \sum_{j=0}^{\infty} q^{\ell(\tau+j)}=\sum_{j=0}^{\infty}  \frac{q^{\tau+j}}{1-q^{\tau+j}}=\frac{\psi_{q}(\tau)+\log(1-q)}{\log(q)} \ .
\end{equation}
The action of the derivatives on the generalized Lambert series is then given by
\begin{equation}
    \sum_{m=0}^\infty \text{csch}\left[\pi \xi (\tau+m) \right]^{2 \Delta_\phi}=
    \frac{2^{2 \Delta_\phi}}{\Gamma(2\Delta_\phi)} \sum_{s=1}^{2\Delta_\phi-1} (-1)^{s+1}\frac{c(\Delta_\phi, s)}{(2 \pi \xi)^{s+1}} \psi_{e^{-2 \pi \xi}}^{(s)}(\tau) 
    \ ,
\end{equation}
where we used the fact that the coefficient $c(\Delta_\phi,0)$ vanishes identically; this retroactively changes the definition of the coefficients to 
\begin{equation} \label{eq: coeff c}
    c(\Delta_\phi, s)=\sum_{r=1}^{2\Delta_\phi-1} \binom{r}{s} S_{2\Delta_\phi-1}^{(r)} (\Delta_\phi-1)^{r-s} \ .
\end{equation}
We conclude by evaluating the sum appearing in the coefficients \eqref{eq: coeff c}, which can be reshaped in terms of the derivatives of a falling factorial\footnote{The falling factorial $a^{(b)}$ is defined as $a^{(b)}=\frac{\Gamma(a+1)}{\Gamma(a-b+1)}$; the falling factorial is implemented in \texttt{Mathematica} as the function \texttt{FactorialPower}$[a,b]$. Notice the difference with the Pochhammer symbol $(a)_b=\frac{\Gamma(a+b)}{\Gamma(a)}$. }
\begin{equation}
    c(\Delta_\phi, s)
    =\frac{1}{s!}\left[\frac{\text{d}^s}{\text{d}x^s}x^{(y)}\right]_{x=\Delta_\phi-1, y=2\Delta_\phi-1} \ .
\end{equation}
The falling factorial can be shown to be a odd function of $x$, when centered around $\Delta_\phi-1$. This means that even derivatives are identically null, and we can simply relabel $s=2 \ell-1$, obtaining 
\begin{equation} 
    c(\Delta_\phi, \ell)
    =\frac{1}{\Gamma(2\ell)}\left[\frac{\text{d}^{2 \ell-1}}{\text{d}x^{2 \ell-1}}x^{(y)}\right]_{x=\Delta_\phi-1 , y= 2\Delta_\phi-1} \ .
\end{equation}
We are ready to write down the final result:
\begin{equation} 
    g_{\text{GFF},\xi} (\Delta_\phi,\tau)
    =
     \frac{1}{\Gamma(2\Delta_\phi)} \sum_{\ell=1}^{\Delta_\phi} \left(4 \pi^2 \xi^2 \right)^{\Delta_\phi-\ell}c(\Delta_\phi, \ell) \left[ \psi_{e^{-2 \pi \xi}}^{(2 \ell-1)}(\tau)+\psi_{e^{-2 \pi \xi}}^{(2 \ell-1)}(1-\tau) \right]  \ .
\end{equation}
We reintroduce the dependence on $\beta$ and $R$, obtaining:
\begin{equation} \label{eq: GFF result}
    g_{\text{GFF},\beta/R} (\Delta_\phi,\tau)
    =
     \frac{1}{\beta^{2 \Delta_\phi}\Gamma(2\Delta_\phi)} \sum_{\ell=1}^{\Delta_\phi} \left(\frac{\beta^2}{R^2} \right)^{\Delta_\phi-\ell}c(\Delta_\phi, \ell) \left[ \psi_{e^{-\beta/R}}^{(2 \ell-1)}\left(\frac{\tau}{\beta}\right)+\psi_{e^{-\beta/R}}^{(2 \ell-1)}\left(1-\frac{\tau}{\beta}\right) \right]  \ .
\end{equation}
\paragraph{The infinite temperature limit.} We now consider the infinite temperature limit, equivalent to $\xi \to 0$ in the chosen conventions. By reversing the definition \eqref{eq: operative} of the $q$-Polygamma function, its derivatives read
\begin{equation}
    \psi_{e^{-2 \pi \xi}}^{(s)}(\tau)=-2 \pi \xi \frac{\text{d}^s}{\text{d}\tau^s}\sum_{n=0}^{\infty} \frac{e^{-2 \pi \xi (n+\tau)}}{1-e^{-2 \pi \xi (n+\tau)}} \ .
\end{equation}
In the limit, this function reduces to 
\begin{equation}
    \lim_{\xi \to 0}\psi_{e^{-2 \pi \xi}}^{(s)}(\tau)
    =(-1)^{s+1}  s! \,  \zeta_H(s+1,\tau) \ .
\end{equation}
In the formula \eqref{eq: GFF result}, the only term in the sum over $\ell$ surviving the limit is $\ell=\Delta_\phi$, hence the whole correlator simplifies to 
\begin{equation}
    \lim_{\xi \to 0}g_{\text{GFF},\xi} (\Delta_\phi,\tau)
    =
     c(n_\phi,n_\phi) \left[ \zeta_H(2 \Delta_\phi,\tau)+\zeta_H(2 \Delta_\phi,1-\tau) \right]  \ .
\end{equation}
Since $c(n_\phi,n_\phi)=1$, after reinstating $\beta$ dependence the evaluation of the infinite temperature limit returns the expected result
\begin{equation}
    g_{\text{GFF}} (\Delta_\phi, \tau)
    = \frac{1}{\beta^{2 \Delta_\phi}} \left[\zeta_H \left(2 \Delta_\phi,\frac{\tau}{\beta}\right)+\zeta_H \left(2 \Delta_\phi,1-\frac{\tau}{\beta} \right) \right]  \ .
\end{equation}

\section{$3d$ vector model at large $N$ on $S_\beta \times S_R^2$}
\label{app:ONModelAtLargeN}

The goal of this appendix is to show that the expansion in~\eqref{eq:ExpansionInGFFtauV} applies to the $\mathrm{O}(N)$ model at large $N$ for $d=3$.
After performing the Hubbard–Stratonovich transformation, the Lagrangian of the model can be written as
\begin{equation}
    \mathcal{L} = \frac{1}{2} (\partial_\mu \phi_i)^2 + \frac{1}{2} \sigma\, \phi_i \phi_i\,,
\end{equation}
where $\sigma$ is the Hubbard–Stratonovich field. 
The corresponding thermal two-point function $\vev{\phi \phi}_{\beta/R}$ reads
\begin{equation}\label{eq:ONmodelcorre}
    g_{\beta/R}(\tau) = \frac{1}{4\pi R^2 \beta} 
    \sum_{n=-\infty}^\infty \sum_{l=0}^\infty 
    \frac{(2l+1)\, e^{\frac{2\pi i n \tau}{\beta}}}
    {\frac{(2\pi n)^2}{\beta^2} + \frac{l(l+1)}{R^2} + m_\text{th}^2(\beta/R)}\,,
\end{equation}
where $m_\text{th}^2(\beta/R) = \vev{\sigma}_{\beta/R}$ is the thermal mass. 

We now establish two facts:  
(i) the two-point function in~\eqref{eq:ONmodelcorre} can be decomposed as a sum over GFF two-point functions;  
(ii) the coefficients of such an expansion can be interpreted as thermal OPE coefficients (at finite volume, thus depending on the dimensionless ratio $\beta/R$) corresponding to the operators appearing in the $\phi \times \phi$ OPE.

\paragraph{GFF decomposition.}
Starting from~\eqref{eq:ONmodelcorre}, one can observe -- by analogy with the infinite volume case -- that the expansion of the correlator in powers of $m_\text{th}^2$ is equivalent to an OPE expansion. 
Indeed,
\begin{equation}\label{eq:GFFexpansion}
    g_{\bR}(\omega_n) 
    = \sum_{m} \frac{m_\text{th}^{2m}(\beta/R)}{\Gamma(2m+1)}\, 
    g_{\text{GFF},\bR}(1-2m,\omega_n)\,,
\end{equation}
where
\begin{equation}
    g_{\text{GFF},\bR}(1-2m,\omega_n)
    = (-1)^m \Gamma(2m+1)
    \sum_{l=0}^\infty 
    \frac{2l+1}
    {\left(\frac{\omega_n^2}{\beta^2} + \frac{l(l+1/2)}{R^2}\right)^{m+1}}\,,
\end{equation}
following the general result obtained in Section 2.4.2 of \cite{Barrat:2025nvu}.

\paragraph{Interpretation of the coefficients.}
The interpretation of the coefficients in~\eqref{eq:GFFexpansion} follows as in the infinite volume case.
Since the spectrum does not change, the operators contributing to the OPE remain the same:
\begin{align}\label{eq:operatorONlargeN}
    &\text{Double-twist operators } [\phi \phi]_{n,J}: && 
    \Delta_{n,J} = 2\Delta_\phi + 2n + J\,, \notag \\[4pt]
    &\text{Scalar operators } \sigma^m: && 
    \Delta_{\sigma^m} = 2m\,.
\end{align}
The identity operator is of course also present. 
Other operators, related to those above by the equations of motion, may also appear.
For example, $\phi \sigma \phi$ is equivalent to $\phi \Box \phi=[\phi \phi]_{1,0}$, since the equations of motion imply $\Box \phi \sim \sigma \phi$. 
Operators such as $\phi \sigma \partial_\mu \partial_\nu \phi$ have dimension $\Delta = 2\Delta_\phi + 2$ and are degenerate with $\phi \Box \partial_\mu \partial_\nu \phi=[\phi \phi]_{1,2}$; even reinstating spatial dependence (and hence spin) does not lift this degeneracy. 
Therefore, the relevant operators can all be recast as those in~\eqref{eq:operatorONlargeN} (plus degenerate descendants). 

Since the double-twist operators do not contribute, since their momentum space blocks vanish, the operators appearing in~\eqref{eq:GFFexpansion} must correspond to $\sigma^m$. 
The corresponding OPE coefficients are directly read from the expansion:
\begin{equation}\label{eq:sigmam}
    a_{\sigma^m}(\beta/R) = \frac{m_\text{th}^{2m}(\beta/R)}{\Gamma(2m+1)}\,.
\end{equation}
The dependence of these one-point functions on the dimensionless ratio $\beta/R$ is encoded in the thermal mass $m_\text{th}$, which acquires a finite volume correction depending on the sphere radius. 
For large $R$, the expansion found in~\cite{David:2024pir} reads
\begin{equation}
    m_\text{th}(\beta/R) = \frac{1}{\beta} 
    \left(\log\varphi^2 + \frac{1}{24 \log\varphi^2}\, \frac{\beta^2}{R^2} + \frac{55 + 32 \sqrt{5}\log\varphi^2}{28800\log^3\varphi^2}\frac{\beta^4}{R^4}+\ldots \right)\,.
\end{equation}
It would be interesting to recover the OPE coefficients~\eqref{eq:sigmam} independently; this has already been done in~\cite{David:2024pir} for the lowest-lying operators.

For completeness, we note that obtaining a closed-form expression for~\eqref{eq:ONmodelcorre} in position space is difficult, even in the infinite volume limit, where only a Bessel function expansion is known. 
However, when the correlator is restricted to the thermal circle, a closed form for $g_{\beta/R}(\omega_n)$ can be derived. 
Finally, we can perform the sum over the spherical modes labelled by $l$. We define $\tilde{m}_{\text{th}}^2=m_{\text{th}}^2-\frac{1}{4 R^2}$ to rewrite the series as follows
\begin{equation}
    \frac{1}{4\pi R^2 \beta}\sum_{l=0}^{\infty}\frac{2l+1}
     {\frac{(2\pi n)^2}{\beta^2} + \frac{(l+1/2)^2}{R^2} + \tilde{m}_\text{th}^2}=\frac{1}{16\pi \beta}\sum_{l=0}^{\infty}\frac{2l+1}
     {\frac{R^2}{4}\left[\frac{(2\pi n)^2}{\beta^2}+ \tilde{m}_\text{th}^2\right] + (2l+1)^2 }
\end{equation}
If we define $a=\frac{R^2}{4}\left[\frac{(2\pi n)^2}{\beta^2}+ \tilde{m}_\text{th}^2\right]$, then the series to solve can be rewritten as 
\begin{equation}
   \frac{1}{4\pi R^2 \beta} \sum_{l=0}^{\infty}\frac{2l+1}
     {\frac{(2\pi n)^2}{\beta^2} + \frac{(l+1/2)^2}{R^2} + \tilde{m}_\text{th}^2}=\frac{1}{16\pi \beta}\sum_{l=0}^{\infty}\frac{2l+1}
     {a + (2l+1)^2} \, .
\end{equation}
We can resum the series by noticing that
\begin{align}
    \sum_{l=0}^{\infty}\frac{2l+1}
     {a + (2l+1)^2}&=\frac{1}{4}\sum_{l=0}^{\infty}\left(\frac{1}
     {l+\frac{1+i \sqrt{a}}{2}}+\frac{1}
     {l+\frac{1-i \sqrt{a}}{2}}\right)\\&=-\frac{1}{4} \left[ \psi\left(\frac{1+i \sqrt{a}}{2} \right)+\psi\left(\frac{1-i \sqrt{a}}{2} \right)\right] \ ,
\end{align}
where $\psi$ is the Digamma function. Putting everything together, the result is 
\begin{equation}
    g_{\beta/R}(\omega_n)=-\frac{e^{ i \omega_n \tau}}{64\pi \beta} \left[ \psi\left(\frac{1}{2}+i\frac{R}{4} \sqrt{\omega_n^2+\tilde{m}_{\text{th}}^2}\right)+\psi\left(\frac{1}{2}-i\frac{R}{4} \sqrt{\omega_n^2+\tilde{m}_{\text{th}}^2}\right)\right] \ .
\end{equation}

\section{The asymptotic model }
\label{app:TheAsymptoticModel}

In this appendix, we reformulate the content of \cite{Buric:2025anb} in the language of Section \ref{sec:BlackBraneBackground}.

\paragraph{Spectrum.}
We focus on the case where the external operators have dimension $\Delta_\phi = 3/2$ and $d=4$.
In this case, the spectrum of the correlator is
\begin{align}
    & \mathds{1}: && \Delta = 0, \quad J= 0\,, \notag \\
    & [T^n]: && \Delta_n = 4n\,, \quad J=0, 2, \ldots, 2n\,, \label{eq:SpectrumAsymptotic} \\
    & [\phi \phi]_{n,J}: && \Delta = 3 + 2n + J\,. \notag
\end{align}
The multi-stress tensor and double-trace operators' conformal dimensions are never equal in this scenario, hence their OPE coefficients are never degenerate in the sense determined by  \eqref{eq:OPEx0}. Most importantly, the poles in the OPE coefficients do not pose any problem in this scenario, hence there is no need to regularize higher contributions. The OPE breaks down into three separate contributions
\begin{equation}
    g(\tau)
     =
    \frac{1}{\tau^3} \left[ 1 + \sum_n a_{[T^n]} \left( \frac{\tau}{\beta}\right)^{4n} + \sum_{\Delta(n,J)} a_{[\phi\phi]_{\Delta}}  \left( \frac{\tau}{\beta}\right)^{\Delta} \right]\,.
    \label{eq:CorrelatorStructureAsymptotic}
\end{equation}
The model is called asymptotic because it assumes this formula for the OPE coefficients of all the multi-stress tensor operators \cite{Ceplak:2024bja,Buric:2025anb}
\begin{equation}
    a_{[T^n]} = - \alpha_0 \frac{96 \sqrt{\pi}}{175} (-4)^n\,,
    \label{eq:OPECoeffsTnAsymptotic}
\end{equation}
with $\alpha_0 \sim 1$.\footnote{The value of $\alpha_0$ is not known but it was shown in~\cite{Ceplak:2024bja} to be close to $1$ at large $\Delta$.}
In a generic holographic model, this formula corresponds to the leading asymptotic for $a_{[T^n]}$ at large $n$.

\paragraph{Expansion into GFF blocks.}
We can apply the dispersion relation to the spectrum in the form of the GFF expansion in Equation~\eqref{eq:ExpansionInGFFtau}.
We find that the identity and all multi-stress tensor operators contribute to the discontinuity, while the double-trace operators drop, explicitly 
\begin{equation}
    g (\tau)
    =
    g_\text{GFF} (3, \tau)
    +
    \sum_{n \geq 1} a_{[T^n]} \beta^{4n} g_\text{GFF} (3 - 4n, \tau)
    +
    g_\text{arcs}(\tau)\,.
    \label{eq:AsymptoticAfterDR}
\end{equation}
It is important to understand that this expression is expected to nevertheless recreate the spectrum of double-trace operators in the OPE~\eqref{eq:thermal OPE}.
Explicitly, the multi-stress tensor sector is given by
\begin{equation}
    \sum_{n \geq 1} a_{[T^n]} \beta^{4n} g_\text{GFF} (3 - 4n, \tau)
    =
    - \frac{\alpha_0}{\beta^3} \frac{96 i \sqrt{\pi}}{175} \left( \frac{1}{\frac{\tau}{\beta} - \frac{1+i}{2}} - \frac{1}{\frac{\tau}{\beta} - \frac{1-i}{2}} \right)\,,
    \label{eq:AsymptoticAfterDRTn}
\end{equation}
which is also found in~\cite{Buric:2025anb} taking images over the single blocks.

\paragraph{Unphysical pole, arcs, and uniqueness.}
The first two terms in the right-hand side of Equation \eqref{eq:AsymptoticAfterDR} cannot be the final answer, since they present poles located at
\begin{equation}
    \tau = \frac{\beta \pm i \beta}{2}\,.
    \label{eq:UnphysicalPole}
\end{equation}
These poles are unphysical, as discussed in \cite{Buric:2025anb}.
In fact, their presence violates our assumption of boundedness in Equation~\eqref{eq:boundess}.
Notice that in \cite{Barrat:2025nvu} the authors proved that $g_\text{arcs}$ is a single constant if the function is bounded and periodic on the real axis (see Appendix A of \cite{Barrat:2025nvu}).
However, as observed in \cite{Buric:2025anb,Ceplak:2024bja}, the sum over multi-stress tensor operators produces an additional pole which prevents $g_\text{arcs}(\tau)$ from being an entire function.
This is a violation of the assumption of \cite{Barrat:2025nvu} and, as a consequence, the arcs cannot be only a constant.
Since the arcs have to present a pole themselves to compensate the unphysical poles, an infinite number of operators has to contribute to $g_\text{arcs}(\tau)$: in accordance with the principle of OPE compatibility, the arc contribution has to allow for an OPE decomposition; however, only the sum of an infinite amount of OPE terms can produce a divergence.
In holographic theories this extra, unphysical pole, is associated with the bouncing singularity in the holographic dual. 

It was noted in \cite{Buric:2025anb} that the problem of determining the arc contribution can be solved by subtracting the spurious pole and summing over images, which results in
\begin{equation}
    g_\text{arcs} (\tau) = \alpha_0 \frac{192 \pi^{3/2}}{175 \beta^3} \left( \frac{1}{1 + e^{\pi (2 i \tau/\beta + 1)}} - \frac{1}{1 + e^{\pi (2 i \tau/\beta - 1)}} \right)\,.
    \label{eq:ArcsAsymptotic}
\end{equation}
This procedure produces a consistent correlator as it is now free of poles and bounded in the sense of \eqref{eq:boundess}.
Moreover, the new correlator satisfies all the properties required and, by using the same argument as in Appendix A of \cite{Barrat:2025nvu}, one can show that additional arcs are at most constant.
This constant can be fixed to zero by using clustering in real time, thus this correlator is \textit{the only one} that satisfies our consistency conditions, given the input.

It is worth emphasizing a major advantage of the method presented here. 
Each multi-stress tensor coefficient corresponds to a distinct block in the GFF expansion. 
Since the arc contributions are associated with the poles generated by the infinite sum over multi-stress tensor coefficients, they depend only on the asymptotic behavior of the OPE data, not on their detailed low-lying structure. 
Consequently, modifying a finite number of OPE coefficients does not affect any fundamental property of the two-point function. 
Indeed, the GFF blocks are intrinsically KMS invariant, their analytic structure is independent of the OPE coefficients, and we have just argued that $g_\text{arcs}(\tau)$ depends solely on the asymptotic behavior of $a_{[T^n]}$. 
This implies that the asymptotic model can, in principle, be systematically improved by replacing the lowest-lying $a_{[T^n]}$ with their exact values, which can be computed using the techniques of~\cite{Ceplak:2024bja,Fitzpatrick:2019zqz}. 
Such refinements can also be incorporated following the method of~\cite{Buric:2025fye}; in fact, a similar strategy (without the asymptotic tail) was successfully implemented in~\cite{Buric:2025anb}.

\paragraph{Implications for asymptotic approximations.}
It was observed in~\cite{Buric:2025anb} that the OPE coefficients associated with the correlator~\eqref{eq:AsymptoticAfterDRTn} do not exhibit the analytic behavior at large conformal dimension $\Delta \to \infty$ predicted in~\cite{Marchetto:2023xap}. 
This discrepancy can already be traced back to the input data: the multi-stress tensor OPE coefficients themselves display alternating signs, in contrast with the monotonic behavior expected from the analysis of~\cite{Marchetto:2023xap}. 

To understand the origin of this difference, it is instructive to revisit the proof of the asymptotic prediction given in~\cite{Barrat:2025nvu}. 
A key assumption in that argument is that the sum over GFF correlators yields a function with the correct analytic structure, so that the arc contributions are at most constant. 
However, this assumption is violated in the asymptotic model of~\cite{Buric:2025anb}, due to the presence of the bouncing singularities: in essence, the expression~\eqref{eq:AsymptoticAfterDR} implicitly includes an infinite number of operators in the arc contribution. 

Nevertheless, the asymptotic behavior derived in~\cite{Marchetto:2023xap} still holds \emph{on average}: 
if one averages the thermal OPE density over a finite window in conformal dimension (of width $\delta \Delta \sim 2$, in the language of~\cite{Marchetto:2023xap}), the result continues to satisfy the expected asymptotic form. 
From the perspective of~\cite{Barrat:2025nvu}, this can be understood as follows: the contribution of double-trace operators also effectively captures the next multi-stress tensor correction, approximating the gap in conformal dimension between the two as negligible. 

On the other hand, the approximation of~\cite{Marchetto:2023xap} coincides -- at least in some examples, such as the three-dimensional Ising model -- with the large-spin perturbation theory of~\cite{Iliesiu:2018fao} (for a detailed explanation of this correspondence, see~\cite{Miscioscia:2025pjh}). 
A breakdown of large-spin perturbation theory could imply that the correlator is not Regge-bounded, and consequently that the OPE coefficients fail to be analytic in spin. 
The alternating sign pattern observed in~\eqref{eq:OPECoeffsTnAsymptotic} could be a manifestation of this phenomenon, although a more detailed analysis at non-zero spatial separation is required to determine whether the dominant contribution to the alternation arises from large or small spin. 
We plan to return to this question in future work.

\section{Numerical solution of the wave equation}
\label{app:NumericalSolutionOfTheWaveEquation}

In this section we review the solution of the wave equation in AdS, with a focus on the thermal scalar two-point function for an integer external dimension. 
The AdS$_{d+1}$ black brane metric is given in \eqref{eq:metric} where we set the AdS radius to one. The wave equation for a massive field in this background takes the form \eqref{eq:EOM}.

\paragraph{Method.}
\label{subsec:Method}

Near the boundary, the field behaves as
\begin{align}
    \Phi(z,\tau,x)=z^{d-\Delta_\phi}\phi_1(\tau,x)+\cdots+z^{\Delta_\phi}\phi_2(\tau,x)+\cdots \,,
\end{align}
and depending on the external dimension and the chosen boundary conditions, the boundary two-point function can be extracted from the ratio of the two leading terms.
The thermal correlator at zero spatial separation is then given by
\begin{equation}
    \vev{\phi(0)\phi(\tau)}_\beta
    =
    \lim_{z \to 0} z^{-\min(\Delta_\phi,\,d-\Delta_\phi)} \Phi(z,\tau,0)\,.
    \label{eq:ThermalCorrelatorFromAdS2}
\end{equation}
Following \cite{Parisini:2023nbd,Buric:2025fye,Buric:2025anb}, we decompose the scalar field as
\begin{align}
    \Phi(z,\tau,x)=\Psi(z,\tau,x)+\tilde G(z,\tau,x)\,,
\end{align}
where $\tilde G$ is built from the bulk-to-boundary propagator in pure AdS,
\begin{align}
    G_{\text{AdS}}(z,\tau,x)=\frac{z^{\Delta_\phi}}{\big(\tau^2+x^2+z^2\big)^{\Delta_\phi}} \ ,
\end{align}
modified by replacing $\tau^2$ with a periodic function $f(\tau)$. For the purposes of this numerical evaluation any periodic function of period $\beta$ and reproducing the correct singularities would work. We used the one also employed in \cite{Buric:2025anb}, which already proved to be reliable in extracting leading OPE coefficients.
This ensures that $\tilde G$ reproduces the correct short-distance divergence. On the boundary ($z=0$) and at zero spatial separation ($x=0$), the contribution to the two-point function contains only the expected UV divergence at small $\tau$, corrected by higher-order terms as proposed in \cite{Buric:2025fye}.

The remaining piece $\Psi$ has the same asymptotics as the full field
\begin{align}
    \Psi(z,\tau,x)=z^{d-\Delta_\phi}\psi_1(\tau,x)+\cdots+z^{\Delta_\phi}\psi_2(\tau,x)+\cdots \,.
\end{align}
Thus, the full field can be written as 
\begin{align}
    \Phi(z,\tau,x)=z^{d-\Delta_\phi}\psi_1(\tau,x)+\cdots+z^{\Delta_\phi}\left[\psi_2(\tau,x)+\left.\left(z^{-\Delta_\phi} \tilde G(z,\tau,x)\right)\right|_{z \to 0}\right]+\cdots \,.
\end{align}
We further define the auxiliary function $H(z,\tau,x)$
\begin{align}
    \Psi(z,\tau,x)=z^{d-\Delta_\phi}H(z,\tau,x) \,.
\end{align}
For $0 \leq d-\Delta_\phi < \Delta_\phi$, the boundary conditions are \cite{Parisini:2023nbd}:
\begin{itemize}
    \item $H|_{z=0}=0$, which enforces $\psi_1=0$;
    \item $\partial_z H|_{z=1}=0$, corresponding to regularity at the tip of the black brane cigar geometry;
    \item $\partial_x H|_{x=0}=0$, imposing regularity at the boundary origin;
    \item $H|_{x\to\infty}=0$, ensuring that the $\delta$-function source vanishes at infinity.
\end{itemize}
For $0 \leq \Delta_\phi < d-\Delta_\phi$, $H(z,\tau,x)$ is instead defined as follows
\begin{align}
    \Psi(z,\tau,x)=z^{\Delta_\phi}H(z,\tau,x) \,,
\end{align}
with boundary conditions
\begin{itemize}
    \item $\partial_z H|_{z=0}=0$, enforcing $\psi_2=0$;
    \item $\partial_z H|_{z=1}=0$, regularity at the tip of the cigar;
    \item $\partial_x H|_{x=0}=0$, regularity at the boundary origin;
    \item $H|_{x\to\infty}=0$, vanishing of the $\delta$-function source at infinity.
\end{itemize}
In both cases, periodic boundary conditions along the $\tau$ direction must also be imposed.

These ingredients allow one to numerically solve the wave equation.\footnote{In practice, it is sometimes convenient to multiply the equation by suitable powers of $z$ and $1-z$ to better enforce the boundary conditions.} 
We employ \texttt{NDSolve} in \texttt{Mathematica}.  
A precise estimate of the numerical errors is beyond the scope of this paper. Nonetheless, since higher conformal dimensions $\Delta_\phi$ require higher derivatives of $H$, and numerical derivatives are less and less accurate, we expect the results for lower $\Delta_\phi$ to be more reliable.

We solve the wave equation in the following cases:
\begin{itemize}
    \item[$\star$] $d=4$, $\Delta_\phi = 1$: the simplest case in four dimensions, corresponding to a free scalar field;
    \item[$\star$] $d=4$, $\Delta_\phi = 3$: the simplest non-trivial case, proposed in this paper as a benchmark. In the main text we show the agreement between our analytic proposal and the numerical solution.
\end{itemize}

\paragraph{$d = 4$ and $\Delta_\phi = 1$.}

We now turn to the case $d=4$ with $\Delta_\phi = 1$. 
We set $f_0 = 1$, which corresponds to choosing $\beta = \pi$, and in this case no derivatives of $H$ are needed. 
The thermal two-point function has the analytic form
\begin{equation}
    g(\tau) = \frac{\pi^2}{\beta^2} \csc^2\left(\frac{\pi \tau}{\beta} \right) \,.
\end{equation}
This result arises naturally from the bootstrap construction but it also independently coincides with the two-point function of a free scalar in four dimensions. 
It is straightforward to show that a scalar operator of dimension $\Delta_\phi=1$ necessarily satisfies the free equation of motion. 
Using the conformal algebra, one finds
\begin{equation}
    \Box \phi(x)\,|0\rangle \;=\; \boldsymbol P_\mu \boldsymbol P^\mu \phi(x)\,|0\rangle 
    \;\sim\; (\Delta-1)\,\phi(x)\,|0\rangle \;=\; 0 \,,
\end{equation}
which implies that $\Box \phi(x) = 0$. 
Thus the field $\phi$ is free.

The comparison between the analytic expression and the numerical solution is shown in Figure~\ref{fig:comparisontpD1}. 
We observe excellent agreement with a relative discrepancy below $0.25\%$.

\begin{figure}[t]
    \centering
        \includegraphics[scale=.25]{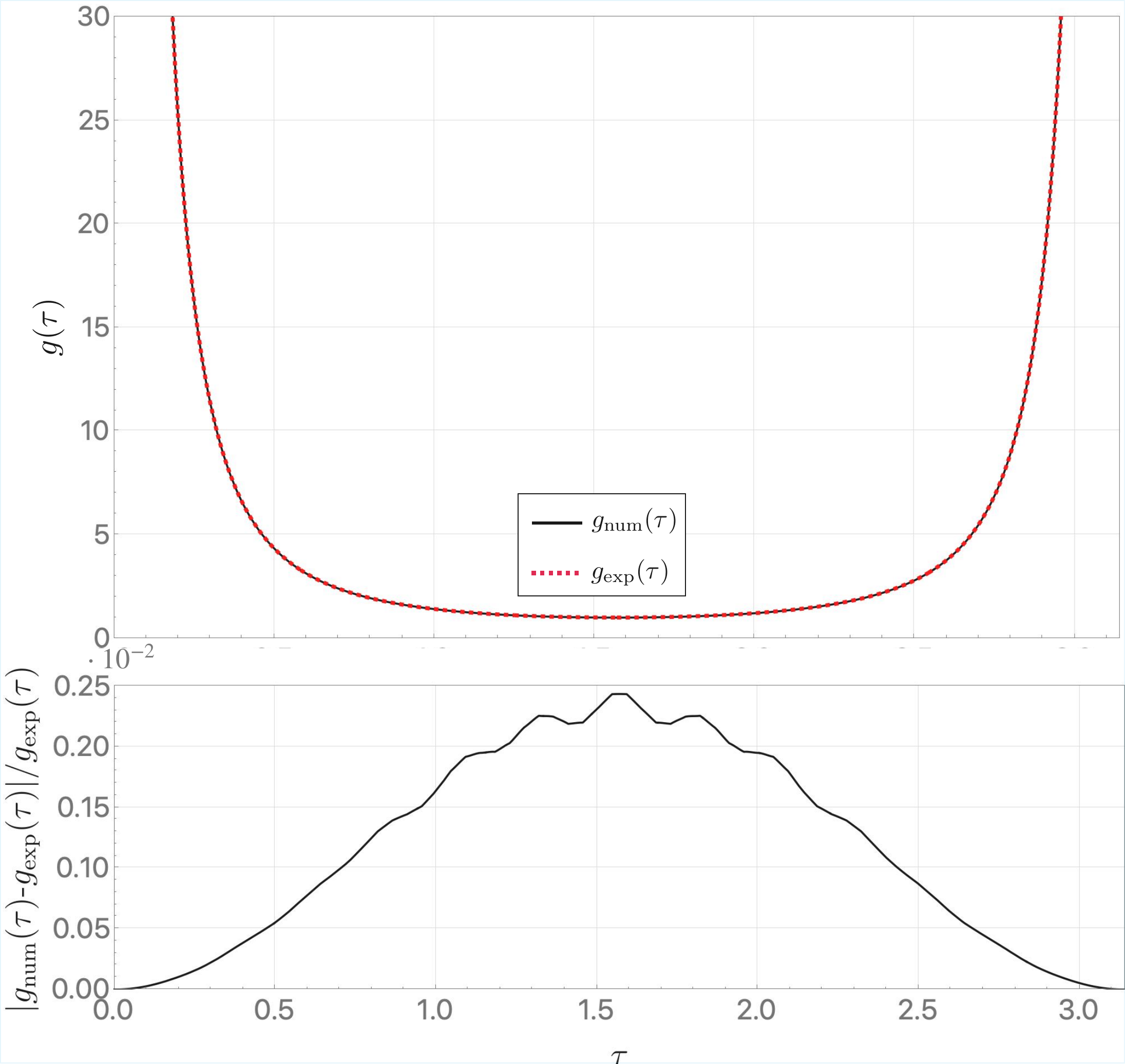}
    \caption{ Comparison between the bootstrapped two-point function for $\Delta_\phi = 1$ ($g_{\text{exp}}(\tau)$) and the boundary limit of the numerical solution of the AdS$_5$ wave equation ($g_{\text{num}}(\tau)$) and relative discrepancy between $g_{\text{exp}}(\tau)$ and $g_{\text{num}}(\tau)$ for $\beta = \pi$. 
    The discrepancy remains below $0.25\%$ for $\tau \in (0,\beta)$.}
    \label{fig:comparisontpD1}
\end{figure} 

\paragraph{Comments on the numerical accuracy.}
Even though a detailed analysis of the numerical errors in solving the wave equation is beyond the scope of this paper, it is still useful to gain some intuition and check that the numerical solution indeed reproduces the expected behaviour. 
A validation case was discussed above ($d = 4$, $\Delta_\phi = 1$), but the lightest non-trivial correlator we are ultimately interested in is $d = 4$, $\Delta_\phi = 3$.\footnote{The case $\Delta_\phi=2$ is in principle non-trivial as well but its principal contribution does not include $a_T$.
Therefore we consider $\Delta_\phi=3$ to be a stronger check.}
For this correlator, knowledge of $H$ alone is not sufficient: its first and second derivatives are also required to compute the thermal two-point function. 

A simple way to estimate the accuracy of the numerical solution is to quantify how well it satisfies the original differential equation. 
Since derivatives of $H$ enter into the computation, we also examine the behaviour of the first two derivatives of this numerical discrepancy. 
The quality of the solution is relatively good, especially away from the singular points of the correlator. 
However, the error increases significantly with the number of derivatives. 
This is expected: even functions that are numerically small may exhibit large derivatives. 

In particular, since the correlator in the case $\Delta_\phi = 3$ in $d = 4$ is a linear combination of the first and second derivatives of $H$, this explains why the simpler case $\Delta_\phi = 1$, which does not require derivatives of $H$, displays a better numerical behaviour. 
This also suggests that the discrepancy of order $\sim 1\%$ shown in the main text is plausibly inside the numerical accuracy of the solution of the wave equation and further terms are therefore difficult to be detected numerically.

\begin{figure}[t]
    \centering
    \begin{minipage}{.48\textwidth}
        \includegraphics[scale=.19]{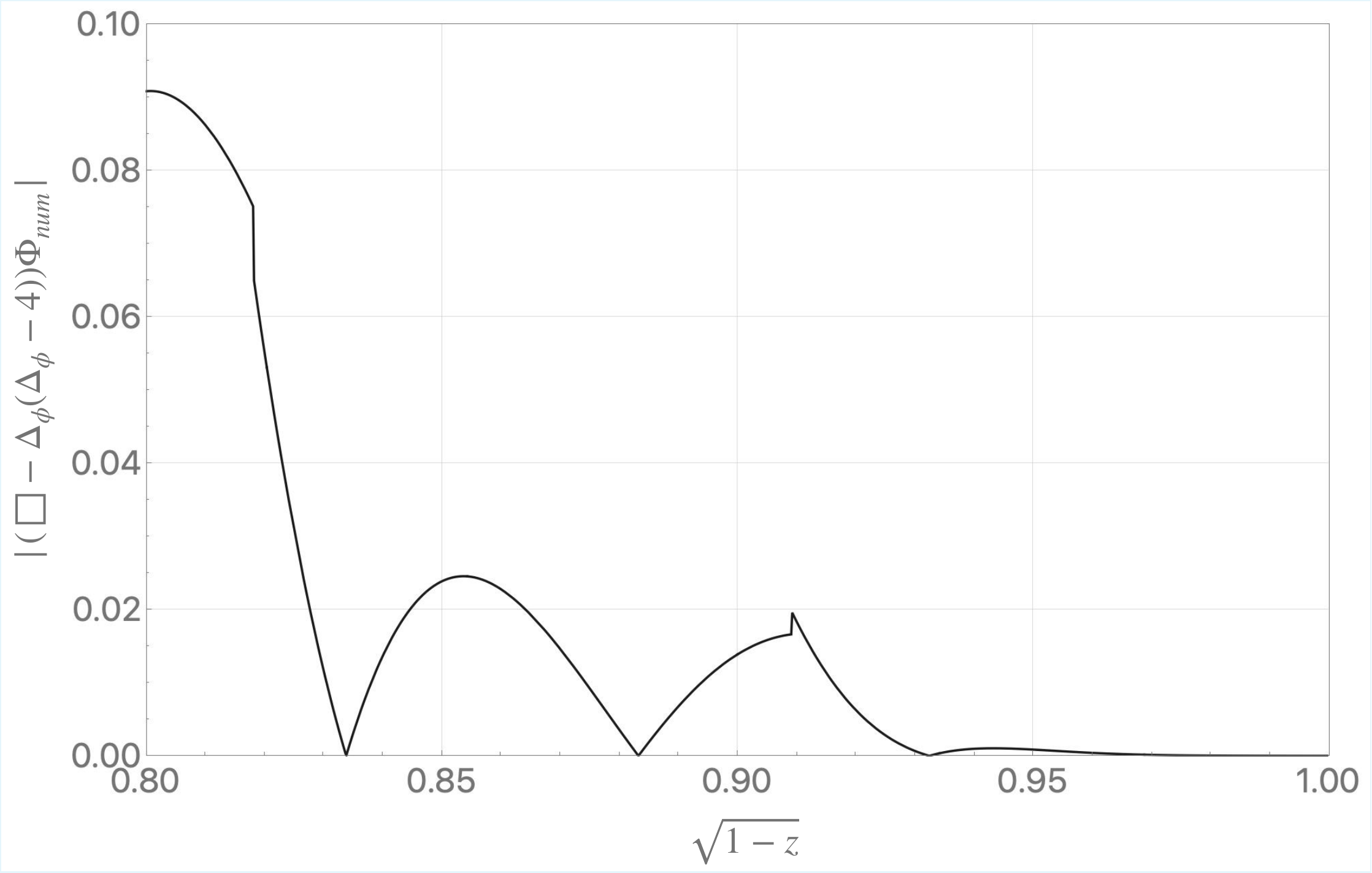}
    \end{minipage}%
    \begin{minipage}{.48\textwidth}
        \includegraphics[scale=.19]{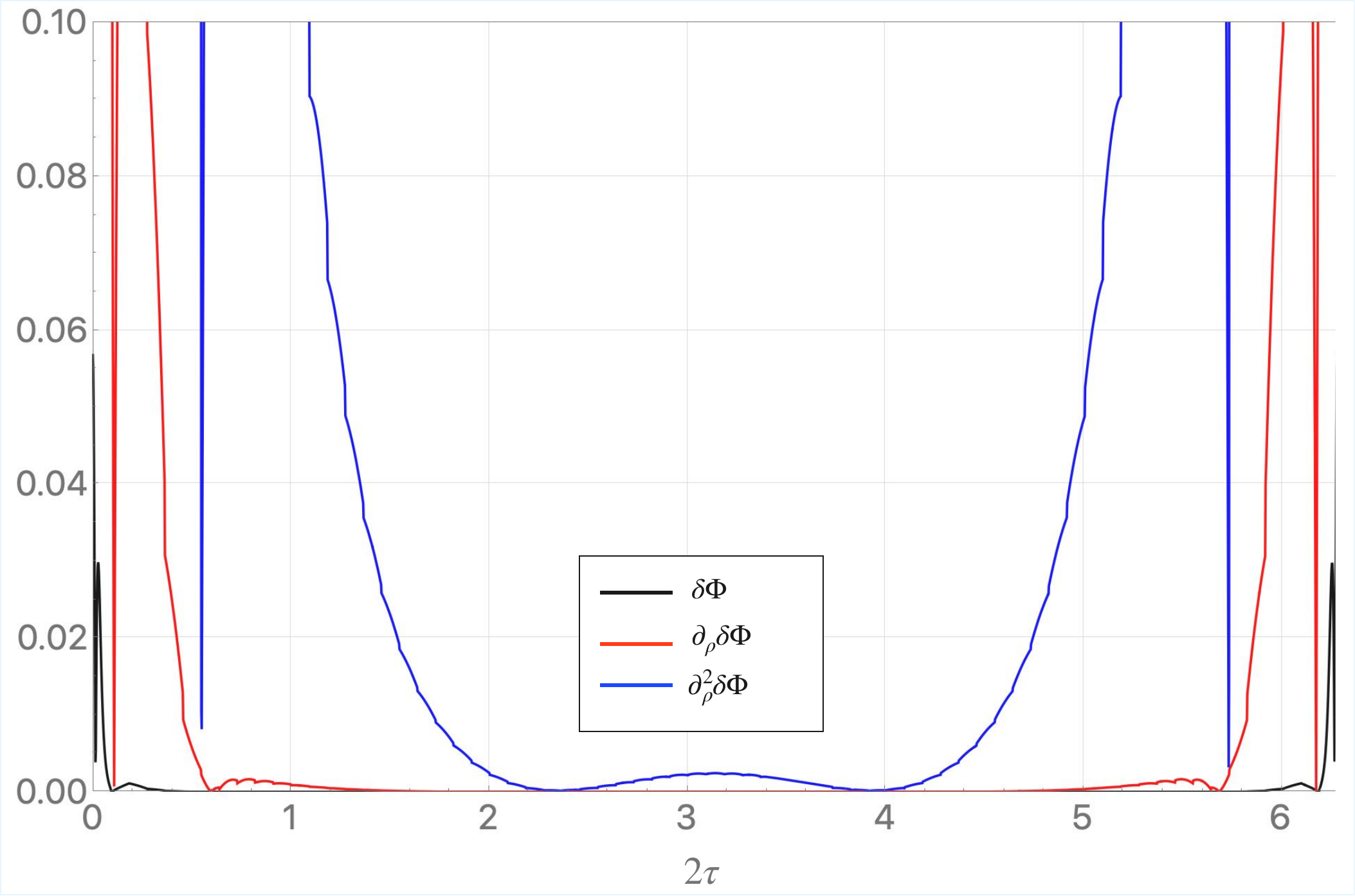}
    \end{minipage}
    \caption{\textbf{Left:} Numerical errors estimated by acting with the differential operator $(\Box-m^2)$ on the numerical solution, shown as a function of the variable $\rho = \sqrt{1-z}$. We are particularly interested in the region close to the boundary ($z \to 0$). 
    \textbf{Right:} Numerical errors estimated by computing $(\Box-m^2)\Phi$ (denoted $\delta \Phi$) and its first two derivatives with respect to $\rho = \sqrt{1-z}$, close to the boundary correlator and at small spatial separation ($x \sim 0.01$, $z \sim 0.01$).}
    \label{fig:errorsnum}
\end{figure}

\paragraph{The spherical black hole case.}

The same procedure can also be applied to the case of the spherical black hole. 
However, in this case $\tilde{f}_0$ must be given by~\eqref{eq:newtildef}, which expresses it as a function of the ratio $\beta/R$, and the metric must be modified accordingly to describe a spherical horizon. 
Moreover, the boundary conditions require an additional adjustment: for $d = 4$ and $\Delta_\phi = 3$, the boundary condition at $R$ must be replaced by the requirement of periodicity on the sphere. 
The resulting problem is a well-defined partial differential equation and can, in principle, be solved numerically. 
Nevertheless, the reality condition on $\tilde{f}_0$ constrains the ratio $\beta/R$ to be bounded from above. 
In addition, numerical analysis indicates that the solution is less accurate than in the planar case -- likely due to the more intricate structure of the differential equation and the subtleties of the boundary conditions -- which results in a poorer overall numerical convergence.

It would be interesting to obtain more precise numerical estimates in order to perform a quantitative comparison with the principal contributions to the full correlators \eqref{eq:CorrelatorFiniteV1}-\eqref{eq:CorrelatorFiniteV5}, and provide an independent confirmation of the results presented in this work.

\section{Details on Witten diagrams}
\label{app:DetailsOnWittenDiagrams}

In this appendix we gather details about the calculation of the Witten diagrams up to $\mathrm{O} (\veps^{2d})$ performed as support for the statements of Section \ref{subsec:BulkInterpretationInWittenDiagramsBB}. We start by defining the zero-temperature bulk-to-boundary and bulk-to-bulk propagators,
\begin{align}
    K^{(\text{AdS})}_\Delta (z_1, \tau_{12}, x_{12})
    &=
    \Cm_\Delta \left( \frac{z_1}{z_1^2 + \tau_{12}^2 + x_{12}^2} \right)^\Delta\,,
    \label{eq:AdSBulkToBoundaryPropagator} \\
    G^{(\text{AdS})}_\Delta (z_{12}, \tau_{12}, x_{12})
    &=
    2^\Delta \tilde{\Cm}_\Delta \xi_{12}^\Delta\, _2F_1 \left( \frac{\Delta}{2}, \frac{\Delta + 1}{2}; \Delta + 1 - \frac{d}{2}; \xi_{12}^2 \right)\,,
    \label{eq:AdSBulkToBulkPropagator}
\end{align}
as well as the normalization constants
\begin{align}
    \left. \Cm_\Delta \right|_{\Delta > d/2}
    &=
    \frac{\Gamma(\Delta)}{\pi^{d/2} \Gamma(\Delta-d/2)}\,,
    \label{eq:Cm} \\
    \Cm_{d/2}
    &= \frac{\Gamma(d/2)}{2 \pi^{d/2}}\,,
    \label{eq:CmSpecial} \\
    \tilde{\Cm}_\Delta
    &=
    \frac{\Gamma(\Delta) \Gamma(\Delta - d/2 - 1/2)}{(4 \pi)^{(d+1)/2} \Gamma(2\Delta -d + 1)}\,,
    \label{eq:tildeCm}
\end{align}
and the kinematic variable
\begin{equation}
    \xi_{12}
    =
    \frac{2 z_1 z_2}{z_1^2 + z_2^2 + \tau_{12}^2 + x_{12}^2}\,,
    \label{eq:xiDefinition}
\end{equation}
which satisfies $|\xi_{12}| \leq 1$. To perform explicit computations, it will be useful to use the Fefferman--Graham (FG) metric, which takes the form
\begin{equation}
    \text{d}s^2
    =
    \frac{\ell^2}{\rho^2}
    \left(
    \text{d}\rho^2 + (1+a)^{4/d-2} (1-a)^2 \text{d}\tau^2 + (1+a)^{4/d} \text{d}x^2
    \right)\,,
    \label{eq:FGMetric}
\end{equation}
where we have defined the variable
\begin{equation}
    \rho
    =
    z
    \left(
    \frac{2}{1+\sqrt{1-f_0 z^d}}
    \right)^{2/d}\,,
    \label{eq:rhoDefinition}
\end{equation}
as well as
\begin{equation}
    a
    =
    \frac{f_0}{4} \rho^d\,.
    \label{eq:aDefinition}
\end{equation}
In the following subsections, we present the details of the metric that appear in the expansion \eqref{eq:ExpandedMetric}, give the integral identities, and  compute the diagrams order by order in $\varepsilon^d$.

\paragraph{Metric identities.}
\label{subsec:MetricIdentities}

We present in this section useful results and identities for the expanded metric in the FG gauge.
In general, the determinant of the metric takes the following form:
\begin{equation}
    \sqrt{g}
    =
    \sqrt{g_\text{AdS}}
    \left[
    1 + \frac{1}{2} \veps^{2d} \left( h^{(2d)} - \frac{1}{2} h^{(d)}_{MN} h^{(d)\, MN} \right) + \ldots
    \right]\,,
    \label{eq:DetExpansion}
\end{equation}
where we have not used an explicit form for $h$ yet but anticipated that the first correction $h^{(d)}$ is traceless.
The determinant of the AdS metric is given by
\begin{equation}
    \sqrt{g_\text{AdS}}
    =
    \left( \frac{\ell}{\rho} \right)^{d+1}\,.
    \label{eq:DetAdS}
\end{equation}
The first-order correction to the AdS metric can be determined by expanding the metric at small temperature.
We find for the first-order correction and its inverse
\begin{align}
    h^{(d)\, MN} &= \left( \frac{\rho}{\ell} \right)^{d+2} \left( \frac{4 \pi}{d} \right)^d \frac{1}{d}\ \text{diag}\, \left( 0, -(d-1), 1, \ldots, 1) \right)^{MN}\,,
    \label{eq:MetricFirstOrder} \\
    h^{(d)}_{MN} &= \left( \frac{\rho}{\ell} \right)^{d-2} \left( \frac{4 \pi}{d} \right)^d \frac{1}{d}\ \text{diag}\, \left( 0, -(d-1), 1, \ldots, 1) \right)_{MN}\,.
    \label{eq:InverseMetricFirstOrder}
\end{align}
Note that, as mentioned above, it is manifest that
\begin{equation}
    h^{(d)} = 0\,.
    \label{eq:FirstOrderTrace}
\end{equation}
It is also useful for our purposes to calculate the product of two $h^{(d)}$:
\begin{equation}
    h^{(d)\, M}_{\phantom{(d)\, M}P}\, h^{(d)\, PN}
    =
    \left( \frac{\rho}{\ell} \right)^{2d+2} \left( \frac{4 \pi}{d} \right)^{2d} \frac{1}{d^2}\ \text{diag}\, \left( 0, (d-1)^2, 1, \ldots, 1 \right)^{MN}\,.
    \label{eq:ProductOfTwoh}
\end{equation}
The second-order correction is given by
\begin{align}
    h^{(2d)\, MN} &= \left( \frac{\rho}{\ell} \right)^{2d+2} \left( \frac{4 \pi}{d} \right)^{2d} \frac{1}{8d^2}\ \text{diag}\, \left( 0, 4(d^2+1)-9d , 4-d, \ldots, 4-d) \right)^{MN}\,,
    \label{eq:MetricSecondOrder} \\
    h^{(2d)}_{MN} &= \left( \frac{\rho}{\ell} \right)^{2d-2} \left( \frac{4 \pi}{d} \right)^{2d} \frac{1}{8d^2}\ \text{diag}\, \left( 0, 4(d^2+1)-9d , 4-d, \ldots, 4-d) \right)_{MN}\,.
    \label{eq:InverseMetricSecondOrder}
\end{align}
This order is not traceless:
\begin{equation}
    h^{(2d)} = \left( \frac{\rho}{\ell} \right)^{2d} \left( \frac{4 \pi}{d} \right)^{2d} \frac{3d-4}{8d}\,.
    \label{eq:SecondOrderTrace}
\end{equation}
Additional useful identities involving $h^{(2d)}$ are
\begin{align}
    \frac{1}{2} \left( h^{(2d)} - \frac{1}{2} h^{(d)}_{MN} h^{(d)\, MN} \right)
    &=
    - \frac{1}{16} \left( \frac{\rho}{\ell} \right)^{2d} \left( \frac{4 \pi}{d} \right)^{2d}\,,
    \label{eq:AdditionalIdentity1} \\
    h^{(d)\, M}_{\phantom{(d)\, M}P} h^{(d)\, PN} - h^{(2d)\, MN}
    &=
    \left( \frac{\rho}{\ell} \right)^{2d+2} \left( \frac{4 \pi}{d} \right)^{2d} \frac{1}{8d^2} \notag \\
    &\phantom{=\ } \times\
    \text{diag}\, \left( 0, 4d^2-7d+4, d+4, \ldots, d+4 \right)^{MN}\,.
    \label{eq:AdditionalIdentity2}
\end{align}

\subsection{Integrals}
\label{subsec:Integrals}

We now give the integrals useful for calculating the diagrams.
To compute the contact diagrams, we need
\begin{equation}
    I_k (\tau_{12}, x_{12})
    =
    \int_\beta \frac{\text{d}^{d+1}y_3}{\rho_3^{d+1}} \left( \frac{\rho_3}{\ell} \right)^k\,
    K^\text{(AdS)}_{\Delta_\phi} (\rho_3, \tau_{13}, x_{13})
    K^\text{(AdS)}_{\Delta_\phi} (\rho_3, \tau_{23}, x_{23})\,.
    \label{eq:FirstIntegralDefinition}
\end{equation}
This is a standard AdS integral \cite{Freedman:1998tz}.
It takes the kinematic form
\begin{equation}
    I_k (\tau_{12}, x_{12})
    =
    \frac{\alpha(d,\Delta_\phi,k)}{\ell^k (\tau_{12}^2 + x_{12}^2)^{(2\Delta_\phi-k)/2}}\,,
    \label{eq:FirstIntegralResult}
\end{equation}
where the coefficient is found to be
\begin{equation}
    \alpha(d,\Delta_\phi,k)
    =
    \frac{\Gamma(\Delta_\phi - k/2) \Gamma^2(k/2) \Gamma(\Delta_\phi + (k-d)/2)}{2 \pi^{d/2} \Gamma(k) \Gamma^2(\Delta_\phi - d/2)}\,.
    \label{eq:FirstIntegralCoefficient}
\end{equation}
In the exchange diagram we encounter a more general version of the same integral:
\begin{equation}
    I_k^{(\Delta_1, \Delta_2)}
    =
    \int_\beta \frac{\text{d}^{d+1}y_3}{\rho_3^{d+1}} \left( \frac{\rho_3}{\ell} \right)^k\,
    K^\text{(AdS)}_{\Delta_1} (\rho_3, \tau_{13}, x_{13})
    K^\text{(AdS)}_{\Delta_2} (\rho_3, \tau_{23}, x_{23})\,.
    \label{eq:IntermediateIntegralDefinition}
\end{equation}
This is also a well-known integral \cite{Freedman:1998tz}, for which the result is
\begin{equation}
    I^{(\Delta_1, \Delta_2)}_k (\tau_{12}, x_{12})
    =
    \frac{\alpha(d, \Delta_1, \Delta_2, k)}{\ell^k (\tau_{12}^2 + x_{12}^2)^{(2\Delta_\phi-k)/2}}\,.
    \label{eq:IntermediateIntegralResult}
\end{equation}
with
\begin{equation}
    \alpha(d, \Delta_1, \Delta_2, k)
    =
    \frac{\Gamma( \Delta_{123}/2 ) \Gamma( \Delta_{231}/2 ) \Gamma( \Delta_{312}/2 ) \Gamma((\Delta_1 + \Delta_2 + \Delta_3 - d)/2)}{2 \pi^{d/2} \Gamma(\Delta_1 - d/2) \Gamma(\Delta_2 - d/2) \Gamma(\Delta_3)}\,.
    \label{eq:IntermediateIntegralCoefficient}
\end{equation}
and
\begin{equation}
    \Delta_{ijk} = \Delta_i + \Delta_j - \Delta_k\,.
    \label{eq:DeltaijkDefinition}
\end{equation}

We now move our attention to the most challenging integral of this work, which is the exchange integral defined as
\begin{equation}
    \begin{split}
    J_k (\tau_{12}, x_{12})
    &=
    \int_\beta \frac{\text{d}^{d+1} y_3}{\rho_3^{d+1}} \left( \frac{\rho_3}{\ell} \right)^k
    K^\text{(AdS)}_{\Delta_\phi} (\rho_3, \tau_{13}, x_{13}) \\
    &\phantom{=\ } \times
    \int_\beta \frac{\text{d}^{d+1} y_4}{\rho_4^{d+1}} \left( \frac{\rho_4}{\ell} \right)^k
    K^\text{(AdS)}_{\Delta_\phi} (\rho_4, \tau_{24}, x_{24})
    G^\text{(AdS)}_{\Delta_\phi} (\rho_{34}, \tau_{34}, x_{34})\,.
    \end{split}
    \label{eq:SecondIntegralDefinition}
\end{equation}
It can be seen to take the kinematic form
\begin{equation}
    J_k (\tau_{12}, x_{12})
    =
    \frac{\beta(d, \Delta_\phi, k)}{\ell^{2k} (\tau_{12}^2 + x_{12}^2)^{\Delta_\phi-k}}\,.
    \label{eq:SecondIntegralResult}
\end{equation}
To determine the coefficient $\beta(d, \Delta_\phi, k)$, we use the split representation of the bulk-to-bulk propagator \cite{DHoker:1999mqo,Costa:2014kfa}:
\begin{equation}
    \begin{split}
    G^\text{(AdS)}_{\Delta_\phi} (\rho_{34}, \tau_{34}, x_{34})
    &=
    \frac{1}{4\pi} 
    \int_{-\infty}^\infty \frac{\text{d}\nu}{\nu^2 + (\Delta - d/2)^2} \\
    &\phantom{=\ } \times
    \int_\beta \text{d}^d x_0\,
    K^\text{(AdS)}_{d/2 + i\nu} (\rho_3, \tau_{03}, x_{03})
    K^\text{(AdS)}_{d/2 - i \nu} (\rho_4, \tau_{04}, x_{04})\,.
    \end{split}
    \label{eq:SplitRepresentation}
\end{equation}
We can then use \eqref{eq:IntermediateIntegralResult}.
The next step is to solve the standard massless Feynman integral
\begin{equation}
    \begin{split}
    \int \frac{\text{d}^d x_0}{(\tau_{03}^2 + x_{03}^2)^A (\tau_{03}^2 + x_{03}^2)^B}
    &=
    \frac{\pi^{d/2} \Gamma(A+B-d/2) \Gamma(d/2-A) \Gamma(d/2-B)}{\Gamma(A) \Gamma(B) \Gamma(d - A - B)} \\
    &\phantom{=\ } \times \frac{1}{(\tau_{12}^2 + x_{12}^2)^{A+B-d/2}}\,.
    \end{split}
    \label{eq:StandardFeynmanIntegral}
\end{equation}
Specializing the calculation to our case of interest ($d=4$, $k=d+2=6$), we find
\begin{equation}
    \begin{split}
    \beta(4,\Delta_\phi,6)
    &=
    \frac{\Gamma(\Delta_\phi-6)}{460800 \pi^2 \Gamma(8 - \Delta_\phi) \Gamma^2(\Delta_\phi - 2)}
    \int_{-\infty}^\infty \frac{\text{d}\nu}{\nu^2 + (\Delta_\phi - 2)^2} \frac{\nu (\nu^2 + (\Delta_\phi - 6)^2) \sinh (\pi \nu)}{\cosh(\pi \nu) - \cos (\pi \Delta_\phi)} \\
    &\phantom{=\ } \times
    \frac{\Gamma(4 - (\Delta_\phi + i\nu)/2) \Gamma(4 - (\Delta_\phi - i\nu)/2) \Gamma^2 ((\Delta_\phi + i\nu + 4)/2) \Gamma^2 ((\Delta_\phi - i\nu + 4)/2)}{\Gamma ((\Delta_\phi + i\nu - 4)/2) \Gamma ((\Delta_\phi - i\nu - 4)/2)}\,.
    \end{split}
    \label{eq:BetaAsAnIntegral}
\end{equation}
We could not solve the remaining one-dimensional integral analytically, but it is observed to agree numerically with the function
\begin{equation}
    \beta(4,\Delta_\phi,6)
    =
    \frac{\Delta_\phi (\Delta_\phi^3 (7 \Delta_\phi - 15) + 8)}{100800 \pi^2 (\Delta_\phi-3) (\Delta_\phi-4) (\Delta_\phi-5) (\Delta_\phi-6)}\,.
    \label{eq:SecondIntegralCoefficient}
\end{equation}
We tested this result for a wide range of values of $\Delta_\phi$ and did not observe a discrepancy.

\subsection{Calculation of diagrams}
\label{subsec:CalculationOfDiagrams}

We are now ready to calculate the diagrams presented in~\eqref{eq:WittenDiagrams}.

\subsubsection{Order $\mathrm{O}(\veps^d)$}
\label{subsubsec:FirstOrder}

The first-order correction consists of the single contact diagram
\begin{equation*}
    \WittenDiagramTwo\ .
\end{equation*}
After performing the Wick contractions, the diagram is given by
\begin{equation}
    \begin{split}
        \WittenDiagramTwo\
        &=
        \frac{\ell^2}{d} \left( \frac{4\pi}{d} \right)^d ((d-1) \pd_{\tau_1} \pd_{\tau_2} - \pd_{x_1} \cdot \pd_{x_2})
        \sum_{m_1 = -\infty}^\infty \sum_{m_2 = -\infty}^\infty \int_0^{\rho_3} \left( \frac{\rho_3}{\ell} \right)^{d+2} \frac{\text{d}\rho_3}{\rho_3^{d+1}} \\
        &\phantom{=\ } \times
        \int_0^\beta \text{d}\tau_3 \int \text{d}^{d-1} x_3\,
        K_{\Delta_\phi} (\rho_3, \tau_{13} + m_1\beta, x_{13}) K_{\Delta_\phi} (\rho_3, \tau_{23}+ m_2\beta, x_{23})\,,
    \end{split}
\end{equation}
where we have defined $\tau = |\tau_{12}|$ and $x = x_{12}$.
Here and thereafter we drop the subscript for the propagators since they are all meant to refer to~\eqref{eq:AdSBulkToBoundaryPropagator}-\eqref{eq:AdSBulkToBulkPropagator}.
One can trade one sum and the $\tau$-integral from $0$ to $\beta$ for a $\tau$-integral from $-\infty$ to $+\infty$.
To see that, shift $\tau_3 \to \tau_3 + m_2 \beta$, absorb $m_2$ in $m_1$, and perform the sum over $m_2$.
We then have
\begin{equation}
    \begin{split}
        \WittenDiagramTwo\
        &=
        \frac{\ell^{1-d}}{d} \left( \frac{4\pi}{d} \right)^d ((d-1) \pd_{\tau_1} \pd_{\tau_2} - \pd_{x_1} \cdot \pd_{x_2}) \\
        &\phantom{=\ } \times \sum_{m = -\infty}^\infty
        \int \text{d}^{d+1} y_3\, \frac{\rho_3}{\ell}
        K_{\Delta_\phi} (\rho_3, \tau_{13} + m\beta, x_{13}) K_{\Delta_\phi} (\rho_3, \tau_{23}, x_{23})\,.
    \end{split}
\end{equation}
We finally obtain
\begin{equation}
    \WittenDiagramTwo\ =
    - \frac{\ell^2}{d} \left( \frac{4 \pi}{d} \right)^d \sum_{m=-\infty}^\infty \left( (d-1) \pd_\tau^2 - \pd_x^2 \right) I_{d+2} (\tau + m \beta, x)\,,
\end{equation}
which exactly matches the form~\eqref{eq:GMIFromWittenDiagrams} for $n=1$ when spelled out explicitly using~\eqref{eq:FirstIntegralDefinition}--\eqref{eq:FirstIntegralCoefficient}.

\subsubsection{Order $\mathrm{O}(\veps^{2d})$}
\label{subsubsec:SecondOrder}

At second order, the correlator~\eqref{eq:WittenDiagrams} consists of two diagrams: one contact and one exchange.

\paragraph{Contact diagram.}
We first compute the contact diagram
\begin{equation}
    \WittenDiagramThree\ = \Lambda_1 + \Lambda_2\,,
\end{equation}
which we have split for convenience into two terms corresponding to the trace and non-trace parts of~\eqref{eq:VertexTwo}.

We now calculate $\Lambda_1$.
Performing the Wick contractions and turning one sum into a zero-temperature integral as for the first-order contact diagram, we obtain
\begin{equation}
    \begin{split}
        \Lambda_1
        &=
        - \frac{\ell^{1-d}}{16} \left( \frac{4\pi}{d} \right)^{2d} \sum_{m=-\infty}^\infty
        \int \text{d}^{d+1} y_3\, \left( \frac{\rho_3}{\ell} \right)^{d-1} \\
        &\phantom{=\ } \times
        \left\lbrace
        g_\text{AdS}^{MN} \pd_M K_{\Delta_\phi} (\rho_3, \tau_{13} + m\beta, x_{13}) \pd_N K_{\Delta_\phi} (\rho_3, \tau_{23}, x_{23}) \right. \\
        &\phantom{=\ } \left.
        +
        m^2 K_{\Delta_\phi} (\rho_3, \tau_{13} + m\beta, x_{13}) K_{\Delta_\phi} (\rho_3, \tau_{23}, x_{23}) 
        \right\rbrace\,.
    \end{split}
\end{equation}
We can use the equation of motion for $K_{\Delta_\phi}$ to eliminate the mass term via
\begin{equation}
    \begin{split}
    g_\text{AdS}^{MN} \pd_M K_{\Delta_\phi} (1,3) \pd_N K_{\Delta_\phi} (2,3) 
    &=
    g_\text{AdS}^{MN} \nabla_{3\, M} \nabla_{3\, N} (K_{\Delta_\phi} (1,3) K_{\Delta_\phi} (2,3)) \\
    &\phantom{=\ }
    - m^2 K_{\Delta_\phi} (1,3) K_{\Delta_\phi} (2,3)\,,
    \end{split}
\end{equation}
where we used the shorthand notation
\begin{equation}
    K_{\Delta_\phi} (1,3)
    =
    K_{\Delta_\phi} (\rho_3, \tau_{13}, x_{13})\,.
\end{equation}
Using integration by parts, we end up with
\begin{equation}
    \begin{split}
        \Lambda_1
        &=
        - \frac{d^2 \ell^{-d-1}}{32} \left( \frac{4\pi}{d} \right)^{2d} \sum_{m=-\infty}^\infty
        \int \text{d}^{d+1} y_3\, \left( \frac{\rho_3}{\ell} \right)^{d-1}
        K_{\Delta_\phi} (\rho_3, \tau_{13} + m\beta, x_{13}) K_{\Delta_\phi} (\rho_3, \tau_{23}, x_{23})\,, 
    \end{split}
\end{equation}
which can be recast as
\begin{equation}
    \Lambda_1 =
    - \frac{d^2}{16} \left( \frac{4\pi}{d} \right)^{2d}
    \sum_{m=-\infty}^\infty I_{2d} (\tau + m\beta, x)\,.
    \label{eq:Lambda1Result}
\end{equation}

The calculation of $\Lambda_2$ is similar to the order $\mathrm{O}(\veps^d)$ presented above, with different expressions for the metric.
We find
\begin{equation}
    \Lambda_2
    =
    \frac{\ell^2}{8d^2} \left( \frac{4\pi}{d} \right)^{2d}
    \sum_{m=-\infty}^\infty \left( -(4d^2-7d+4) \pd_\tau^2 - (d+4) \pd_x^2 \right) I_{2d+2} (\tau+m\beta, x)\,.
    \label{eq:Lambda2Result}
\end{equation}

\paragraph{Exchange diagram.}
We now compute the exchange diagram
\begin{equation}
    \WittenDiagramFour\,.
\end{equation}
After performing the Wick contractions, we have
\begin{equation}
    \begin{split}
        \WittenDiagramFour\ &=
        \ell^4 \sum_{m_i} \int_0^\infty \frac{d\rho_3}{\rho_3^{d+1}} \int_0^\infty \frac{d\rho_4}{\rho_4^{d+1}} h_{3 \mu \nu}^{(d)} h_{4 \rho \sigma}^{(d)} \\
        &\phantom{=\ } \times
        \int_0^\beta d\tau_3 \int d^{d-1}x_3 \int_0^\beta d\tau_4 \int d^{d-1}x_4\
        \pd_3^\mu K_{\Delta_\phi} (\rho_3, \tau_{13} + m_1 \beta, x_{13}) \\
        &\phantom{=\ } \times
        \pd_4^\sigma K_{\Delta_\phi} (\rho_4, \tau_{24} + m_2 \beta, x_{24}) \pd_3^\mu \pd_4^\rho G_{\Delta_\phi} (\rho_{34}, \tau_{34} + m_3 \beta, x_{34})\,.
    \end{split}
\end{equation}
Using similar tricks as for the other integrals, this can be rewritten as
\begin{equation}
    \WittenDiagramFour\ =
    \frac{\ell^4}{d^2} \left( \frac{4\pi}{d} \right)^{2d}
    \sum_{m=-\infty}^\infty \left( -(d-1) \pd_\tau^2 + \pd_x^2 \right)^2 J_{d+2} (\tau+m\beta,x)\,,
    \label{eq:ExchangeResult}
\end{equation}
with $J_k(\tau,x)$ defined in~\eqref{eq:SecondIntegralDefinition} and solved in~\eqref{eq:SecondIntegralCoefficient} for our case of interest.
Remarkably, summing~\eqref{eq:Lambda1Result},~\eqref{eq:Lambda2Result}, and~\eqref{eq:ExchangeResult} we obtain the second order version of Equation~\eqref{eq:GMIFromWittenDiagrams}, i.e., for $n=2$.

\bibliography{Draft.bib}
\bibliographystyle{./auxi/JHEP}

\end{document}